\documentclass[a4paper,11pt]{article}
\pdfoutput=1 

\usepackage{jheppub} 

\usepackage[T1]{fontenc} 
\usepackage{amsthm}
\usepackage{bm}
\usepackage{physics}
\usepackage{mathtools}

\usepackage{feynmp-auto}
\definecolor{myGreen}{rgb}{0.196, 0.603, 0.298}

\newcommand{\ra}{\rightarrow}
\newcommand{\Cov}{\text{Cov}}

\newcommand{\reffig}[1]{Fig.~\ref{#1}}
\newcommand{\refsec}[1]{Sec.~\ref{#1}}
\newcommand{\refapp}[1]{App.~\ref{#1}}
\newcommand{\eqrefeq}[1]{Eq.~\eqref{#1}}

\arxivnumber{2305.14092} 

\title{\boldmath Approaches to inclusive semileptonic $B_{(s)}$-meson decays from Lattice QCD}

\author[a,b,c]{Alessandro Barone,}
\author[c,d]{Shoji Hashimoto,}
\author[a,b,e]{Andreas J\"uttner,}
\author[c,d,f]{Takashi Kaneko,}
\author[c,d]{Ryan Kellermann}

\affiliation[a]{School of Physics and Astronomy, University of Southampton, Southampton SO17 1BJ, UK}
\affiliation[b]{STAG Research Center, University of Southampton, Southampton SO17 1BJ, UK}
\affiliation[c]{High Energy Accelerator Research Organization (KEK), Ibaraki 305-0801, Japan}
\affiliation[d]{School of High Energy Accelerator Science, SOKENDAI (The Graduate University for Advanced Studies), Ibaraki 305-0801, Japan}
\affiliation[e]{CERN, Theoretical Physics Department, Geneva, Switzerland}
\affiliation[f]{Kobayashi-Maskawa Institute for the Origin of Particles and the Universe, Nagoya University, Aichi 464–8602, Japan}

\emailAdd{a.barone@soton.ac.uk}
\emailAdd{shoji.hashimoto@kek.jp}
\emailAdd{andreas.juttner@cern.ch}
\emailAdd{takashi.kaneko@kek.jp}
\emailAdd{kelry@post.kek.jp}

\abstract{
We address the nonperturbative calculation of the inclusive decay rate of semileptonic 
$B_{(s)}$-meson decays from lattice QCD. Precise Standard-Model predictions are key
ingredients in searches for new physics, and this type of computation may eventually provide new insight
into the long-standing tension between the inclusive and exclusive determinations of the Cabibbo-Kobayashi-Maskawa (CKM)
matrix elements $|V_{cb}|$ and $|V_{ub}|$. We present results from a pilot lattice computation for $B_s \rightarrow X_c\, l \nu_l$,
where the initial $b$ quark described by the
relativistic-heavy-quark (RHQ) formalism on the lattice and the other valence quarks discretised
with domain-wall fermions are simulated approximately
at their physical quark masses.
We compare two different methods for 
computing the decay rate from lattice data of Euclidean $n$-point functions, namely Chebyshev and Backus-Gilbert approaches.
We further study how much the
ground-state meson dominates the inclusive decay rate and indicate 
our strategy towards a computation with a more comprehensive 
systematic error budget.
}

\begin{document} 

\begin{flushright}
 KEK-CP-0394\hspace{1em}CERN-TH-2023-087
\end{flushright}

\maketitle
\flushbottom

\section{Introduction}
\label{sec:intro}

The study of the $b$-quark sector of particle physics remains an exciting arena of precision physics, in which intriguing
tensions between observations and Standard-Model (SM) predictions 
have been found~\cite{LHCb:2013ghj, LHCb:2014vgu, BaBar:2012obs, BaBar:2013mob,
  Belle:2015qfa, LHCb:2015gmp}. Scrutinising these findings and better controlling and reducing experimental and theoretical error budgets therefore remain a crucial task.
Any such anomaly could be an indicator of new effects: while new particles
may be too heavy to be produced with energies achievable by current experimental facilities, quantum effects could leave detectable traces in flavour-physics processes.
One of these long-standing tensions involves the measured values of the CKM matrix elements $|V_{cb}|$ and $|V_{ub}|$ between exclusive and inclusive decays. 
Apart from leptonic decays, these can be determined through the exclusive semileptonic decay of a $B$  into a $D^{(*)}$ (or $\pi$), or through the measurement of the 
inclusive decay rate, respectively. For example, one of the most recent determination of $|V_{cb}|$ finds
\begin{align*}
 |V_{cb}| &= (42.19\pm 0.78) \times 10^{-3} \quad \text{inclusive\,\,\cite{HeavyFlavorAveragingGroup:2022wzx,Gambino2016}} ,\\
 |V_{cb}| &= (39.36\pm 0.68) \times 10^{-3} \quad \text{exclusive\,\,\cite{Aubert:2009ac,Glattauer:2015teq,MILC:2015uhg,Na:2015kha,Aoki2021}}\, .
\end{align*}
Lattice computations provide crucial nonperturbative input to the exclusive determination and the required techniques in this case are well established (see reviews~\cite{Aoki2021,Kaneko:2022eev}). The
existing results for the inclusive decay are based on perturbative QCD. First viable theoretical proposals for how to accomplish the computation of the inclusive decay rate on the lattice have appeared only recently~\cite{Hashimoto2017}. The idea relies
on the extraction of a forward-scattering matrix element through analytic continuation of lattice results obtained in an unphysical kinematical region.
In~\cite{Hansen2017} it was then proposed to address decay and transition rates of multi-hadron processes through
finite-volume Euclidean four-point functions provided that a method to extract the associated spectral function exists.

In this paper, we present work towards an improved understanding of the calculation of the inclusive decay rate by means of a pilot study of semileptonic decays of $B_s$ mesons into charmed particles, namely $B_{s}\rightarrow X_{c}\,l\nu_{l}$, following~\cite{Gambino2020}, where the extraction of the spectral function is bypassed and the decay rate is evaluated directly.
Preliminary work has been presented in~\cite{Barone:2022gkn,Kellermann:2022mms}.
In particular, we improve and compare two existing methods, namely 
Chebyshev~\cite{Barata:1990rn,Bailas2020a,Gambino2020} and
Backus-Gilbert~\cite{Hansen2019,Gambino:2022dvu} reconstructions.
Our work uses the relativistic-heavy-quark action (RHQ)~\cite{RHQFermilab,RHQColumbia1,RHQColumbia2} to simulate the bottom-valence quark at its physical mass, while the strange- and charm-valence quarks are treated with a domain-wall fermion action~\cite{Shamir1993,Furman1994,Brower2017,Cho2015}, and their masses are tuned to values close to
the ones found in nature.

The structure of this paper is as follows: in \refsec{sec:theory} we describe the theoretical framework, extending the formalism introduced in
\cite{Gambino2020}.
We also address the ground-state limit and its connection with the corresponding exclusive processes. In \refsec{sec:inclusiveLattice} we describe
some details of the lattice implementation.
In \refsec{sec:analysis} we report on our analysis strategies; to keep the discussion fluent we refer to~\refapp{sec:app_cheb}, \ref{sec:app_bg} and \ref{sec:app_fit} for technical details. Finally, we discuss the details of the
simulation in \refsec{sec:setup}  and present our results in \refsec{sec:results}. We summarise our findings and discuss future prospects in \refsec{sec:concl}.

\section{Theoretical framework}
\label{sec:theory}

\subsection{The inclusive decay rate}
\label{sec:Theoretical foundations}

\begin{figure}[t]
 \centering
 \begin{fmffile}{inclusive}
  \begin{fmfgraph*}(180, 80)
    \fmfset{arrow_len}{10}
    \fmfstraight
    \fmfleft{i4,i3,i2,i1}
    \fmfright{o4,o3,o2,o1}
    \fmffreeze
    \fmf{fermion}{i1,o1}
    \fmf{fermion}{o2,v2,i2}
    \fmffreeze
    \fmf{fermion,tension=1.5}{o3,v4,o4}
    \fmf{phantom,tension=1.8}{i4,v4}
    \fmflabel{$s$}{i1}
    \fmflabel{$\bar{b}$}{i2}
    \fmflabel{$\bar{c}$}{o2}
    \fmflabel{$s$}{o1}
    \fmflabel{$l^{+}$}{o3}
    \fmflabel{$\nu_l$}{o4}
    \fmf{boson,label=$W^{+}$,label.side=left}{v4,v2}
    \fmfipair{B,X}
    \fmfiequ{B}{(-.2w,.55h)}
    \fmfiequ{X}{(1.2w,.55h)}
    \fmfiv{l=$B_s\hspace{1pt}\Bigg\{$, l.a=90}{B}
    \fmfiv{l=$\Bigg\}\hspace{1pt}X_c$, l.a=90}{X}
  \end{fmfgraph*}
 \end{fmffile}
 \caption{Feynman diagram for $B_s \ra X_c \, l\nu_l$.}
 \label{fig:BtoXdiag}
\end{figure}
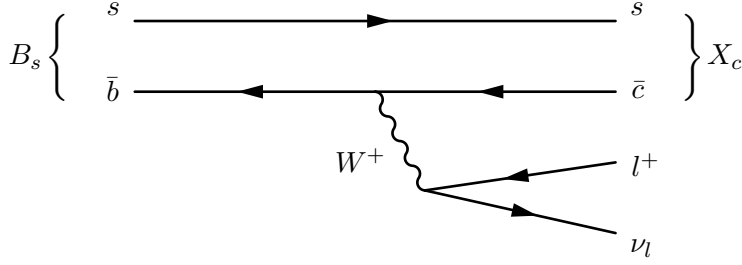

We start by reviewing the formalism to calculate the decay rate of inclusive semileptonic processes \cite{ManoharWise,Blok:1993va}.
Here, we focus on the decay $B_s \rightarrow X_c\, l \nu_l$ illustrated in \reffig{fig:BtoXdiag},
but the formalism is more generally applicable to other channels such as, e.g., $B \rightarrow X l\nu_l$ or  $D_{(s)} \rightarrow X l \nu_l$.
The final state $X_c$ represents all possible charmed-meson final states allowed by flavour, spin and parity quantum numbers. The ground-state contribution to $X_c$ in the vector channel is given by the $D_s$ meson.
The leading order weak Hamiltonian for the $\bar{b}\rightarrow \bar{c}$ process is given by
\begin{align}
 H_{W} = \frac{4G_F}{\sqrt{2}} V_{cb} \left[ \bar{b}_L \gamma^{\mu} c_L \right] \left[ \bar{\nu}_{lL}\gamma_{\mu} l_L \right] \, ,
\end{align}
where $G_F$ is the Fermi constant and $V_{cb}$ is the CKM matrix element for the charged-current flavour-changing quark transition. 
The electroweak quark current for this process is then $J_{\mu} =  \bar{b}_L \gamma^{\mu} c_L = \bar{b} \gamma_{\mu}(1-\gamma_5) c$, which we can also write as
$J_{\mu} = V_{\mu}-A_{\mu}$ with $V_\mu = \bar{b} \gamma_{\mu} c$ and $A_\mu = \bar{b} \gamma_{\mu}\gamma_5 c$.

The differential decay rate for the inclusive process depends on
three kinematical variables, i.e. one more than the corresponding exclusive decay
due to the freedom in the mass of the outgoing hadrons.
Neglecting QED corrections it reads
\begin{align}
 \frac{\dd \Gamma}{\dd q^2 \dd q_0 \dd E_{l}} = \frac{G^2_F |V_{cb}|^2}{8\pi^3} L_{\mu\nu} W^{\mu\nu} \, .
\end{align}
The lepton contribution is given in terms of the leptonic tensor
\begin{align}
 L^{\mu\nu} = p_{l}^{\mu}p_{\nu_l}^{\nu} +  p_{l}^{\nu}p_{\nu_l}^{\mu} - g^{\mu\nu} p_{l}\cdot p_{\nu_l} -i\epsilon^{\mu \alpha \nu \beta} p_{l\,,\alpha }p_{\nu_l\,,\beta} \, ,
\end{align}
where $p_l$ and $p_{\nu_l}$ are the four-momenta of the lepton and the neutrino, respectively.
The hadronic tensor $W^{\mu\nu}$ is defined as
\begin{align}
 \begin{split}
 W^{\mu\nu}(p_{B_s}, q) = &  \frac{1}{2E_{B_s}} \int \dd^4 x \, e^{iq\cdot x} \bra{B_s(\bm{p}_{B_s})} J^{\mu\dagger}(x) J^{\nu}(0) \ket{B_s(\bm{p}_{B_s})} \\
 =&\frac{1}{2E_{B_s}}\sum_{X_c}(2\pi)^3 
  \delta^{(4)}(p_{B_s}-q-p_{X_c})  \\
 &\times \bra{B_s(\bm{p}_{B_s})} J^{\mu\dagger}(0)\ket{X_{c}(\bm{p}_{X_c})}  \bra{X_{c}(\bm{p}_{X_c})} J^{\nu}(0) \ket{B_s(\bm{p}_{B_s})} \, ,
 \end{split}
\label{eq:hadronicTensor}
\end{align}
where in the second line we have inserted the sum $\sum_{X_c} \ket{X_{c}(\bm{p}_{X_c})}  \bra{X_{c}(\bm{p}_{X_c})}$ over a complete set of states,
which is understood to include an integration over all possible momenta $\bm{p}_{X_c}$ under a Lorentz invariant phase-space integral, and 
$q=p_{B_s}-p_{X_c}=p_l+p_{\nu_l}$ is the transferred momentum between the initial and final hadronic states. Note that we will consider only the case of the 
 $B_s$ meson at rest, i.e. $\bm{p}_{B_s}=(0,0,0)$,
 and will henceforth suppress the corresponding momentum label.
The hadronic tensor can be decomposed into five scalar structure functions $W_i\equiv W_i(q^2, v\cdot q )$ as 
\begin{align}
 W^{\mu\nu} = -g^{\mu\nu} W_1 + v^{\mu}v^{\nu}W_2 - i\epsilon^{\mu\nu\alpha\beta}v_{\alpha}q_{\beta}W_3 + q^{\mu}q^{\nu}W_4 + (v^{\mu}q^{\nu}+v^{\nu}q^{\mu})W_5 \, ,
\end{align}
where $v = p_{B_s}/M_{B_s}=(1,0,0,0)$ is the velocity of the initial $B_s$ meson at rest,
and $q=(q_0, \bm{q})=(M_{B_s}-\omega, -\bm{p}_{X_c})$. From now on, we will indicate with $\omega=E_{X_c}$ the energy of the final-state hadron.
The individual components of the hadronic tensor can be expressed conveniently in terms of the structure functions,
\begin{align}
\label{eq:W00}
W_{00} &= - W_1 + W_2 +q_0^2W_4 + 2q_0 W_5 \, ,\\
\label{eq:Wij}
W_{ij} &= \delta_{ij}W_1 + q_iq_j W_4 -i \epsilon_{ij0k}q^{k}W_3\, ,\\
\label{eq:W0i}
W_{0i} &= W_{i0} = q_i (q_0W_4+W_5) \, ,
\end{align}
where $i,j,k$ refers to the spatial indices $1,2,3$.
We note that contracting the spatial indices with the three-momentum components $q_i$, we can invert these relations and find expressions 
for the structure functions in terms of the hadronic tensor and $\bm{q}$.

Integrating over the lepton energy $E_l=p_{l,0}$ and assuming $m_{l}= 0$ we obtain the expression for the decay rate
\begin{align}\label{eq:Gamma_Xbar}
 \Gamma &= \frac{G_F^2 |V_{cb}|^2}{24\pi^3} \int_{0}^{\bm{q}^2_{\rm max}} \dd \bm{q}^2 \, \sqrt{\bm{q}^2} \bar{X}(\bm{q}^2) \, , 
\end{align}
where the integration over $\omega$ is contained in 
%
\begin{align}\label{eq:Xbar def}
 \bar{X}(\bm{q}^2) = \sum_{l=0}^{2}\bar{X}^{(l)}(\bm{q}^2) \, , \qquad
 \bar{X}^{(l)}(\bm{q}^2) \equiv \int_{\omega_{\rm min}}^{\omega_{\rm max}} \dd \omega X^{(l)}(\bm{q}^2) \, ,
\end{align}
and where we defined
\begin{align}
 \begin{split}
 X^{(0)}(\bm{q}^2)&= \bm{q}^2 W_{00} + \sum_i (q_i^2-\bm{q}^2) W_{ii} + \sum_{i\neq j}q^{i}W_{ij}q^{j} \, , \\
 X^{(1)}(\bm{q}^2) &= -q_0 \sum_i q^{i} (W_{0i}+W_{i0})  \, ,\\
 X^{(2)}(\bm{q}^2) &= q_0^2 \sum_i W_{ii} \, .
 \end{split}
 \label{eq:X(l)}
\end{align}
Recalling that the $D_{s}$ meson is the lightest final state in this inclusive decay process, and imposing four-momentum conservation we obtain
$\bm{q}^{2}_{\rm max} = \left(M_{B_s}^2-M_{D_s}^2\right)^2/(4M_{B_s}^2)$, 
$\omega_{\rm min}=\sqrt{M_{D_s}^{2}+\bm{q}^2}$ and $\omega_{\rm max}=M_{B_s}-\sqrt{\bm{q}^2}$
for the integral limits.
$X^{(l)}$ and $\bar{X}^{(l)}$ depend only on $\bm{q}^2$ and not on individual components of $\bm{q}$, as can be seen after substituting Eqs.~\eqref{eq:W00}, \eqref{eq:Wij} and \eqref{eq:W0i} into the expressions \eqref{eq:X(l)}. 

Starting from the decomposition of the hadronic tensor  $W^{\mu\nu} = W^{\mu\nu}_{VV} + W^{\mu\nu}_{AA} - W^{\mu\nu}_{VA}-W^{\mu\nu}_{AV}$, the $X^{(l)}$ can also be rewritten in a way that exposes the $V-A$ nature of the charged current, namely  
\begin{equation}\label{eq:X V-A decomposition}
    X^{(l)} = X^{(l)}_{VV}+X^{(l)}_{AA}-X^{(l)}_{VA}-X^{(l)}_{AV}\,,
\end{equation} and similarly for $\bar{X}^{(l)}$.

\subsection{Ground-state limit}
\label{sec:theory-ground}

In this section we consider a hypothetical world in which  only the lowest-mass final state
 $D_s$ contributes to  the inclusive decay, i.e.,
\begin{align}
 W_{\mu\nu} \, \rightarrow \,  \delta(\omega-E_{D_s})\frac{1}{4E_{B_s}E_{D_s}} \bra{B_s(\bm{p}_{B_s})}  V^{\dagger}_{\mu} \ket{D_s(\bm{p}_{D_s})} \bra{D_s(\bm{p}_{D_s})} V_{\nu} \ket{B_s(\bm{p}_{B_s})} \, .
 \label{eq:Wmunu_ground}
\end{align}
In this limit we can reconstruct the inclusive decay rate from lattice simulations of the exclusive decay, allowing us to compute the ground-state contribution.
We will also use results in this limit to devise  consistency checks of the inclusive-decay setup.
The required hadronic form factors $f_+(q^2)$ and $f_-(q^2)$ parametrising the corresponding matrix element 
\begin{align}
\bra{D_s(\bm{p}_{D_s})} V_{\mu} \ket{B_s(\bm{p}_{B_s})} = 
  f_{+}(q^2) (p_{B_s}+p_{D_s})_{\mu} + 
  f_{-}(q^2) (p_{B_s}-p_{D_s})_{\mu} \, ,
 \label{eq:matrix_f+f-}
\end{align}
of the exclusive decay $B_s\to D_s \, l\nu_l$
can be computed separately on the lattice using more conventional methods~\cite{McLean:2019qcx, Blossier:2021xvl,Flynn2021}.

In order to compute the inclusive decay rate in this limit we now establish the relation 
between the vector form factor $f_+(q^2)$ and $\bar{X}_{VV}=\sum_{l=0}^{2}\bar{X}_{VV}^{(l)}$ defined 
in Eqs.~\eqref{eq:Xbar def} and (\ref{eq:X(l)}) using the decomposition in Eq.~\eqref{eq:X V-A decomposition}.
Let us first decompose $\bar{X}_{VV} = \bar{X}_{VV}^{\parallel}+\bar{X}_{VV}^{\perp}$ into
longitudinal and transverse components in terms of the projectors $\Pi^{\perp}_{\mu\nu}=g^{\mu\nu}-q^{\mu}q^{\nu}/q^2$ and
$\Pi^{\parallel}_{\mu\nu}=q^{\mu}q^{\nu}/q^2$,
 where
\begin{align}
 \begin{split}
 X_{VV}^{\parallel} &= q^2W_1 + \bm{q}^2 W_2 \, , \\
 X_{VV}^{\perp} & = 2 q^2 W_1 \, ,
 \end{split}
\end{align}
which, inverting~\eqrefeq{eq:W00}-\eqref{eq:W0i} and considering $\bm{q}^2\neq 0$, can be expanded as
\begin{align}
 X_{VV}^{\parallel} &= \bm{q}^2W^{00}_{VV}-q_0 \sum_{i}q_i (W^{0i}_{VV}+W^{i0}_{VV}) + \frac{q_0^2}{\bm{q}^2} \sum_{i,j} q_i W^{ij}_{VV} q_j \label{eq:Xbar_VV^||}\, , \\
 X_{VV}^{\perp} &= (q_0^2 - \bm{q^2})\sum_{i}W^{ii}_{VV} + \sum_{i,j}\left(1-\frac{q_0^2}{\bm{q}^2} \right) q_i W^{ij}_{VV}q_j \\ \notag
                &= \sum_{i} \left(1-\frac{q_0^2}{\bm{q}^2}\right) \left(q_i^2 - \bm{q}^2 \right)  W^{ii}_{VV} 
                   +\sum_{i\neq j}\left(1-\frac{q_0^2}{\bm{q}^2} \right) q_i W^{ij}_{VV} q_j \, .
\end{align}
Inserting the expression~\eqrefeq{eq:Wmunu_ground} into~\eqrefeq{eq:Xbar_VV^||}, we obtain 
\begin{align}
 \bar{X}_{VV}^{\parallel} &= \frac{M_{B_s}}{E_{D_s}} \bm{q}^2 |f_{+}(q^2)|^2 \, .
 \label{eq:XVVpar}
\end{align}
In~\refsec{sec:results-ground} we will use this relation to devise a cross-check of our method for the computation of the inclusive decay rate by comparing with the 
exclusive decay to the ground state. 
Note that because of the Dirac delta in~\eqref{eq:Wmunu_ground} the integral over $\omega$ just selects the ground-state energy for the $D_s$ meson with a given momentum. This then implies that $\bar{X}^{(l)}= X^{(l)}$ up to $\delta (\omega-E_{D_s})$. Further details on the ground-state limit can also be found in the Appendix of \cite{Gambino2022}.

\subsection{Inclusive decays on an Euclidean space-time lattice}
\label{sec:inclusiveLattice}

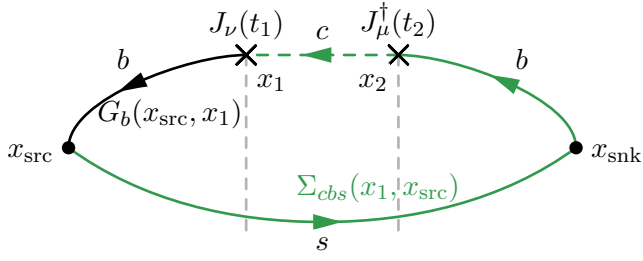
\begin{figure}
 \vspace{-0.7cm}
 \centering
 \begin{fmffile}{4pt}
  \begin{fmfgraph*}(190, 70)
    \fmfipair{tr,tc,tl,br,bc,bl}
    \fmfiequ{tl}{(0,h)}
    \fmfiequ{tc}{(.5w,h)}
    \fmfiequ{tr}{(w,h)}
    \fmfiequ{bl}{(0,-h)}
    \fmfiequ{bc}{(.5w,-h)}
    \fmfiequ{br}{(w,-h)} 
    \fmfipair{src,snk,t,tt,td,ttd,vm}
    \fmfiequ{src}{(0,0)}
    \fmfiequ{snk}{(w,0)}
    \fmfiequ{t}{(.35w,.5h)}
    \fmfiequ{tt}{(.65w,.5h)}
    \fmfiequ{td}{(.35w,-.5h)}
    \fmfiequ{ttd}{(.65w,-.5h)}
    \fmfiequ{vm}{(.5w,-.4h)}
    \fmfipair{G,Gseq,xu,xd}
    \fmfiequ{G}{(.2w,.02h)}
    \fmfiequ{Gseq}{(.61w,-.38h)}
    \fmfiequ{xu}{(.40w,.22h)}
    \fmfiequ{xd}{(.60w,.22h)}
    \fmfi{fermion, label=$b$}{t{left}  .. tension 1.5 .. {down}src}
    \fmfi{dashes_arrow, label=$c$, foreground=(0.196,, 0.603,, 0.298)}{tt .. t}
    \fmfi{fermion, label=$b$, foreground=(0.196,, 0.603,, 0.298)}{snk{up} .. tension 1.5 .. {left}tt}
    \fmfi{fermion, label=$s$, foreground=(0.196,, 0.603,, 0.298)}{src{bc-src} .. 1[src,vm] .. {snk-bc}snk}
    \fmfi{dashes, foreground=(0.7,,0.7,,0.7)}{t .. td}    
    \fmfi{dashes, foreground=(0.7,,0.7,,0.7)}{tt .. ttd}
    \fmfiv{d.sh=circle,d.f=1,d.siz=2thick,l=$x_{\rm src}$}{src}
    \fmfiv{d.sh=circle,d.f=1,d.siz=2thick,l=$x_{\rm snk}$}{snk}
    \fmfiv{d.sh=cross,d.f=1,d.siz=5thick, l=$J_{\mu}^{\dagger}(t_2)$, l.a=90}{tt}
    \fmfiv{d.sh=cross,d.f=1,d.siz=5thick, l=$J_{\nu}(t_1)$, l.a=90}{t}
    \fmfiv{l=$G_b(x_{\rm src},,x_{1})$, l.a=90}{G}
    \fmfiv{l=\textcolor{myGreen}{$\Sigma_{cbs}(x_1,, x_{\rm src})$}, l.a=90}{Gseq}
    \fmfiv{l=$x_1$, l.a=90}{xu}
    \fmfiv{l=$x_2$, l.a=90}{xd}
  \end{fmfgraph*}
 \end{fmffile}
 \vspace{1.5 cm}
\caption{Diagram of the four-point correlator. Two propagators used for the contraction are depicted in the picture. The black one, $G_b(x_{\rm src}, x_1)$, is a propagator
for the $b$ quark from $x_1$ to $x_{\rm src}$. The green one, $\Sigma_{cbs}(x_1, x_{\rm src})$, is a sequential propagator that propagates the $s$ quark from $x_{\rm src}$ to $x_{\rm snk}$, 
the $b$ quark from $x_{\rm snk}$ to $x_2$ and the $c$ quark from $x_2$ to $x_1$.}
\label{fig:4pt}
\end{figure}

We now address the strategy for the computation of the inclusive decay rate on the lattice, which follows~\citep{Hashimoto2017,Gambino2020,Gambino2022}.
The key quantity is the hadronic tensor in \eqref{eq:hadronicTensor}
\begin{align}
 W^{\mu\nu}(q)  = \frac{1}{2M_{B_s}} \int \dd^{4} x \, e^{iq \cdot x}\bra{B_s} J^{\mu\dagger}(x) J^{\nu}(0) \ket{B_s} \, .
 \label{eq:Wmatrixel}
\end{align}
The matrix element in~\eqrefeq{eq:Wmatrixel} can be extracted from the time dependence of the Euclidean four-point function
\begin{equation}
 C_{\mu \nu}^{S J J S}\left(\bm{q},t_{\rm snk}, t_{2}, t_{1}, t_{\rm src}\right) \stackrel{t_2\geq t_1}{=} \sum_{\bm{x}_{\rm snk},\bm{x}_{\rm src}}  
 \left\langle 
  \mathcal{O}_{B_s}^{S}\left(x_{\rm snk}\right) \tilde{J}_{\mu}^{\dagger}\left(\bm{q}, t_{2}\right) \tilde{J}_{\nu}\left(\bm{q}, t_{1}\right) 
   \mathcal{O}_{B_s}^{S \dagger}\left(x_{\rm src} \right)  \right\rangle  \, ,\label{eq:CmunuSJJS}
\end{equation}
where $\mathcal{O}^{S}_{B_s}$ is an interpolating operator with quantum numbers of the $B_s$ meson
and the currents are projected onto three-momentum by a discrete Fourier transform $\tilde{J}_{\nu}(\bm{q},t) = \sum_{\bm{x}} e^{-i\bm{q}\cdot \bm{x}}J_{\nu}(\bm{x},t)$. 
In this setup the $B_s$ meson is created with zero momentum at source position
$x_{\rm src}$ 
and annihilated at sink position $x_{\rm snk}$. 
In  \reffig{fig:4pt} we show the corresponding quark-flow diagram:
the black line, $G_b(x_{\rm src}, x_1)$, is a propagator
for the $b$ quark from $x_1$ to $x_{\rm src}$ whereas the green one, $\Sigma_{cbs}(x_1, x_{\rm src})$,
is a sequential propagator that propagates the $s$ quark from $x_{\rm src}$ to $x_{\rm snk}$, 
the $b$ quark from $x_{\rm snk}$ to $x_2$ and the $c$ quark from $x_2$ to $x_1$.

The matrix element in~\eqrefeq{eq:Wmatrixel} can be extracted in the window $t_{\rm snk}-t_2 \gg 0$, $t_1-t_{\rm src} \gg 0$ and $t_2 > t_1$, where excited states
of the $B_s$ meson have decayed sufficiently.
By increasing the overlap of the operator $\mathcal{O}^S_{B_s}$ with the ground-state $B_s$ state the size of this window can be enlarged. This can be 
achieved by means of operator smearing, to be detailed later. We use a superscript $S$ in case of smearing and $L$ in case of no smearing.

Within the window we expect
\begin{align}
 C_{\mu \nu}^{S J J S}\left(\bm{q},t_{\rm snk}, t_{2}, t_{1}, t_{\rm src}\right) 
 = & \frac 1{4M_{B_s}^2}
 \langle 0|\mathcal{O}_{B_s}^{S}|B_s\rangle 
  \langle B_s|\tilde{J}_{\mu}^{\dagger}\left(\bm{q}, t_{2}\right) \tilde{J}_{\nu}\left(\bm{q}, t_{1}\right) |B_s\rangle
   \langle B_s|\mathcal{O}_{B_s}^{S \dagger}|0\rangle  \, .
\end{align}
In order to extract the $B_s$ forward-scattering matrix element in~\eqrefeq{eq:Wmatrixel} we cancel the 
smeared $B_s$ wave function factors $\langle B_s|\mathcal{O}^{S\dagger}_{B_s}|0\rangle$ and
$\langle 0|\mathcal{O}_{B_s}^{S}|B_s\rangle $
by constructing suitable ratios with $B_s$ meson 
two-point functions with zero momentum
\begin{align}
\begin{split}
C^{SL}(t_2, t_1) &= \sum\limits_{\bm{x}_2,\bm{x}_1}\langle \mathcal{O}_{B_s}^S(x_2)\mathcal{O}_{B_s}^{L\dagger}(x_1)\rangle \\
&\stackrel{\mathclap{\substack{t_2 -t_1\gg 0}}}{=} \quad
    \frac{1}{2M_{B_s}} \bra{0} \mathcal{O}_{B_s}^{S} \ket{B_s} \bra{B_s} \mathcal{O}_{B_s}^{L\dagger} \ket{0} \, e^{-(t_2-t_1)M_{B_s}}\,.
\end{split}
\end{align}
Our choice of ratio is
\begin{equation}
 \frac{C_{\mu\nu}^{SJJS}(\bm{q},t_{\rm snk},t_2 , t_1, t_{\rm src})}{C^{SL}(t_{\rm snk}, t_2)C^{LS}(t_1, t_{\rm src})} \, \quad \longrightarrow \quad
 \frac{\frac{1}{2M_{B_s}} \bra{B_s} \tilde{J}_\mu^{\dagger}(\bm{q}, t_2) \tilde{J}_\nu(\bm{q}, t_1)  \ket{B_s} }{\frac{1}{2M_{B_s}}|\bra{0} \mathcal{O}_{B_s}^{L} \ket{B_s} |^2 } \, ,
\label{eq:ratioC4/C2}
\end{equation}
where we cancel the residual factor $|\bra{0} \mathcal{O}_{B_s}^{L} \ket{B_s} |^2/2M_{B_s}$ with its value obtained from fits to the time-dependence of, e.g., the  $C^{LL}$ two-point function.
This leads us to define the key observable
\begin{align}
 C_{\mu\nu}(\bm{q}, t)  
  &= \frac{1}{2M_{B_s}} \bra{B_s} \tilde{J}_{\mu}^{\dagger}(\bm{q},0)  e^{-\hat{H}t} \tilde{J}_\nu (\bm{q},0) \ket{B_s} \, ,
\end{align}
where we have used time-translation invariance $t=t_2-t_1$. It is related to the hadronic tensor defined
in~\eqrefeq{eq:Wmatrixel} through a  Laplace transform
\begin{align}
 \begin{split}\label{eq:Cmunu}
 C_{\mu\nu} (\bm{q}, t)
 &= \int_{0}^{\infty} \dd \omega \, \frac{1}{2M_{B_s}}
    \bra{B_s} \tilde{J}_{\mu}^{\dagger}(\bm{q},0) \delta (\hat{H} - \omega) \tilde{J}_\nu (\bm{q},0) \ket{B_s} e^{-\omega t} \\
 &=  \int_{0}^{\infty} \dd \omega \, W_{\mu\nu}(\bm{q}, \omega) e^{-\omega t} \, ,
 \end{split}
\end{align}
where 
\begin{align}
 W_{\mu\nu}(\bm{q}, \omega) = \frac{1}{2M_{B_s}} \sum_{X_c} \delta (\omega - E_{X_{c}}) \bra{B_s} \tilde{J}^{\dagger}_{\mu}(\bm{q},0) \ket{X_c} \bra{X_c} \tilde{J}_{\nu}(\bm{q},0) \ket{B_s} 
 \label{eq:WmunuSpectral}
\end{align}
corresponds to the spectral representation of $C_{\mu\nu}(\bm{q},t)$.
By means of~\eqrefeq{eq:ratioC4/C2} we can compute $C_{\mu\nu}$ on the lattice from a combination of meson two- and four-point functions for a finite and discrete set of Euclidean times $t$. The determination of the hadronic tensor by means of inversion of the integral equation~\eqrefeq{eq:Cmunu} therefore constitutes an ill-posed inverse problem, similar to the
extraction of hadronic spectral densities from Euclidean correlators: while the reconstruction of $C_{\mu\nu}$ from $W_{\mu\nu}$ is straightforward, the other way around is a very difficult task. 

Fortunately, in order to compute the inclusive decay rate~\eqrefeq{eq:Gamma_Xbar}, we do not have to compute the hadronic tensor itself, but only integrals $\bar X^{(l)}(\bm{q}^2)$,
where the hadronic tensor is \emph{smeared} with the leptonic tensor integrated over the lepton energy, as defined in Eqs.~\eqref{eq:Gamma_Xbar}-\eqref{eq:X(l)}.
In general, we can write
\begin{align}
 \bar{X}^{(l)}(\bm{q}^2) = \int_{\omega_{\rm min}}^{\omega_{\rm max}} \dd \omega \, W^{\mu\nu}(\bm{q}, \omega) k^{(l)}_{\mu\nu}(\bm{q}, \omega) \, ,
\end{align}
where $k^{(l)}_{\mu\nu}(\bm{q}, \omega)$ is a known kinematic factor that depends only on the energy and three-momentum. 
Introducing a step function $\theta (\omega_{\rm max} -\omega)$ and extending the limit of integration as $\omega_{\rm max}\rightarrow \infty$ and
$\omega_{\rm min}\rightarrow \omega_0$, with $\omega_0 \leq \omega_{\rm min}$ we can rewrite
\begin{align}
 \begin{split}
 \bar{X}^{(l)}(\bm{q}^2)
 &= \int_{\omega_0}^{\infty} \dd \omega \, W^{\mu\nu}(\bm{q}, \omega) k^{(l)}_{\mu\nu}(\bm{q}, \omega) \theta (\omega_{\rm max} -\omega) \\
 &=  \int_{\omega_0}^{\infty} \dd \omega \, W^{\mu\nu}(\bm{q}, \omega) K^{(l)}_{\mu\nu}(\bm{q}, \omega) \, , 
 \end{split}
\end{align}
defining the \emph{kernel function} $K^{(l)}_{\mu\nu}(\bm{q}, \omega)=k^{(l)}_{\mu\nu}(\bm{q}, \omega)\theta (\omega_{\rm max} -\omega)$.
Note that $\omega_0$ can be chosen freely in $0 \leq \omega_0 \leq \omega_{\rm min}$ as there are no states below the ground state energy $\omega_{\rm min}$, as seen from \eqref{eq:WmunuSpectral}. 
For instance, for $B_s\to X_c\,l\nu_l$ we expect $\omega_{\rm min}=M_{D_s}$ for the contribution from the vector channel at vanishing transferred momentum ${\bm q}$. We will later exploit this freedom in the choice of $\omega_0$.

Let us now discuss how to obtain $\bar X^{(l)}$ from lattice data for $C_{\mu\nu}(\bm{q},t)$. First we
introduce a smoothing of the kernel $K_{\mu\nu}^{(l)}$ by replacing the step function by a sigmoid of the 
form
\begin{align}
 \theta_{\sigma}(x) = \frac{1}{1+e^{-x/\sigma}} \, .
 \label{eq:sigmoid}
\end{align}
While we eventually have to take the limit $\sigma\to 0$ in order to obtain the physical decay rate,
smoothing is useful to control and understand the systematic effects involved in the
strategy to compute the decay rate.
Following~\cite{Gambino2020}, we now expand the smoothed kernel $K_{\sigma,\mu\nu}^{(l)}(\bm{q},\omega)$ 
as a polynomial of $e^{-a\omega}$ (we will set $a=1$ for simplicity) up to some order $N$, i.e., 
\begin{align}
 K^{(l)}_{\sigma, \mu\nu}(\bm{q}, \omega) \simeq c^{(l)}_{\mu\nu, 0}(\bm{q}; \sigma) + c^{(l)}_{\mu\nu,1}(\bm{q}; \sigma) e^{-\omega} 
 + \dots + c^{(l)}_{\mu\nu,N}(\bm{q}; \sigma) e^{-\omega N} \, ,
\end{align}
with $N$ coefficients $c_{\mu\nu,k}^{(l)}(\bm{q};\sigma)$. 
In this way, the target quantity $\bar X^{(l)}_\sigma(\bm{q}^2)$, which now also depends on the smearing parameter $\sigma$, can be computed as 
\begin{align}
 \bar{X}_\sigma^{(l)}(\bm{q}^2) &= \int_{\omega_0}^{\infty} \dd \omega \, W^{\mu\nu}(\bm{q}, \omega) e^{-2 \omega t_0} K^{(l)}_{\sigma,\mu\nu}(\bm{q}, \omega; t_0) \notag \\
 			  &\simeq c^{(l)}_{\mu\nu, 0} \int_{\omega_0}^{\infty} \dd \omega \, W^{\mu\nu}(\bm{q}, \omega) e^{-2 \omega t_0}
                     + c^{(l)}_{\mu\nu,1} \int_{\omega_0}^{\infty} \dd \omega \, W^{\mu\nu}(\bm{q}, \omega)e^{-2 \omega t_0} e^{-\omega} + \dots \notag \\
                     & \hspace{1 em}+ c^{(l)}_{\mu\nu,N} \int_{\omega_0}^{\infty} \dd \omega \, W^{\mu\nu}(\bm{q}, \omega)e^{-2 \omega t_0} e^{-\omega N} \, .
\end{align}
The factor $e^{-2 \omega t_0}$ has been introduced, and compensated for in $K^{(l)}_{\sigma,\mu\nu}(\bm{q}, \omega; t_0)= \break e^{2\omega t_0} K^{(l)}_{\sigma,\mu\nu}(\bm{q}, \omega)$, in order
to avoid the equal-time matrix element $t_1=t_2$, see~\eqrefeq{eq:CmunuSJJS}, which contains contributions from the opposite time ordering corresponding
to unphysical $\bar{b}sc\bar{b}$ final states.
We will discuss suitable choices for the free parameter $t_0$ together with the discussion of the analysis of actual simulation data.
Inserting now~\eqrefeq{eq:Cmunu} we arrive at the compact expression
\begin{align}
 \bar{X}_\sigma^{(l)}(\bm{q}^2) = \sum_{k=0}^{N} c^{(l)}_{\mu\nu,k} C^{\mu\nu}(\bm{q},k+2t_0) \, ,
 \label{eq:Xl_approx}
\end{align}
which relates $C^{\mu\nu}(\bm{q},t)$, which can be computed on the lattice, to $\bar{X}_\sigma^{(l)}(\bm{q}^2)$.
The expression is understood to be an approximation of $\bar{X}_\sigma^{(l)}(\bm{q}^2)$ due the truncation to a finite value $N$; we use the same convention
for all similar quantities that we address in the following sections.
Note that the order $N$ of the polynomial approximation is now directly related to the
separation  in Euclidean time of the two charged currents in the four-point function in~\eqrefeq{eq:CmunuSJJS}. What remains to be done
towards the computation of the decay rate for a given value of $\sigma$, is to carry out the 
 phase-space integration in~\eqrefeq{eq:Gamma_Xbar}.

Before we close this section, let us list the explicit expressions for the kernels $K_{\sigma,\mu\nu}^{(l)}$:
\begin{align}
 \label{eq:K0_00}
 K^{(0)}_{\sigma, 00}(\bm{q}, \omega; t_0) &= e^{2\omega t_0} \bm{q}^{2} \, \theta_{\sigma}\left(\omega_{\rm max}-\omega\right) \, , \\
 K^{(0)}_{\sigma, ii}(\bm{q}, \omega; t_0) &= e^{2\omega t_0} (q_i^2 - \bm{q}^{2}) \,\theta_{\sigma}\left(\omega_{\rm max}-\omega\right) \, , \\
 K^{(0)}_{\sigma, ij}(\bm{q}, \omega; t_0) & \stackrel{i\neq j}{=} e^{2\omega t_0} q_{i}q_{j} \,\theta_{\sigma}\left(\omega_{\rm max}-\omega\right) \, , \\
 K^{(1)}_{\sigma, 0i}(\bm{q}, \omega; t_0) &= -e^{2\omega t_0} q_i q_0 \,\theta_{\sigma}\left(\omega_{\rm max}-\omega\right) \, , \\
 \label{eq:K2}
 K^{(2)}_{\sigma, ii}(\bm{q}, \omega; t_0) &= e^{2\omega t_0} q_0^2 \,\theta_{\sigma}\left(\omega_{\rm max}-\omega\right) \, .
\end{align}
For the parallel and perpendicular components at $\bm{q}^2 \neq 0$, as defined in \refsec{sec:theory-ground}, we have
\begin{align}
 K^{\parallel}_{\sigma, 00}(\bm{q}, \omega; t_0) &= e^{2\omega t_0} \bm{q}^{2} \, \theta_{\sigma}\left(\omega_{\rm max}-\omega\right) \, , \\
 K^{\parallel}_{\sigma, 0i}(\bm{q}, \omega; t_0) &= -e^{2\omega t_0} q_0q_i \,\theta_{\sigma}\left(\omega_{\rm max}-\omega\right) \, , \\
 K^{\parallel}_{\sigma, ij}(\bm{q}, \omega; t_0) &= e^{2\omega t_0} \frac{q_0^2}{\bm{q}^2}q_{i}q_{j} \,\theta_{\sigma}\left(\omega_{\rm max}-\omega\right) \, , \\
 K^{\perp}_{\sigma, ii}(\bm{q}, \omega; t_0) &= e^{2\omega t_0} (q_i^2-\bm{q}^2) \left( 1-\frac{q_0^2}{\bm{q}^2} \right) 
                                                \,\theta_{\sigma}\left(\omega_{\rm max}-\omega\right) \, , \\
 K^{\perp}_{\sigma, ij}(\bm{q}, \omega; t_0) & \stackrel{i\neq j}{=} e^{2\omega t_0}  q_i q_j \left( 1-\frac{q_0^2}{\bm{q}^2} \right)
                                               \,\theta_{\sigma}\left(\omega_{\rm max}-\omega\right) \, .
\end{align}
All other index combinations vanish.


\subsection{Data analysis}
\label{sec:analysis}

In the previous section we reduced the problem of computing the inclusive decay rate to that of
finding a suitable polynomial approximation for the kernel $K_{\sigma,\mu\nu}^{(l)}(\bm{q}, \omega;t_0)$. Here we  describe two 
separate methods that we follow (and later compare in~\refsec{sec:results}), for determining the expansion coefficients $c_{\mu\nu,k}^{(l)}$ given lattice data for the ratio of correlation functions in~\eqrefeq{eq:ratioC4/C2}. 

The analysis has to deal with the statistical noise from the data and also  systematic errors, e.g. those associated with the polynomial approximation. 
Here we consider data for a single lattice spacing and lattice volume, leaving discretisation and finite-volume errors for future studies.

In principle, $\bar X^{(l)}_\sigma(\bm{q}^2)$ as defined in~\eqrefeq{eq:Xl_approx}, 
could be computed straightforwardly from lattice data for $C_{\mu\nu}(\bm{q},t)$.
For a given order $N$, the coefficients $c_{\mu\nu,k}^{(l)}$ in the power series for the 
analytically known kernel $K_{\sigma,\mu\nu}^{(l)}(\bm{q},\omega)$ could, for instance, be determined via linear regression,
allowing to construct $\bar X^{(l)}_\sigma(\bm{q}^2)$ from the data for $C_{\mu\nu}(\bm{q},t)$. The order of the expansion is limited by the
number of time slices in the window where $C_{\mu\nu}(\bm{q},t)$ can be extracted from the lattice data. 
Unfortunately, the exponential deterioration of the signal-to-noise ratio with increasing Euclidean time separation $t$
makes a meaningful signal for the decay rate difficult to extract. 
What is needed is some form of regulator that provides balance between
statistical noise and systematic error due to the truncation.
We  proceed with outlining two methods that achieve this: one based on Chebyshev polynomials
and the other based on the modified Backus-Gilbert method.

For the sake of readability we introduce the following notation
\begin{align}
\begin{split}
 \bar{X}_\sigma^{(l)}(\bm{q}^2) 
  &= \int_{\omega_0}^{\infty} \dd \omega \, W^{\mu\nu}(\bm{q}, \omega) e^{-2\omega t_0} K^{(l)}_{\sigma, \mu\nu}(\bm{q}, \omega; t_0) \\
  &= \frac{1}{2 M_{B_s}} \int_{\omega_0}^{\infty} \dd \omega \, K^{(l)}_{\sigma, \mu\nu}(\bm{q}, \omega; t_0)            
     \bra{B_s} \tilde{J}^{\mu\dagger}(\bm{q},0) 
     e^{-\omega t_0}\delta(\hat{H}-\omega)e^{-\omega t_0}
     \tilde{J}^\nu (\bm{q},0) \ket{B_s}  \\
  &= \bra{\psi^{\mu}(\bm{q})} K^{(l)}_{\sigma, \mu\nu}(\bm{q}, \hat{H}; t_0) \ket{\psi^{\nu}(\bm{q})}\,,
 \end{split}
\end{align}
where we made use of~\eqrefeq{eq:WmunuSpectral} and defined $\ket{\psi^{\nu}(\bm{q})} = e^{-\hat H t_0} \tilde{J}^{\nu}(\bm{q},0) \ket{B_s} /\sqrt{2M_{B_s}}$.
Note that the kernel has been promoted to an operator, 
$K_{\sigma,\mu\nu}^{(l)}(\bm{q}, \hat{H}; t_0)$.

\subsubsection{Chebyshev-polynomial approximation}

Chebyshev polynomials $T_k(\omega)$ defined on $-1\le \omega \le1$
provide an optimal approximation of functions under the L$_\infty$-norm.
We provide a summary of basic properties in~\refapp{sec:app_cheb}. 
For the case at hand
we define shifted Chebyshev polynomials $\tilde T_k(\omega)$, which are defined 
in the interval $\omega_0\le \omega\le \infty$. Here, $\tilde{T}_k (\omega)=T_k(h(\omega))$, and $h(\omega)=Ae^{-\omega}+B$ is a map $h: [\omega_0,\infty)\rightarrow [-1,1]$, where 
expressions for the coefficients $A$ and $B$ can be found in~\eqrefeq{eq:MAPcoefficients}.
The kernel function from the previous section can then be expanded up to order $N$ as
\begin{align}
 K_{\sigma, \mu\nu}^{(l)}(\bm{q}, \omega; t_0) = \frac{1}{2} \tilde{c}^{(l)}_{\mu\nu, 0} \tilde{T}_{0}(\omega) +
 \sum_{k=1}^{N}\, \tilde{c}^{(l)}_{\mu\nu, k} \tilde{T}_{k}(\omega) \, ,
\end{align}
where $\tilde{T}_{0}(\omega)=1$ by definition, and 
\begin{equation}
 \tilde T_k(\omega)=\sum_{j=0}^{k} \tilde{t}_j^{(k)}e^{-j\omega}\,,
\end{equation} 
with coefficients $\tilde{t}_j^{(k)}$ defined and discussed in App.~\ref{app:Shifted Chebysehv polynomials}. Making use of the Chebyshev polynomials' orthogonality properties, the coefficients $\tilde c_{\mu\nu,k}^{(l)}$ are defined by projection as in \eqrefeq{eq:chebyshev_ck_projection}, 
\begin{align}
 \tilde{c}^{(l)}_{\mu\nu, k} = \int_{\omega_0}^{\infty} \dd \omega \, K_{\sigma,\mu\nu}^{(l)}(\bm{q},\omega;t_0) \tilde{T}_k(\omega) \Omega_h (\omega) \, ,
\end{align}
where the weight function $\Omega_{h}(x)$ is defined in \refapp{sec:app_cheb}.
In this way, the expectation value of the kernel operator is
\begin{align}
 \bra{\psi^\mu} K^{(l)}_{\sigma, \mu\nu}(\bm{q}, \hat{H}; t_0)\ket{\psi^\nu} = 
 \frac{1}{2}\tilde{c}^{(l)}_{\mu\nu,0} \bra{\psi^\mu} \tilde{T}_0(\hat{H})  \ket{\psi^\nu } + 
 \sum_{k=1}^{N}\,\tilde{c}^{(l)}_{\mu\nu,k} \bra{\psi^\mu} \tilde{T}_k(\hat{H})  \ket{\psi^\nu } \, .
\end{align}
By construction, in particular thanks to the condition of \eqrefeq{eq:AB_chebCoeff_general}, shifted Chebyshev polynomials are bounded, $|\tilde{T}_k(\omega) | \leq 1$.
As we will discuss later, this a crucial ingredient in the data analysis: in order to make use of this property, we divide
the terms $\bra{\psi^\mu} \tilde{T}_k(\hat{H})  \ket{\psi^\nu }$
by a normalisation factor $\bra{\psi^\mu }\ket{\psi^\nu}= C^{\mu\nu}(2t_0)$.
For a more compact notation we define
\begin{align}
 \langle K^{(l)}_{\sigma}\rangle_{\mu\nu} \equiv \frac{\bra{\psi_\mu} K^{(l)}_{\sigma,\mu\nu}(\bm{q}, \hat{H}; t_0)\ket{\psi_\nu }}{\bra{\psi_\mu }\ket{\psi_\nu}} \, , \qquad
 \langle \tilde{T}_{k}\rangle_{\mu\nu} \equiv \frac{\bra{\psi_\mu} \tilde{T}_k(\hat{H})  \ket{\psi_\nu }}{\bra{\psi_\mu }\ket{\psi_\nu}} \, ,
\end{align}
such that
\begin{align}
 \langle K^{(l)}_{\sigma}\rangle_{\mu\nu} = \frac{1}{2}\tilde{c}^{(l)}_{\mu\nu,0} \langle \tilde{T}_{0}\rangle_{\mu\nu} +
 \sum_{k=1}^{N} \, \tilde{c}^{(l)}_{\mu\nu,k} \langle \tilde{T}_{k}\rangle_{\mu\nu} \, ,
\end{align}
where in this case there is no summation on $\mu, \nu$. We refer to $\langle\tilde{T}_k \rangle_{\mu\nu}$ as the \textit{Chebyshev matrix elements},
for which, thanks to the normalisation, $| \langle\tilde{T}_k \rangle_{\mu\nu} | \leq 1$.
In terms of the Chebyshev expansion the expression for  $\bar{X}^{(l)}_\sigma(\bm{q}^2)$ now reads
\begin{align}
 \bar{X}_\sigma^{(l)}(\bm{q}^2) &= \sum_{\{\mu,\nu\}}  \bra{\psi_\mu }\ket{\psi_\nu}  \langle K_{\sigma}^{(l)}\rangle_{\mu\nu} \, ,
\end{align}
and explicitly
\begin{align}
\bar{X}_{\sigma}^{(0)} &= C_{00}(2t_0) \langle K_{\sigma}^{(0)}\rangle_{00}
   				 + \sum_{i} C_{ii}(2t_0)\langle K_{\sigma}^{(0)}\rangle_{ii} 
   				 + \sum_{i\neq j} C_{ij}(2t_0) \langle K_{\sigma}^{(0)}\rangle_{ij} \, ,\\
\bar{X}_{\sigma}^{(1)} &= \sum_i \left( C_{0i}(2t_0) \langle K_{\sigma}^{(1)}\rangle_{0i}
   				 + C_{i0}(2t_0) \langle K_{\sigma}^{(1)}\rangle_{i0} \right) \, ,\\
\bar{X}_{\sigma}^{(2)} &= \sum_{i} C_{ii}(2t_0) \langle K_{\sigma}^{(2)}\rangle_{ii} \, .
\label{eq:Xbar(l)_Cmunu<K>}
\end{align}
The Chebyshev matrix elements can be constructed directly from the lattice data using
\begin{align}
 \frac{\bra{\psi_\mu} e^{-\hat{H}t}  \ket{\psi_\nu }}{\bra{\psi_\mu }\ket{\psi_\nu}} =  \frac{C_{\mu\nu}(t+2t_0)}{C_{\mu\nu}(2t_0)} \equiv \bar{C}_{\mu\nu}(t)\, .
\label{eq:Cbar}
\end{align}
Using the properties of shifted Chebyshev polynomials as detailed in ~\refapp{app:Shifted Chebysehv polynomials}, we can directly relate the
matrix element $\langle \tilde{T}_k \rangle_{\mu\nu}$ to the correlator $\bar{C}_{\mu\nu}$. 
In particular,
\begin{align}
 \langle \tilde{T}_{k}\rangle_{\mu\nu}
 &= \frac{\bra{\psi_\mu} \tilde{T}_k(\hat{H})  \ket{\psi_\nu }}{\bra{\psi_\mu }\ket{\psi_\nu}}  
    = \sum_{X_c} \frac{\bra{\psi_\mu}\tilde{T}_k(\hat{H})\ket{X_c}\bra{X_c}  \ket{\psi_\nu }}{\bra{\psi_\mu }\ket{\psi_\nu}}\nonumber  \\
 &= \sum_{X_c} \sum_{j=0}^k \tilde{t}^{(k)}_j e^{-jE_{X_c}} \frac{\bra{\psi_\mu} \ket{X_c}\bra{X_c}  \ket{\psi_\nu }}{\bra{\psi_\mu }\ket{\psi_\nu}} \nonumber\\
 &= \sum_{j=0}^{k} \tilde{t}_{j}^{(k)}\bar{C}_{\mu\nu}(j) \, , \label{eq:linear system for T}
\end{align}
where we have inserted the identity $I=\sum_{X_c} \ket{X_c}\bra{X_c}$ and $\tilde{t}^{(k)}_j$ are defined in \eqref{eq:tilde_tn}. 
Overall the full Chebyshev expansion of the kernel reads
\begin{align}
 \begin{split}
 \langle K_{\sigma}^{(l)} \rangle_{\mu\nu}
 & =  \frac{1}{2}\tilde{c}_{\mu\nu, 0}^{(l)} \langle \tilde{T}_{0}\rangle_{\mu\nu} +
      \sum_{k=1}^{N} \tilde{c}_{\mu\nu, k}^{(l)} \langle \tilde{T}_{k}\rangle_{\mu\nu}  \\
 & = \sum_{k=0}^{N} \bar{C}_{\mu\nu}(k) \sum_{j=k}^{N} \tilde{c}_{\mu\nu,j}^{(l)}\left(1-\frac{1}{2}\delta_{0j}\right) \tilde{t}_k^{(j)} \,,
 \end{split}
\end{align}
where we emphasise once more that the analytical expressions for the coefficients $\tilde{c}_{\mu\nu,j}^{(l)}$ and $\tilde{t}_k^{(j)}$ 
are known and can be evaluated.
Collecting the coefficients into
\begin{align}
 \bar{c}_{\mu\nu,k}^{(l)} \equiv \sum_{j=k}^{N} \tilde{c}_{\mu\nu,j}^{(l)} \tilde{t}_k^{(j)} \left(1-\frac{1}{2}\delta_{0j}\right) \, ,
\end{align}
we arrive at the compact expression
\begin{align}
  \langle K_{\sigma}^{(l)} \rangle_{\mu\nu}  = \sum_{k=0}^{N} \bar{c}_{\mu\nu,k}^{(l)} \bar{C}_{\mu\nu}(k) \, .
  \label{eq:kernel_cbar}
\end{align}
While $\bar{c}_{\mu\nu,k}^{(l)}$ is known in terms of solvable 
analytical expressions, $\bar{C}_{\mu\nu}(k)$ needs to be computed on the lattice using Monte-Carlo methods. The resulting
statistical error on $\bar{C}_{\mu\nu}(k)$ can lead to violations of the
bound $|\langle \tilde{T}_k\rangle_{\mu\nu}|\le 1$ when solving 
the linear system in Eq.~(\ref{eq:linear system for T}).
This can however be avoided in a Bayesian analysis of the correlator data,
imposing the bound in terms of priors. One way to impose 
the constraint is to
use a Gaussian prior on some internal parameters $\langle\tilde \tau_k\rangle_{\mu\nu}\sim\mathcal{N}(0,1)$
and convert it to a flat prior on the interval $[-1,1]$ using the map $f(x)={\rm erf}(x/\sqrt{2})$
such that $\langle \tilde{T}_k\rangle_{\mu\nu} = f(\langle\tilde \tau_k\rangle_{\mu\nu})$. 
We refer to~\refapp{sec:app_fit} for a thorough discussion on the fitting procedure that we adopt.

\subsubsection{Backus-Gilbert}
\label{sec:analysis-BG}

A different approach to determine the polynomial approximation of the kernel is given by a variant of the Backus-Gilbert method \cite{Backus1968} proposed in \cite{Hansen2019,Bulava2021}.
In this work, we consider a more general scenario to allow the use of different polynomial bases following ~\cite{ExtendedTwistedMassCollaborationETMC:2022sta}.
Note that, although what we propose is mathematically equivalent to the approach in~\cite{ExtendedTwistedMassCollaborationETMC:2022sta}, our formulation may have the advantage of avoiding some of the numerical technicalities that arise in the original version.
Indeed, while the latter requires the inversion of an ill-conditioned matrix with the help of arbitrary precision arithmetic, our approach relies on the inversion of an equivalent diagonal matrix in the case where an orthogonal polynomial basis is chosen, at least as far as the systematics are concerned.
We briefly present the idea below and refer to \refapp{sec:app_bg} for a more detailed discussion.
Note that we adopt a different notation with respect to the original works (we use $F$ instead of $W$ for the final functional to avoid confusion with the hadronic tensor).

The central idea is to address the reconstruction of the (smeared) kernel $K_{\sigma, \mu\nu}^{(l)}$ of the form
\begin{align}
 K_{\sigma, \mu\nu}^{(l)}(\bm{q}, \omega; t_0) = \sum_{k=0}^{N} g^{(l)}_{\mu\nu,k} \tilde{P}_{k}(\omega) \, ,
\end{align}
where $\tilde{P}_{k}(\omega) = \sum_{j=0}^{k} \tilde{p}^{(k)}_{j} e^{-j\omega}$ are a basis of functions
defined on $[\omega_0, \infty)$, and $g_{\mu\nu,k}^{(l)}\equiv g_{\mu\nu,k}^{(l)}(\bm{q}, \sigma; t_0)$ is a set of coefficients to be determined.
In order to compute them, the strategy is to minimise the functional
\begin{align}
 F^{(l)}_{\mu\nu,\lambda}[g] = (1-\lambda)\frac{A_{\mu\nu}^{(l)}[g]}{A_{\mu\nu}^{(l)}[0]} + \lambda B_{\mu\nu}^{(l)}[g] \, ,
\end{align}
where
\begin{align}
A^{(l)}_{\mu\nu}[g] &= \int_{\omega_0}^{\infty} \dd \omega \, \Omega(\omega)
                 \left[ K_{\sigma, \mu\nu}^{(l)}(\bm{q}, \omega; t_0) - \sum_{k=0}^{N} g^{(l)}_{\mu\nu,k} \tilde{P}_{k}(\omega) \right]^2 
\end{align}
is the L$_2$-norm of the difference between the target kernel function and its reconstruction, weighted with a smooth function $\Omega(\omega)$,
and
\begin{align}
B^{(l)}_{\mu\nu}[g] &= \sum_{j,k=0}^{N} g^{(l)}_{\mu\nu, j} \Cov[\bar{C}^{P}_{\mu\nu}(j), \bar{C}^{P}_{\mu\nu}(k)] g^{(l)}_{\mu\nu, k} 
\label{eq:BG_B_functional}
\end{align}
is the variance of the corresponding channel 
$\bar{X}^{(l)}_{\mu\nu}$, with 
$\bar{C}^{P}_{\mu\nu}(k) = \sum_{j=0}^{k} \tilde{p}^{(k)}_{j} \bar{C}_{\mu\nu}(j)$.
The functional $F_{\mu\nu,\lambda}^{(l)}$ encodes the information about both systematic and statistical error, whose interplay is controlled
by the parameter $\lambda \in [0, 1)$, which in principle can be chosen by hand.
The values of the coefficients $g_{\mu\nu,k}^{(l)}(\lambda)$ for each $\lambda$ are given by the variational principle, i.e.
\begin{align}
 g_{\mu\nu,k}^{(l)}(\lambda) \quad \leftrightarrow \quad \frac{\partial F^{(l)}_{\mu\nu,\lambda}}{\partial g_{\mu\nu,k}^{(l)}} = 0 \, .
\end{align}
We can now devise a method to find the optimal $\lambda^{*}$. Following \cite{Bulava2021}, we can simply
evaluate the functional $F_{\mu\nu,\lambda}^{(l)}$ at its minimum
i.e. $F_{\mu\nu}^{(l)}(\lambda) = F^{(l)}_{\mu\nu,\lambda}[g(\lambda)] $, which then becomes a function of $\lambda$,
and require that $\lambda^{*}$ maximises $F_{\mu\nu}^{(l)}(\lambda)$, $\dv{F_{\mu\nu}^{(l)}(\lambda)}{\lambda}\bigg|_{\lambda^{*}}=0$. 
It is clear that this choice corresponds to
$A_{\mu\nu}^{(l)}[g^{*}]/A_{\mu\nu}^{(l)}[0]=B_{\mu\nu}^{(l)}[g^{*}]$, i.e. an optimal balance between statistical and systematic errors.
This is the prescription we follow and take $g_{\mu\nu,k}^{*(l)} \equiv g_{\mu\nu,k}^{(l)}(\lambda^{*})$.

Following the steps for the Chebyshev approach we get for the kernel
\begin{align}
 \langle K^{(l)}_{\sigma} \rangle_{\mu\nu} & = 
 \sum_{k=0}^{N} g^{*(l)}_{\mu\nu, k} \langle \tilde{P}_{k} \rangle_{\mu\nu} \, , \\
 \langle \tilde{P}_{k} \rangle_{\mu\nu} = \frac{\bra{\psi_\mu} \tilde{P}_{k}(\hat{H}) \ket{\psi_\nu}}{\bra{\psi_\mu}\ket{\psi_\nu}} &=
  \sum_{j=0}^{k} \tilde{p}^{(k)}_{j} \frac{\bra{\psi_\mu} e^{-j\hat{H}} \ket{\psi_\nu}}{\bra{\psi_\mu}\ket{\psi_\nu}} = \bar{C}^{P}_{\mu\nu}(k)\, .
\end{align}
In particular, considering the domain $[\omega_0, \infty)$, we focus on two choices:
\begin{itemize}
    \item \emph{exponential Backus-Gilbert:}  $\tilde{P}_k(\omega) = e^{-k\omega}$ and $\Omega(\omega)=1$ (and set $g_{\mu\nu,0}^{(l)}=0$ by hand, as in the original proposal \cite{Hansen2019});
    \item \emph{Chebyshev Backus-Gilbert:} $\tilde{P}_k(\omega)=\tilde{T}_k(\omega)$, i.e. the shifted Chebyshev polynomials with $\Omega(\omega)=1/\sqrt{e^{a(\omega-\omega_0)}-1}$ being the weight that enters in the definition of the scalar product as in \eqref{eq:cheb_scalar_product}.
\end{itemize}

\section{Numerical setup}
\label{sec:setup}

We perform a pilot study using a $24^3\times 64$ lattice with 2+1-flavour domain-wall fermion (DWF)~\cite{Shamir:1993zy,Furman:1994ky}
gauge-field ensembles with the Iwasaki gauge action~\cite{Iwasaki:1983iya}
taken from the RBC/UKQCD Collaboration \cite{Allton2008}
 at
lattice spacing $a\simeq 0.11 \,\text{fm}$ and pion mass $M_{\pi}\simeq 330 \,\text{MeV}$.
The correlation functions analysed in this paper have been generated with the Grid \cite{Grid,GridProc,Yamaguchi:2022feu} and Hadrons \cite{HadronsZenodo} software packages. Part of the fits in the analysis have been
performed using lsqfit \cite{lsqfit, Lepage:2001ym}.

We use the same simulation parameter RBC/UKQCD is using in the heavy-light meson projects on exclusive semileptonic $B_{(s)}$ meson decays~\cite{Flynn2018,Flynn2019,Flynn2021,Flynn:2023ufa}.
In particular, the valence-strange quark is simulated using DWF, whereas the valence-charm quark is simulated by using the M\"obius DWF action \cite{Cho2015,Brower2017}.
Their masses are tuned such that mesons containing bottom, charm and strange valence quarks have masses close to the physical ones.
The bottom quark has been
simulated at its physical mass using the Columbia formulation of the relativistic-heavy-quark (RHQ) action \cite{RHQColumbia1,RHQColumbia2},
which is based on the Fermilab heavy quark action \cite{RHQFermilab}. In particular, this
formulation allows to reduce the $b$-quark discretisation effects of order $\mathcal{O}((m_0a)^n)$, $\mathcal{O}(\bm{p}a)$ and $\mathcal{O}((\bm{p}a)(m_0a)^n)$
by tuning three nonperturbative parameters, one of them being the bare mass $m_0$.

For the computation we average over 120 statistically independent gauge configurations, and on each configuration the measurements are performed on 8 different linearly spaced source time planes.
We use $\mathbb{Z}_2$ wall sources~\cite{Foster:1998vw,McNeile:2006bz,Boyle:2008rh} to improve the signal.
We induce 10 different momenta in the four-point functions in~\eqrefeq{eq:CmunuSJJS}
using twisted boundary conditions \cite{DeDivitiis,Sachrajda2004} with the same momentum in all three spatial directions. Considering 
$\bm{q} = 2\pi \bm{\theta}/L $ in lattice units we have
$\bm{\theta} \equiv (\theta, \theta, \theta)$, where $\theta$ indicates the twist. We choose them such that all the momenta are linearly spaced in $\bm{q}^2$: $\theta_k= 1.90 \, \sqrt{\frac{k}{3}}$ for $k=0,1,\dots,7$, where the factor $1.90$ is determined by the value of $\bm{q}^2_{\rm max}=1.83$ in lattice units.
We also take $\theta=1.90 \, \sqrt{\frac{1}{9}}$ and $\theta=1.90 \, \sqrt{\frac{2}{9}}$ to increase the resolution in $\bm{q}^2$ 
for small momenta. 

We compute two-point functions for both $B_s$ and $D_s$. As discussed in \refsec{sec:inclusiveLattice}, for $B_s$ we consider three cases at zero momentum 
$C^{LS}_{B_s}(t, t_{\rm src})$, $C^{SL}_{B_s}(t, t_{\rm src})$ and $C^{SS}_{B_s}(t, t_{\rm src})$ with different smearing combinations,
as indicated by the superscripts ``$L$'' (local) and ``$S$'' (smeared).
The smeared-smeared $C^{SS}_{B_s}(t, t_{\rm src})$ is also used to determine the renormalisation constant together with the three-point functions.
The sources are smeared gauge-invariantly using Jacobi iteration~\cite{Alford:1995dm,Lichtl:2006dt} using the same parameters 
as in RBC/UKQCD's study of exclusive semileptonic decays in~\cite{Flynn:2015mha,Flynn:2023ufa,PhysRevD.86.116003}.

\begin{figure}[t!]
 \centering
 \includegraphics[scale=0.4]{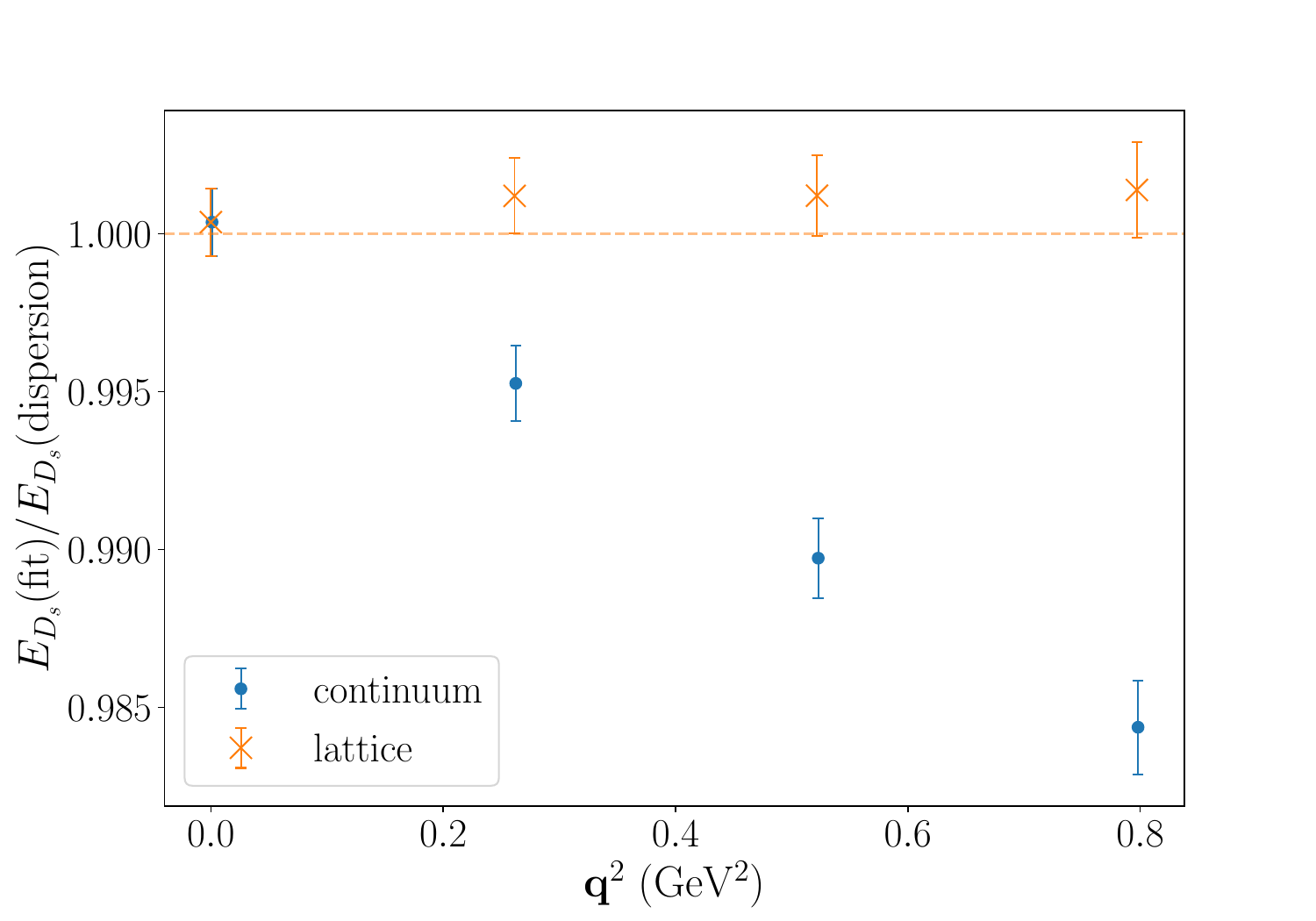}
 \caption{Speed-of-light plot for the $D_{s}$ meson. The numerator is the energy of the ground state mass for a given momentum as extracted from
 a fit to the data. The denominator is given by either the lattice dispersion relation
 or the continuum one,
 where the $D_s$ mass has been determined from a fit to the data.}
 \label{fig:MDs_dispersion}
\end{figure}

The $D_s$ correlators are relevant mainly for the analysis of the ground-state limit in Sec.~\ref{sec:results-ground}. We consider again three different combinations of smearing at
source and sink and we induce momenta for the $c$ quark with the available twists. We show the speed of light from the fitted masses of the $D_s$ for 
the smallest momenta, comparing with the
continuum dispersion relation
and the lattice dispersion relation
in \reffig{fig:MDs_dispersion}. The latter shows excellent agreement
with the fitted energies.

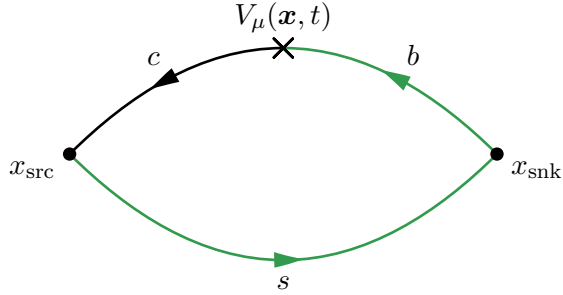
\begin{figure}[h!]
  \centering
\begin{fmffile}{3pt}
  \begin{fmfgraph*}(160, 80)
    \fmfipair{tr,tc,tl,br,bc,bl}
    \fmfiequ{tl}{(0,h)}
    \fmfiequ{tc}{(.5w,h)}
    \fmfiequ{tr}{(w,h)}
    \fmfiequ{bl}{(0,-h)}
    \fmfiequ{bc}{(.5w,-h)}
    \fmfiequ{br}{(w,-h)} 
    \fmfipair{src,snk,v,vm}
    \fmfiequ{src}{(0,0)}
    \fmfiequ{snk}{(w,0)}
    \fmfiequ{v}{(.5w,.5h)}
    \fmfiequ{vm}{(.5w,-.5h)}
    \fmfipair{G,Gseq}
    \fmfiequ{G}{(.4w,.05h)}
    \fmfiequ{Gseq}{(.6w,-.4h)}
    \fmfi{fermion, label=$c$}{v{left} .. {src-tc}src}  
    \fmfi{fermion, label=$b$, foreground=(0.196,, 0.603,, 0.298)}{snk{tc-snk} .. {left}v}
    \fmfi{fermion, label=$s$, foreground=(0.196,, 0.603,, 0.298)}{src{bc-src} .. 1.[src,vm] .. {snk-bc}snk}
    \fmfiv{d.sh=circle,d.f=1,d.siz=2thick,l=$x_{\rm src}$}{src}
    \fmfiv{d.sh=circle,d.f=1,d.siz=2thick,l=$x_{\rm snk}$}{snk}
    \fmfiv{d.sh=cross,d.f=1,d.siz=5thick, l=$V_{\mu}(\bm{x},,t)$, l.a=90}{v}
  \end{fmfgraph*}
\end{fmffile}
  \vspace{16mm}
  \caption{Three-point correlator diagram for the exclusive channel $B_s \rightarrow D_s \,l\nu_l $.
  }
  \label{fig:3pt_diag}
\end{figure}

We also compute three-point correlators for the $B_s \rightarrow D_s \,  l\nu_l$ process
\begin{align}
 C^{SS}_{B_s D_s, \mu}(\bm{q}, t_{\rm snk}, t, t_{\rm src}) = 
 \sum_{\bm{x}_{\rm snk}, \bm{x}} \langle {\mathcal{O}_{B_s}^{S}(x_{\rm snk})} V_{\mu}(\bm{x}, t) {\mathcal{O}_{D_s}^{S \dagger}(x_{\rm src})}  \rangle \, .
\end{align}
Following the analysis of~\citep{Flynn2018,Flynn2019,Flynn2021},
we extract its form factors and compare with our inclusive results. The momentum is carried by the charm quark through twisted 
boundary conditions, $\bm{q}=2\pi \bm{\theta} / L$. We use a source-sink separation of $t_{\rm snk}-t_{\rm src}=20$ in lattice units. The corresponding quark-flow diagram is depicted in \reffig{fig:3pt_diag}.

We now move to the four-point correlators defined in Eq.~\eqref{eq:CmunuSJJS}, which are the building blocks in the computation of inclusive processes.
We use the same source-sink separation as for the three-point functions, i.e., $t_{\rm snk}-t_{\rm src}=20$ in lattice units.
The current $J_{\mu}^{\dagger}$ is fixed at the time slice $t_2=t_{\rm src}+14$,
such that the time dependence is enclosed in $0\leq t \leq 14$ with $t=t_2-t_1$. For this choice we find ground state saturation at the points where we insert the currents.
In practice, referring to \reffig{fig:4pt}, the contractions are performed between a $b$-quark propagator 
$G_b(x_1, x_{\rm src})$ and a sequential propagator $\Sigma_{cbs}(x_1, x_{\rm src})$. For the latter,
we first propagate the $s$ quark to point $x_{\rm snk}$, starting from a $\mathbb{Z}_2$ wall source at $t_{\rm src}$; we then use it 
as a sequential source at fixed $t_{\rm snk}$ with zero momentum to propagate the $b$ quark.
The $b$ quark is propagated to point $x_2$, and it is then used again as a source 
with a specific choice of gamma matrix corresponding to the current $J_{\mu}^{\dagger}(x_2)$ and the momentum insertion to propagate the $c$ quark. 

As before, the momentum $\bm{q}$ induced through twisted boundary conditions is carried by the $c$ quark.
Given that we are dealing with $(V-A)$ currents, we consider all possible combinations of $J^{\dagger}_{\mu}(x_2)$ and $J_{\nu}(x_1)$,  
i.e. $V^{\dagger}_{\mu}V_{\nu}, V^{\dagger}_{\mu}A_{\nu}, A^{\dagger}_{\mu}V_{\nu}, A^{\dagger}_{\mu}A_{\nu}$.
However, in the limit of massless leptons the combinations $A^{\dagger}_{\mu}V_{\nu}$ and $V^{\dagger}_{\mu}A_{\nu}$ do no contribute to the total decay rate.
Indeed, these terms are related to the structure function $W_3$ as $W_{ij}^{AV} + W_{ij}^{VA} = i\epsilon_{ij0k}q^{k}W_3$, as can be seen analysing parity in \eqrefeq{eq:Wij}, which does not contribute to
the total decay rate for $m_l = 0$.

\begin{figure}[t!]
 \centering
 \hbox{
  \includegraphics[scale=0.3]{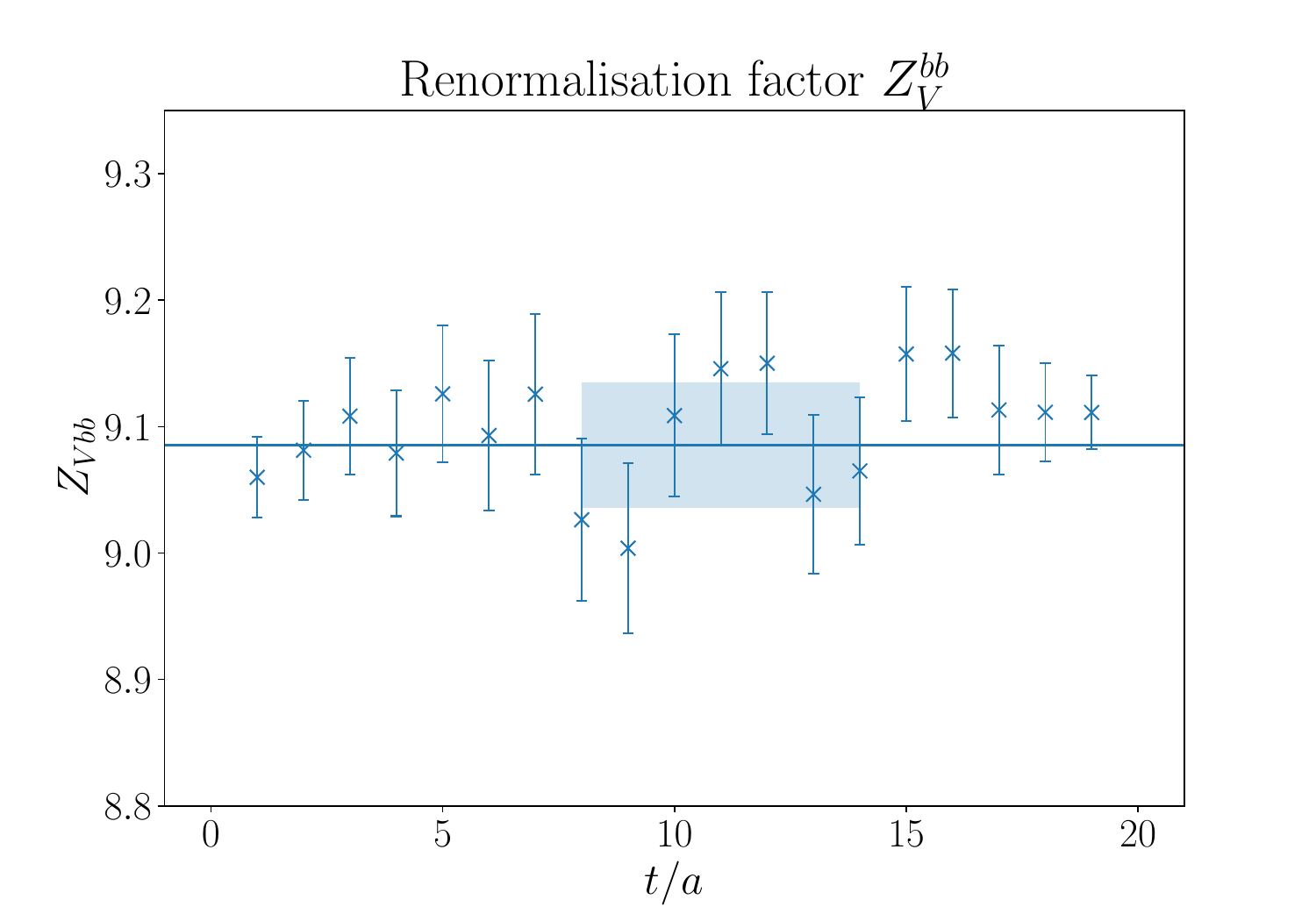}
 \includegraphics[scale=0.3]{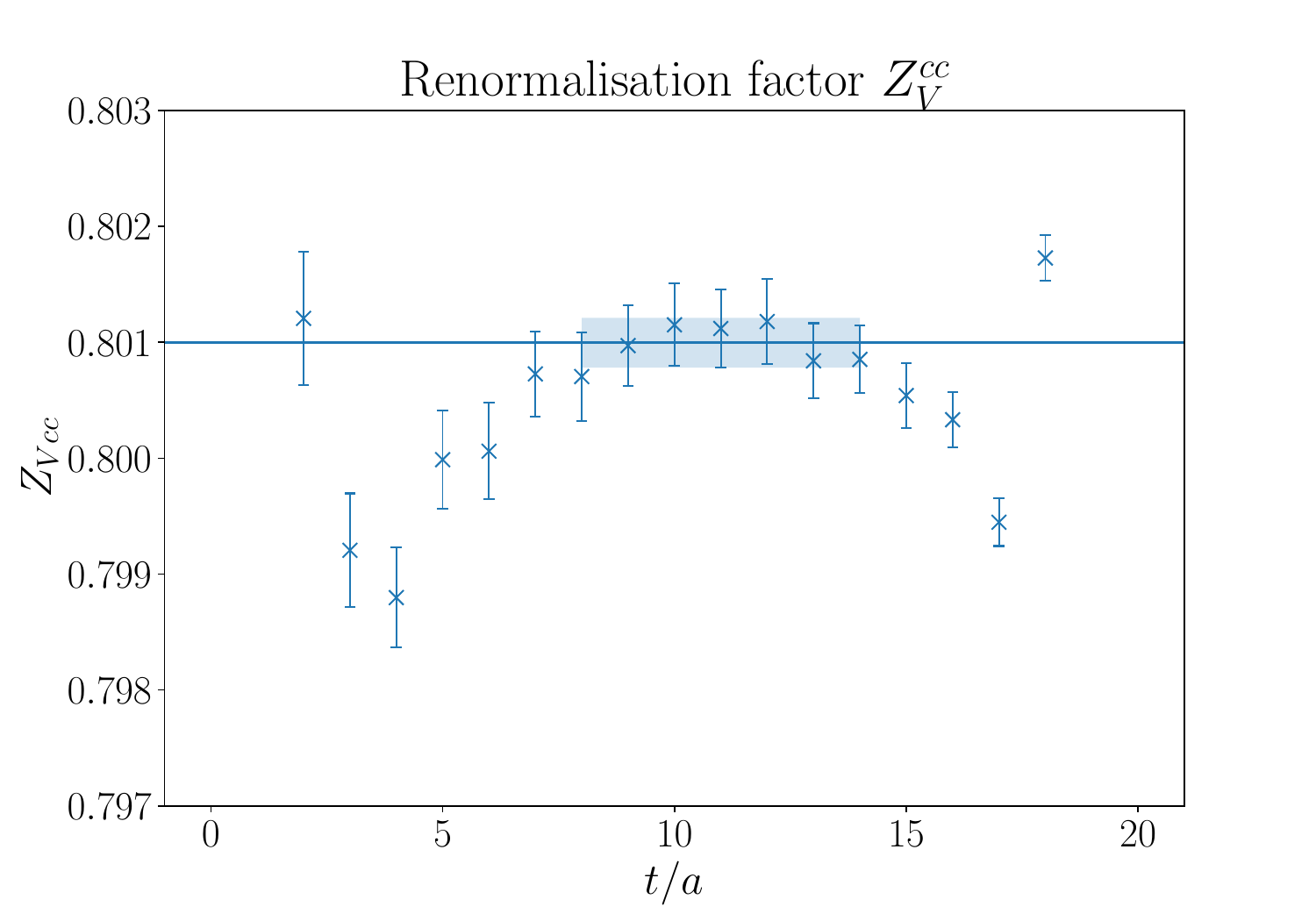} 
 }
 \caption{Determination of renormalisation $Z_V^{bb}$ (left) and $Z_V^{cc}$ (right) from the ratio 
 of two- and three-point functions defined in~Eq.~\eqref{eq:ZVbb}.}
 \label{fig:ZV}
\end{figure}

The local vector and axial-vector currents used in our lattice calculation receive a finite 
renormalisation. We use the \emph{almost nonperturbative} prescription of~\cite{El-Khadra:2001wco},
whereby
\begin{equation}
V_\mu=\rho_{V}^{bc}\sqrt{Z_V^{cc}Z_V^{bb}}\left(V_\mu\right)_{\rm bare}\qquad {\rm and}\qquad
A_\mu=\rho_{A}^{bc}\sqrt{Z_V^{cc}Z_V^{bb}}\left(A_\mu\right)_{\rm bare}\,.
\end{equation}
The subscript ``bare'' indicates the bare, unrenormalised heavy-light
vector or axial-vector current. $Z_V^{cc}$ is the vector-current renormalisation
constant for domain-wall fermions. Due to the approximate chiral symmetry
of domain-wall fermions, $Z_V^{cc}=Z_A^{cc}$ up to residual 
chiral-symmetry-breaking effects.
The renormalisation constants $Z_V^{bb}$ and $Z_{V}^{cc}$
are computed from the \emph{charge} of the heavy-light mesons, and are defined as
\begin{align}
 Z_{V}^{bb} = \frac{C^{SS}_{B_s}(t_{\rm snk}, t_{\rm src})}{C^{SS}_{B_s B_s,0}(t_{\rm snk}, t, t_{\rm src})}  \quad{\rm and}\qquad
 Z_{V}^{cc} = \frac{C^{SS}_{D_s}(t_{\rm snk}, t_{\rm src})}{C^{SS}_{D_s D_s,0}(t_{\rm snk}, t, t_{\rm src})}\,,
 \label{eq:ZVbb}
\end{align}
where both the two- and three-point functions are zero-momentum projected.
The results for $Z_{V}^{bb}=9.085(50)$ and $Z_{V}^{cc}=0.80099(21)$ are reported in \reffig{fig:ZV}.
The coefficient $\rho_{V/A}^{bc}$ is expected to be close to unity and 
can be computed in perturbation theory. Here we set
it to its tree-level value, i.e. $\rho_{V/A}^{bc}=1$.
This is sufficient for the qualitative study aimed at here, where
no attempt is made at taking the continuum limit.

For all the three-point and four-point functions we always average over the spatial directions given that the momentum is the same in all three directions. Note
in particular that for the four-point correlators we have to average separately over $J_i^{\dagger}J_i$ and $J_i^{\dagger}J_k$ with $i\neq k$, which can be seen from~\eqrefeq{eq:Wij}.

\section{Results}
\label{sec:results}

In this section we present and discuss the main results of our work. We first discuss how well the kernels $K^{(l)}_{\mu\nu,\sigma}$ are approximated by the polynomials and then discuss the reconstruction via Chebyshev and Backus-Gilbert methods.
Eventually we combine various analysis steps for a prediction of the inclusive decay rate.
Towards the end of this section we compare our results with the ground-state contribution.
We emphasise that the work presented here focuses on a qualitative understanding of the methods aiming at developing reliable techniques, which in future work can be used to make phenomenologically relevant predictions.

\subsection{Polynomial approximation of the kernel}

In this section we discuss the key aspects of the polynomial approximation. The two ingredients to optimise the
approximation are the choice of the starting point of the approximation $\omega_0$, and the value of $t_0$ in \eqref{eq:Cbar}. In particular, we choose
$t_0=1/2$ in lattice units, such that the exponential growth of the term $e^{2\omega t_0}$ in the kernels \eqref{eq:K0_00}-\eqref{eq:K2} is minimal,
and the number of data points we can use is maximised. We study two values of $\omega_0$, i.e.
$\omega_0=0$ and $\omega_0=0.9\,\omega_{\rm min}$ for each momentum $\bm{q}^2$.
Note that this section deals purely with the approximation of the kernel with no connection to the data; for the  Backus-Gilbert method this means
that we set  $\lambda=0$. 

\begin{figure}[t!]
 \hbox{
 \includegraphics[scale=0.3]{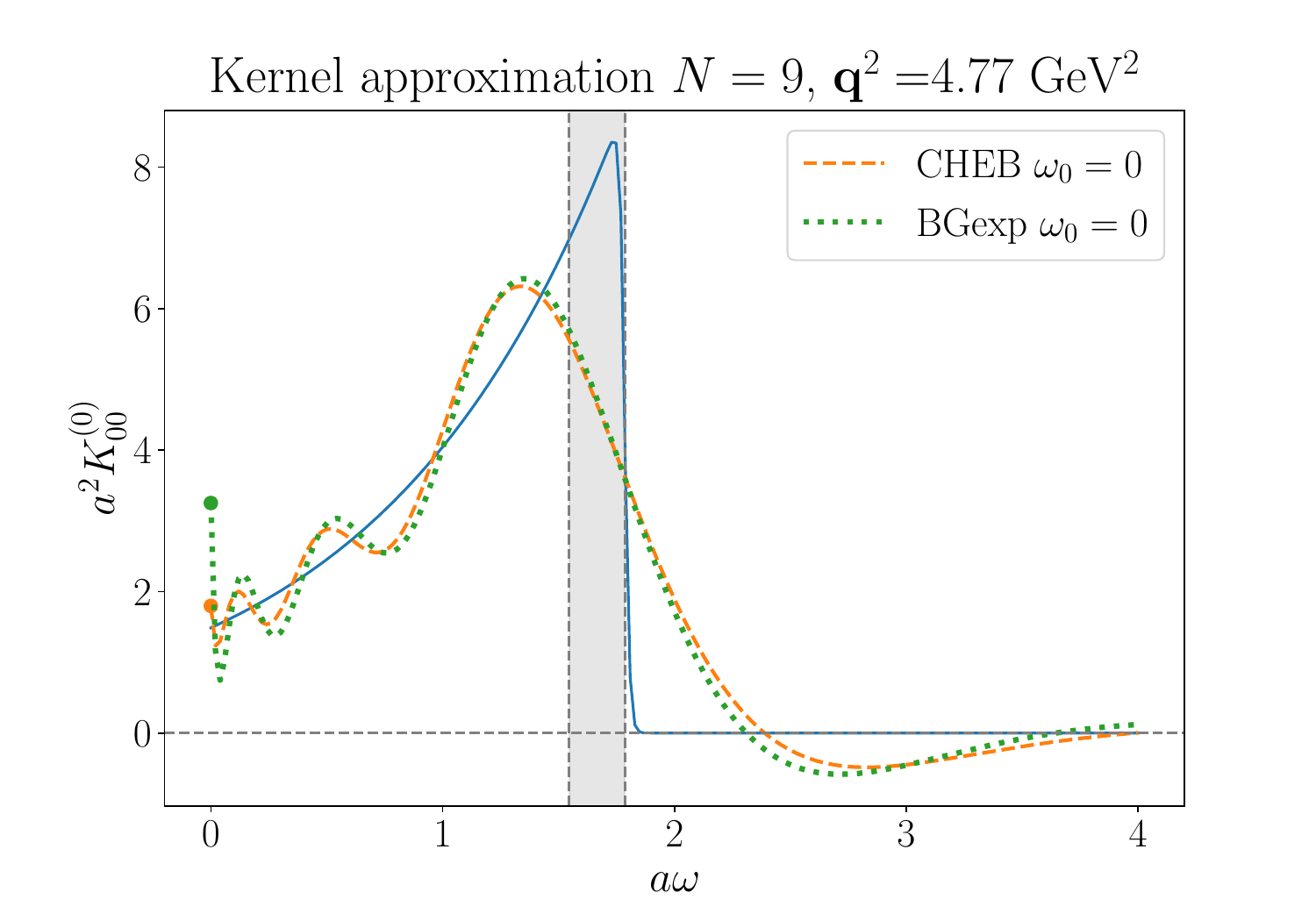}
 \hspace{-0.5 cm}
 \includegraphics[scale=0.3]{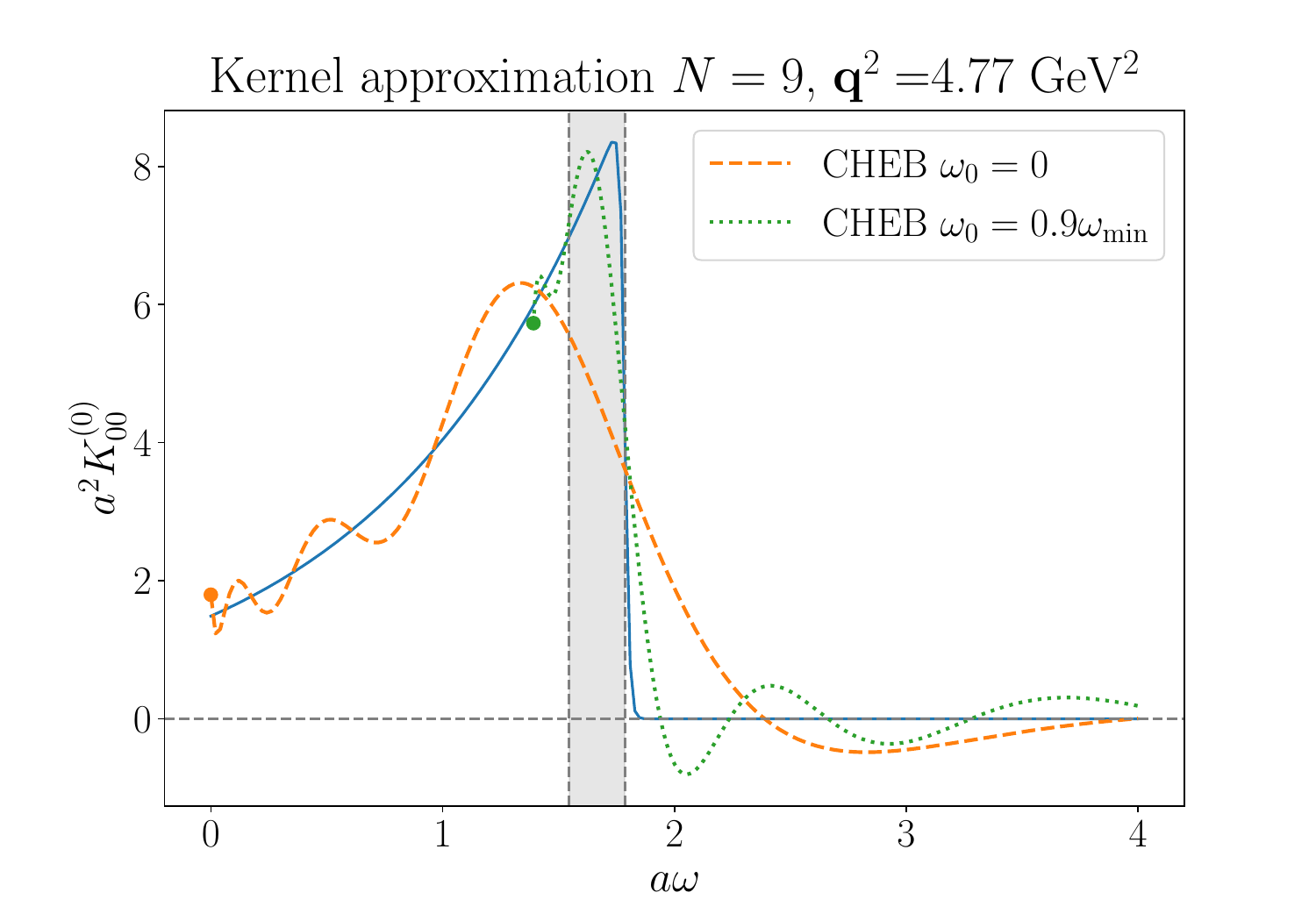}
 }
 \caption{Comparison between Chebyshev and Backus-Gilbert (with exponential basis) approach with $N=9$ at $\omega_0=0$ (left)
 and comparison between Chebyshev approach with different values of $\omega_0$ (right) for kernel $K_{00}^{(0)}$ at $\bm{q}^2=4.77\text{\,GeV}^2$. The solid
 blue line shows the target kernel function with a smearing $\sigma=0.02$.}
 \label{fig:kernelZoom}
\end{figure}

\begin{figure}[h!]
 \hbox{
 \includegraphics[scale=0.3]{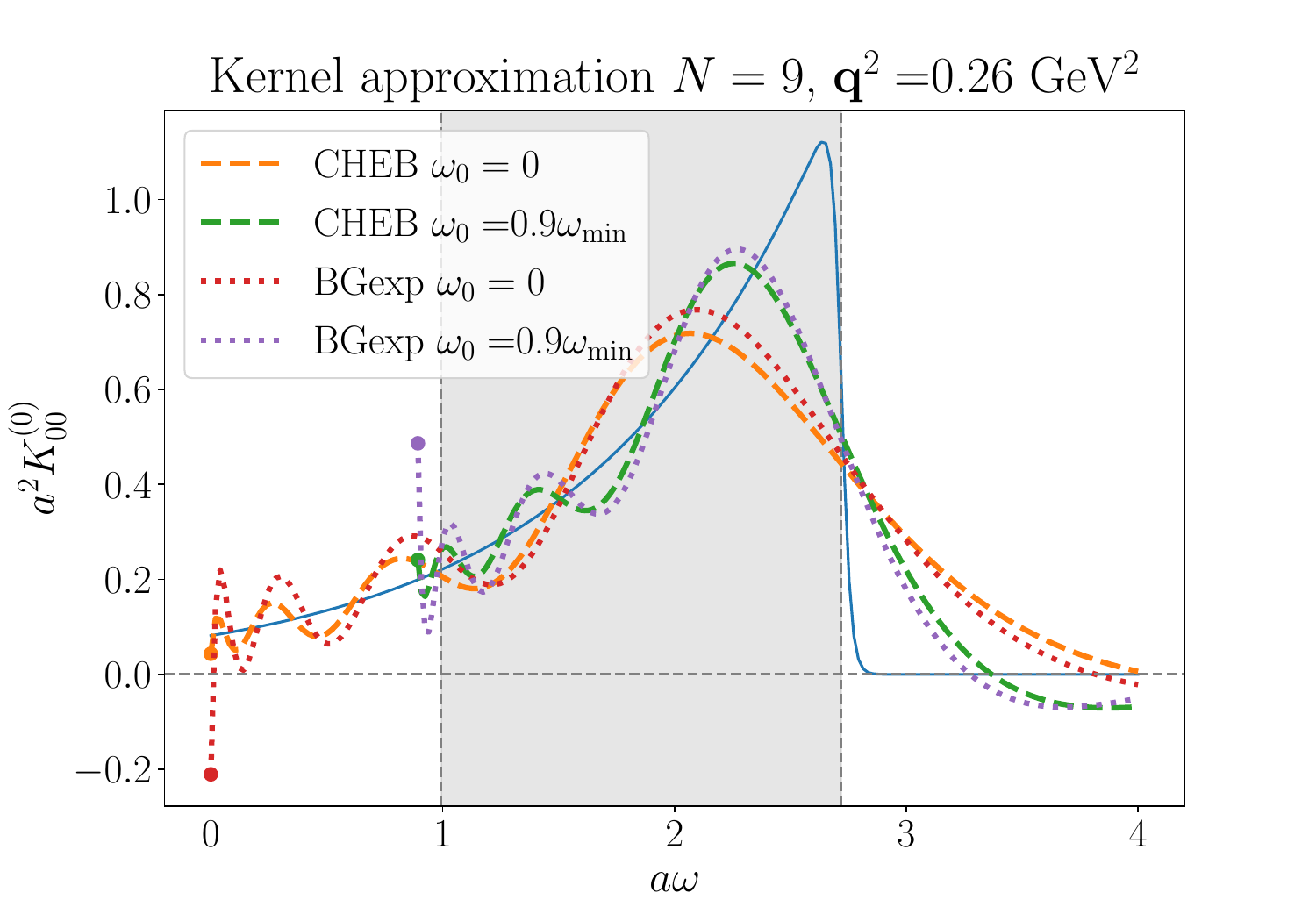}
 \hspace{-0.5 cm}
 \includegraphics[scale=0.3]{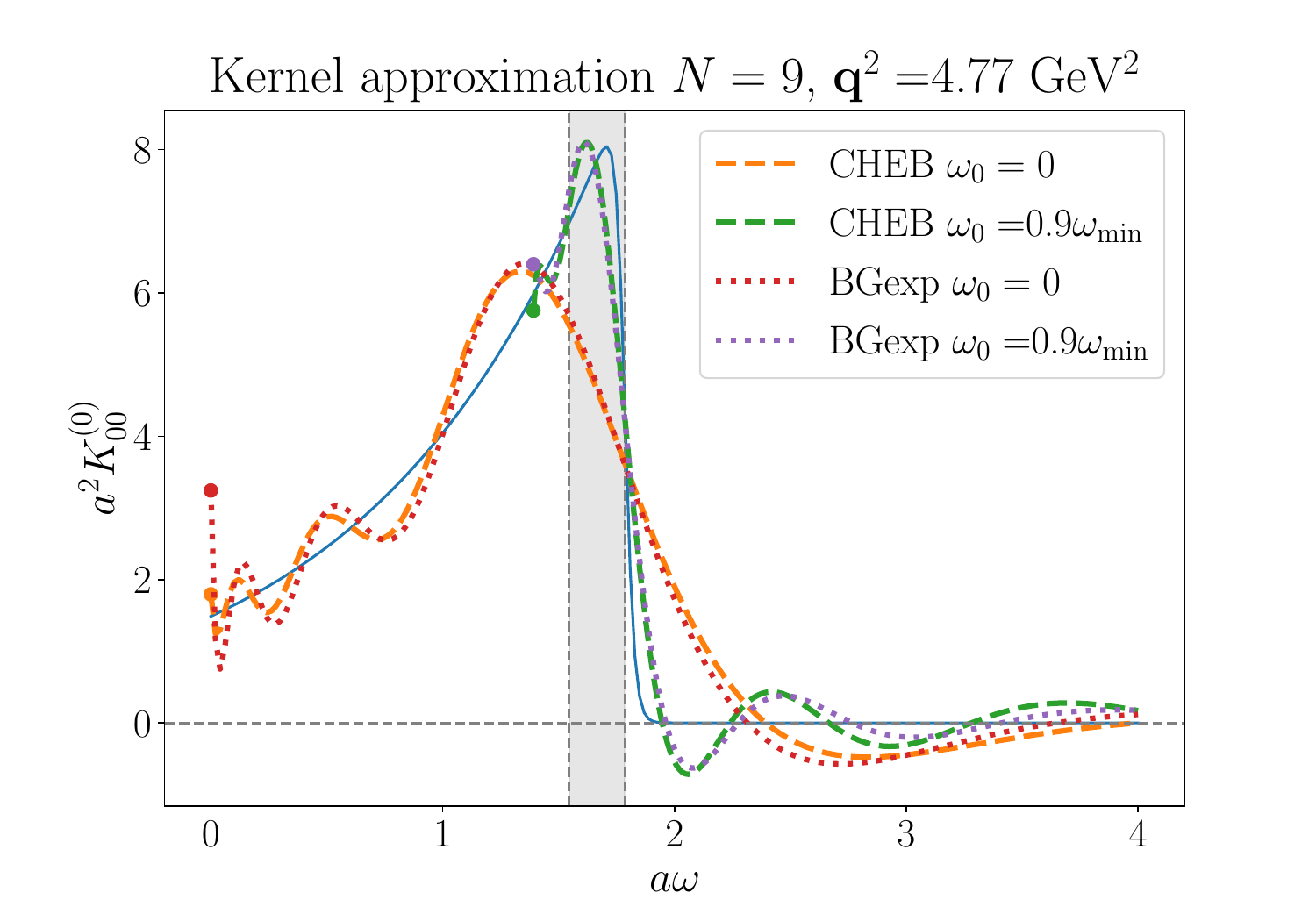}
 }
 \hbox{
 \includegraphics[scale=0.3]{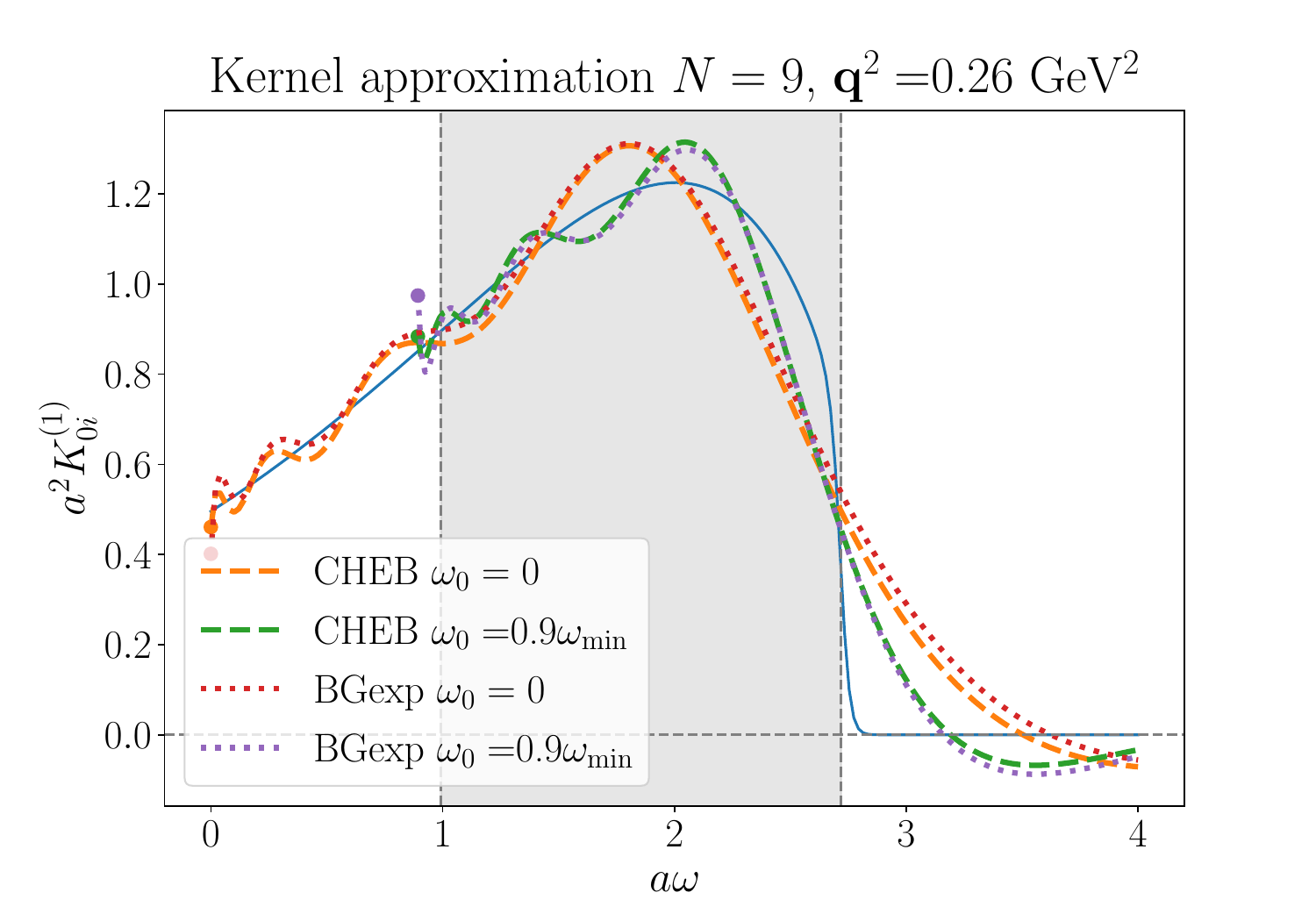}
 \hspace{-0.5 cm}
 \includegraphics[scale=0.3]{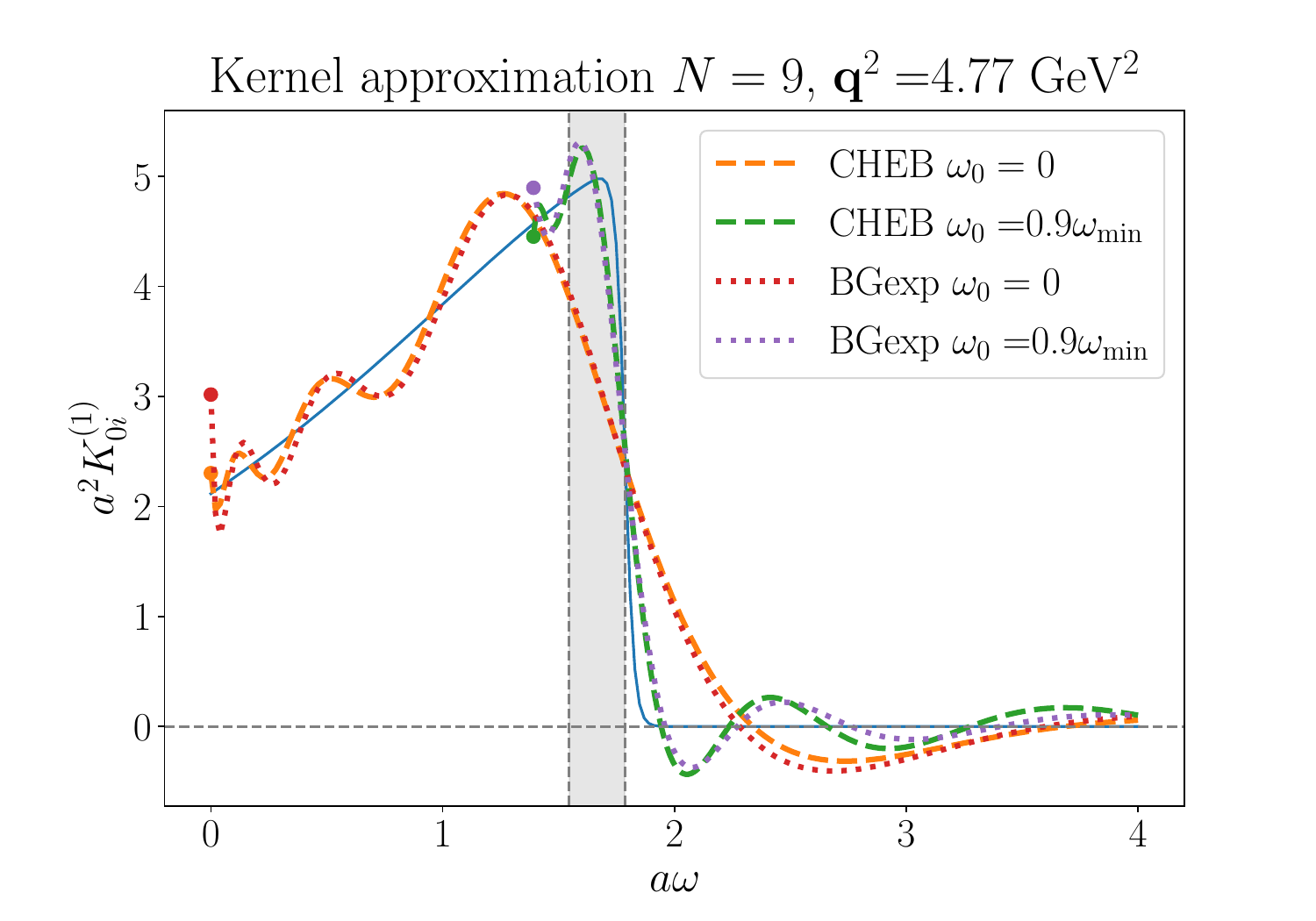}
 }
 \hbox{
 \includegraphics[scale=0.3]{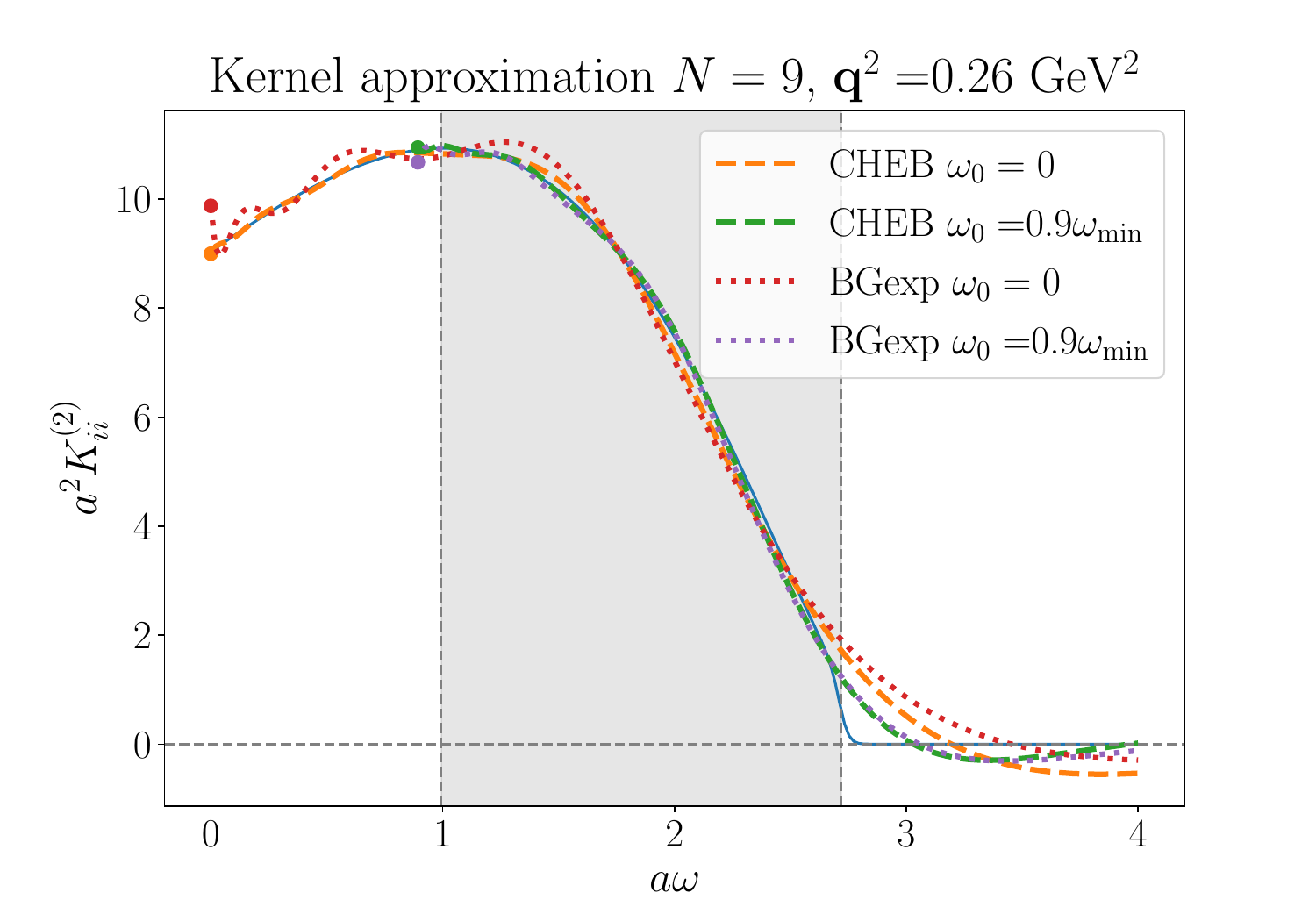}
 \hspace{-0.5 cm}
 \includegraphics[scale=0.3]{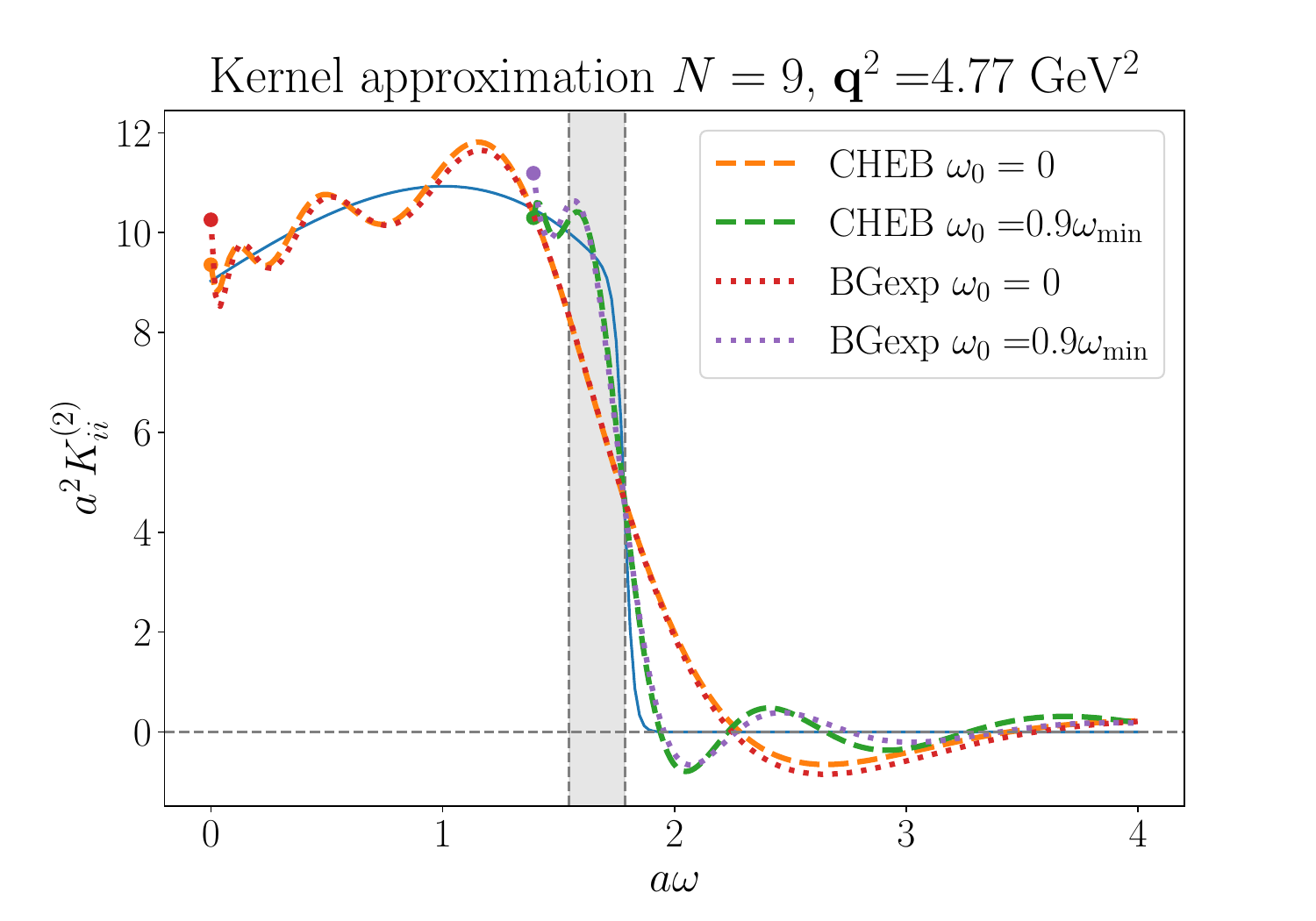}
 }
 \caption{
 Polynomial approximation at order $N=9$ of the kernel $K_{\mu\nu, \sigma}^{(l)}(\bm{q}, \omega; 2t_0)$, for $l=0$ (first row), $l=1$ (second row)
 and $l=2$ (third row) with $t_0=1/2$ and $\sigma=0.02$. The left column shows the case of the smallest $\bm{q}^2=0.26 \, \text{GeV}^2$, whereas the right column shows one of the largest momentum $\bm{q}^2=4.77 \, \text{GeV}^2$.
 The grey area corresponds to the kinematically
 allowed range $\omega_{\rm min} \leq \omega \leq \omega_{\rm max}$ for the given $\bm{q}^2$. The solid lines show the target function; the dashed lines show the approximation
 with the Chebyshev approach, whereas the dotted ones
 show the approximation with Backus-Gilbert with an exponential base and $\lambda=0$.
 }
 \label{fig:kernel}
\end{figure}

In \reffig{fig:kernelZoom} we highlight some of the key features of our approach
and in \reffig{fig:kernel} we show the approximation for different kernels $K^{(l)}_{\mu\nu,\sigma}$ with $l=0,1,2$.
The plots are for the smallest and one of the largest $\bm{q}^2$ computed, respectively.
Here we illustrate the case of $\sigma=0.02$, which smoothes the step function only mildly. Later we will also discuss the case of larger values
of $\sigma$.

Some comments are in order. First of all, we point out that with the current data set, the polynomial order $N=9$ is the maximum value available. This depends on the size of the lattice
and the choice of $t_{\rm src}$, $t_2$ and $t_{\rm snk}$ in the four-point correlator. In particular, setting $a=1$,
the available time slices are $2t_0\leq t < t_2-t_{\rm src}$, which in our case
correspond to $1\leq t < 14$. On top of that, we need to make sure that $t\ll t_2-t_{\rm src}$, i.e. $t_1-t_{\rm src}\gg 0$:
the choice $N=9$ corresponds to a separation $t_1-t_{\rm src}=4$.
Of course, with an improved data set $N$ could be chosen larger and the differences between the two approaches would reduce further.

We also notice that the kernel with $l=0$ is the most delicate to treat, as it is the one that shows
the sharpest drop to zero at the threshold. Note also that for the case $l=0$ we plotted only $K^{(0)}_{00}$ as all the other kernels are the same up to a constant factor.
Secondly, as shown in \reffig{fig:kernelZoom} (left) the results for Chebyshev and Backus-Gilbert agree very well and the quality of the approximation seems comparable.

The quality of the approximation varies with $\omega_0$: as shown in Fig.~\ref{fig:kernelZoom} (right),
starting the approximation as close as possible to $\omega_{\rm min}$ gives the best result,
as the nodes of the interpolation
(the points where the target function and its polynomial reconstruction meet) are denser in the allowed phase space in energy (the grey shaded area).
This is most evident in the case of large $\bm{q}^2$, as $\omega_{\rm min}$ is moved further away from 0.
This is then the region where we expect larger deviations
for the values of $\bar{X}^{(l)}(\bm{q}^2)$ between the two choices of $\omega_0$.
Note also that a value slightly below $\omega_{\rm min}$ (e.g. $0.9 \omega_{\rm min}$)
safeguards against statistical fluctuations in the ${D_s}$-meson mass.

\subsection{Chebyshev polynomials and Backus-Gilbert in practice}
\begin{figure}
 \centering
 \includegraphics[scale=0.19]{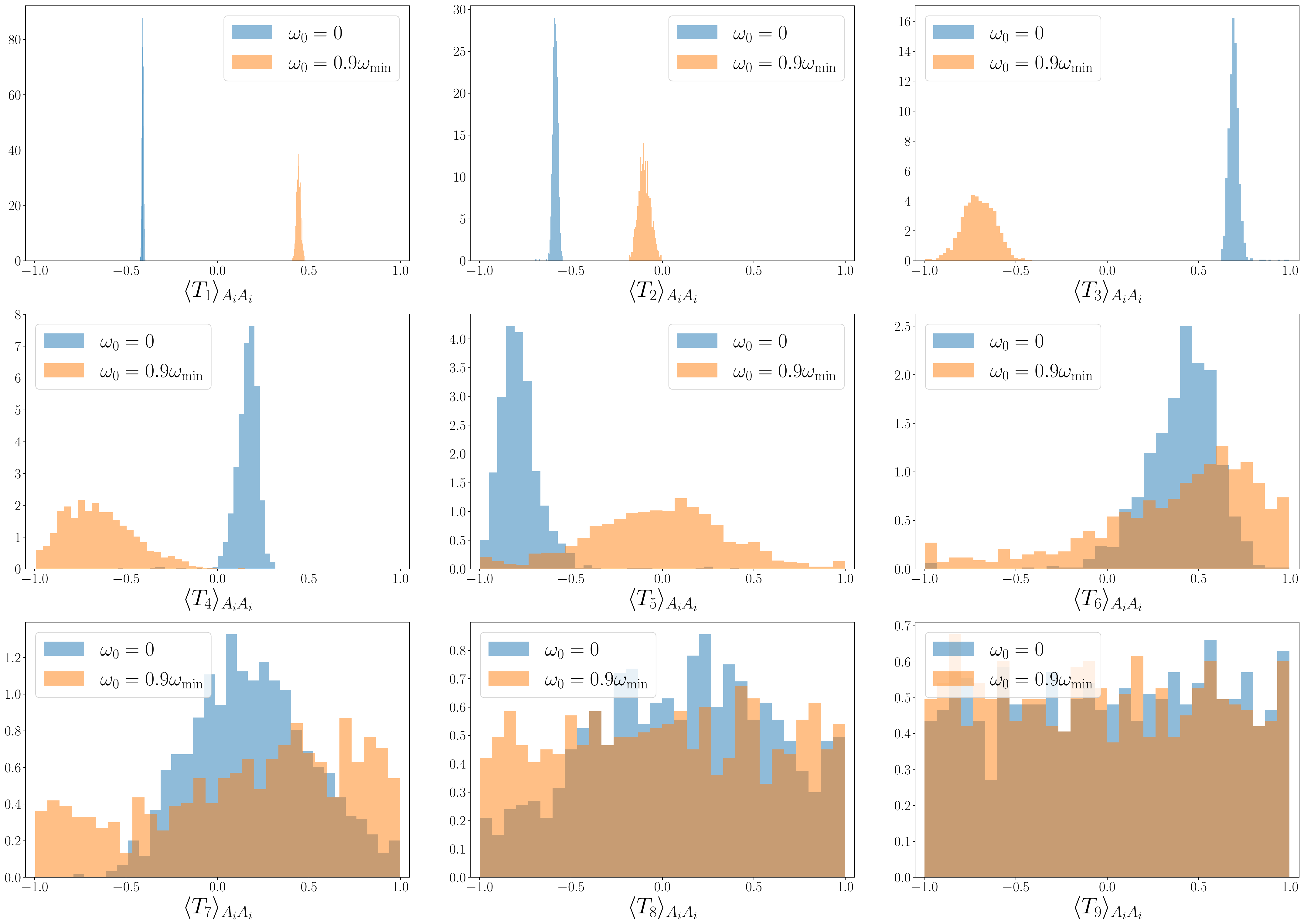}
 \caption{Histogram of the Chebyshev matrix elements $\langle \tilde{T}_k \rangle_{A_iA_i}$ for $k=1,2, \dots, N$ with $N=9$ for two values $\omega_0=0$ (blue) and
 $\omega_0 = 0.9 \omega_{\rm min}$ (orange) at $\bm{q}^2=0.26\, \text{GeV}^2$. The matrix element $\langle \tilde{T}_0 \rangle_{A_iA_i}=1$
 by definition and is therefore not shown. This channel is one of the most precise: we find that in both cases the fitting
 procedure is able to determine the matrix elements up to order $N\simeq 7$, after which the distribution of the bootstrap bins
 remains flat.}
 \label{fig:Tn_fit_AiAi}
\end{figure}

\begin{figure}[h!]
 \centering
 \includegraphics[scale=0.19]{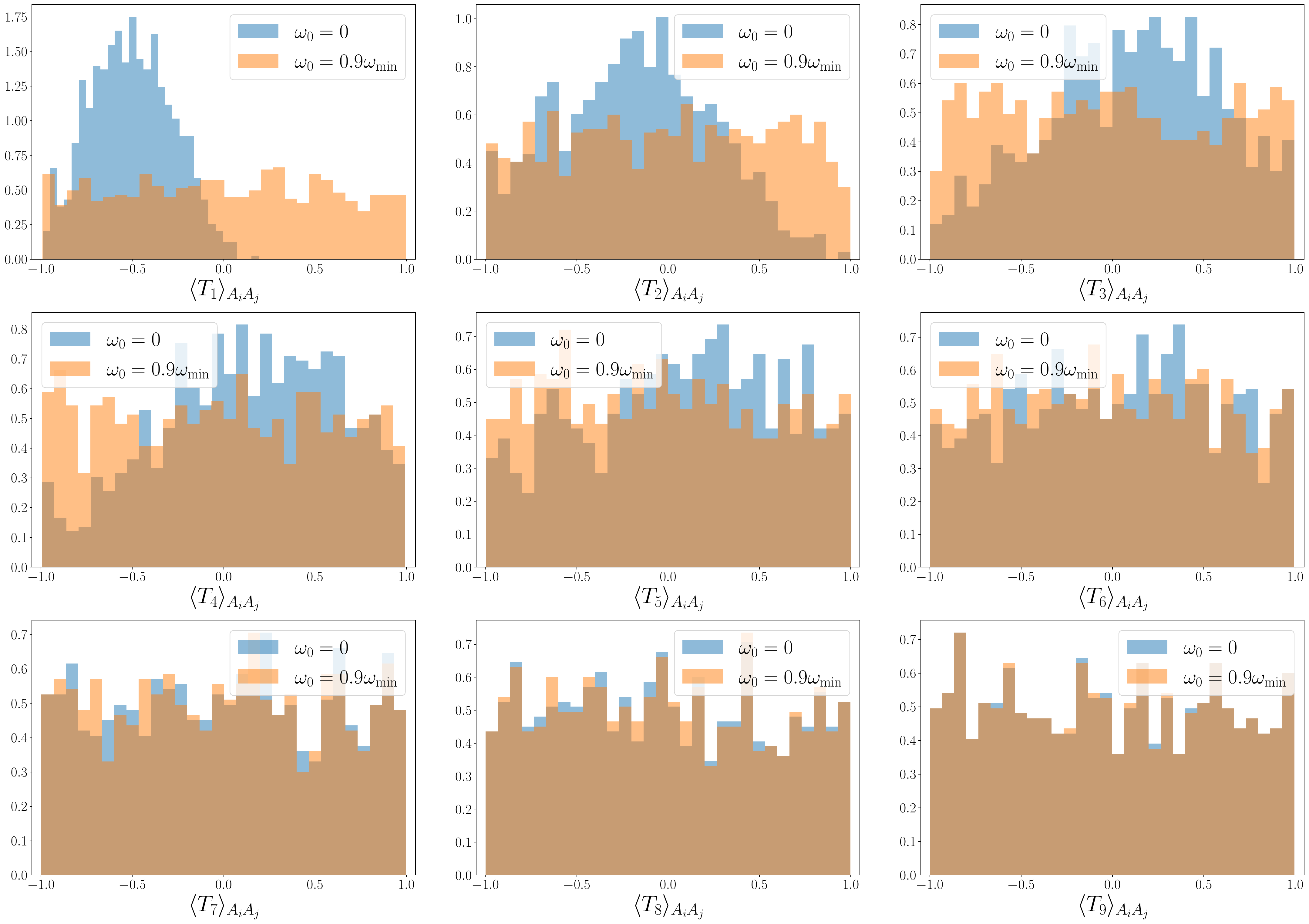}
 \caption{Histogram of the Chebyshev matrix elements $\langle \tilde{T}_k \rangle_{A_iA_j}$ with $i\neq j$ for $k=1,2, \dots, N$ with $N=9$ for two values $\omega_0=0$ (blue) and
 $\omega_0 = 0.9 \omega_{\rm min}$ (orange) at $\bm{q}^2=0.26\, \text{GeV}^2$. The results for $\langle \tilde{T}_k \rangle_{A_iA_j}$ are less well constrained than the ones for  
 $A_i A_i$ shown in Fig.~\ref{fig:Tn_fit_AiAi}. The minimum of the $\chi^2$ is determined almost entirely
 by the uniform priors.}
 \label{fig:Tn_fit_ij}
\end{figure}

We now discuss the quality of the data analysis as outlined in \refsec{sec:analysis}.
Focusing  first on the Chebyshev-polynomial approach, the correlator data are traded with the fitted Chebyshev matrix elements as
\begin{align}
 \bar{C}^{\rm fit}_{\mu\nu}(k) = \sum_{j=0}^{k} \tilde{a}^{(k)}_j \langle \tilde{T}_j \rangle_{\mu\nu} \, ,
 \label{eq:Cfit_Tn}
\end{align}
where the coefficients $\tilde{a}_{j}^{(k)}$ are given by the power representation of the Chebyshev polynomials, see \refapp{sec:app_cheb}.
Following \eqref{eq:kernel_cbar}, the kernel with fitted Chebyshev matrix elements can be written as
\begin{align}
\langle K^{(l)}_{\sigma} \rangle_{\mu\nu} 
 = \frac{\tilde{c}^{(l)}_{\mu\nu,0} }{2}\langle \tilde{T}_0 \rangle_{\mu\nu} + \sum_{k=1}^{N}\tilde{c}^{(l)}_{\mu\nu,k} \langle \tilde{T}_k \rangle_{\mu\nu}
 = \sum_{k=0}^{N}\bar{c}^{(l)}_{\mu\nu,k} \bar{C}^{\rm fit}_{\mu\nu}(k) \, .
\end{align}
An example of the Chebyshev matrix elements obtained from the fits can be seen in \reffig{fig:Tn_fit_AiAi}, 
where we compare two different extractions according to the starting point of the approximation $\omega_0$.
The plots show the distribution of each order of the Chebyshev matrix elements obtained through the fitting procedure described
in \ref{sec:app_fit}: each histogram plots values obtained for all the 1000 bootstrap bins.
We show the axial channel $A_iA_i$, as its signal turns out to be particularly clean.
In \reffig{fig:Tn_fit_ij} we show results for the $A_iA_j$ channel, with $i\neq j$, which is found to be the noisiest channel.
Here, only few terms can be determined meaningfully by the lattice data. Higher-order terms just follow the flat prior distribution in $[-1,1]$.

In both cases we observe that a larger number of Chebyshev matrix elements can be determined meaningfully for $\omega_0=0$ than for $\omega_0=0.9\, \omega_{\rm min}$.
For example, in the $A_iA_i$ channel the distribution of the former is close to the prior distribution, which is flat between $-1$ and $+1$, for  $N= 9$, whereas the latter start flattening at $N\gtrsim 7$. 
A possible explanation is as follows: as can be seen from \eqref{eq:app_a-t_relation},
$\tilde{a}^{(k)}_j |_{\omega_0=0}= e^{-0.9\omega_{\rm min}k} \tilde{a}^{(k)}_j |_{\omega_0=0.9 \omega_{\rm min}}$.
The additional exponential factor largely cancels the ground-state exponential decay in the correlation function in~\eqrefeq{eq:Cfit_Tn}.
Hence, the polynomial approximation has less structure to describe and higher-order terms become less relevant.
Nevertheless, in both cases the $\chi^2$ of the fits are acceptable and the reconstruction of the data as in \eqrefeq{eq:Cfit_Tn}
gives comparable results.

\begin{figure}[h!]
 \hbox{
 \includegraphics[scale=0.3]{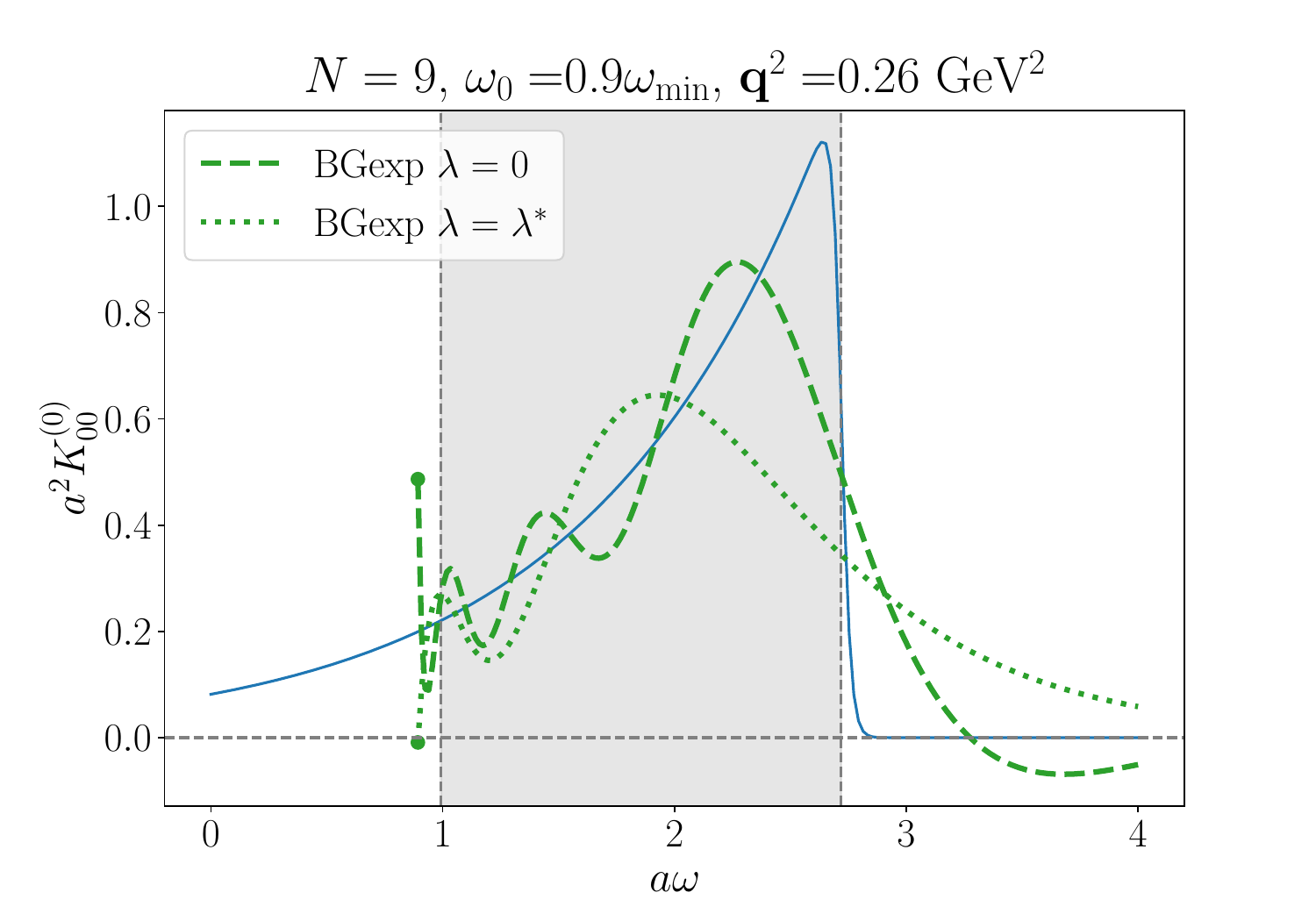}
 \hspace{-0.5 cm}
 \includegraphics[scale=0.3]{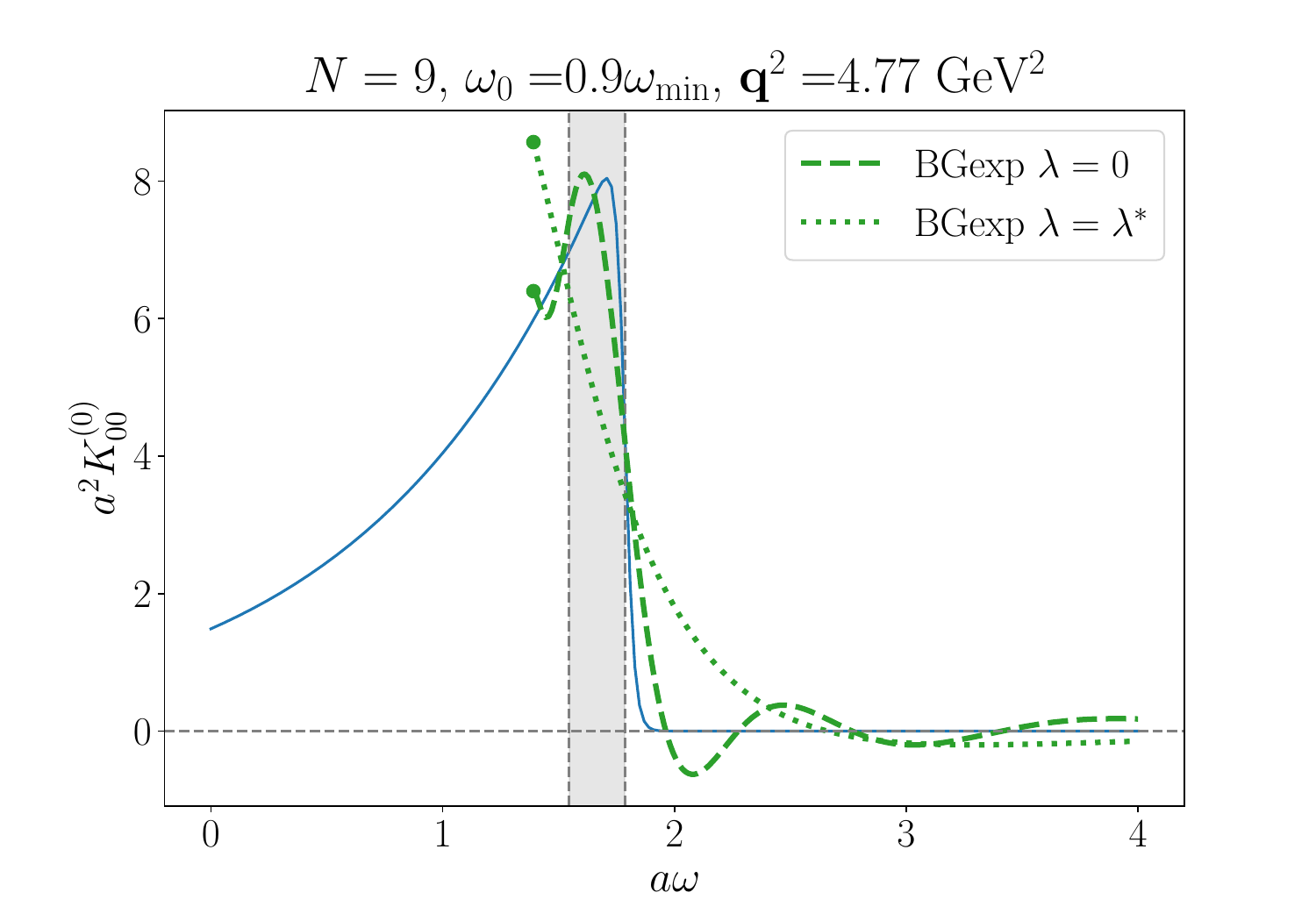}
 }
 \hbox{
 \includegraphics[scale=0.3]{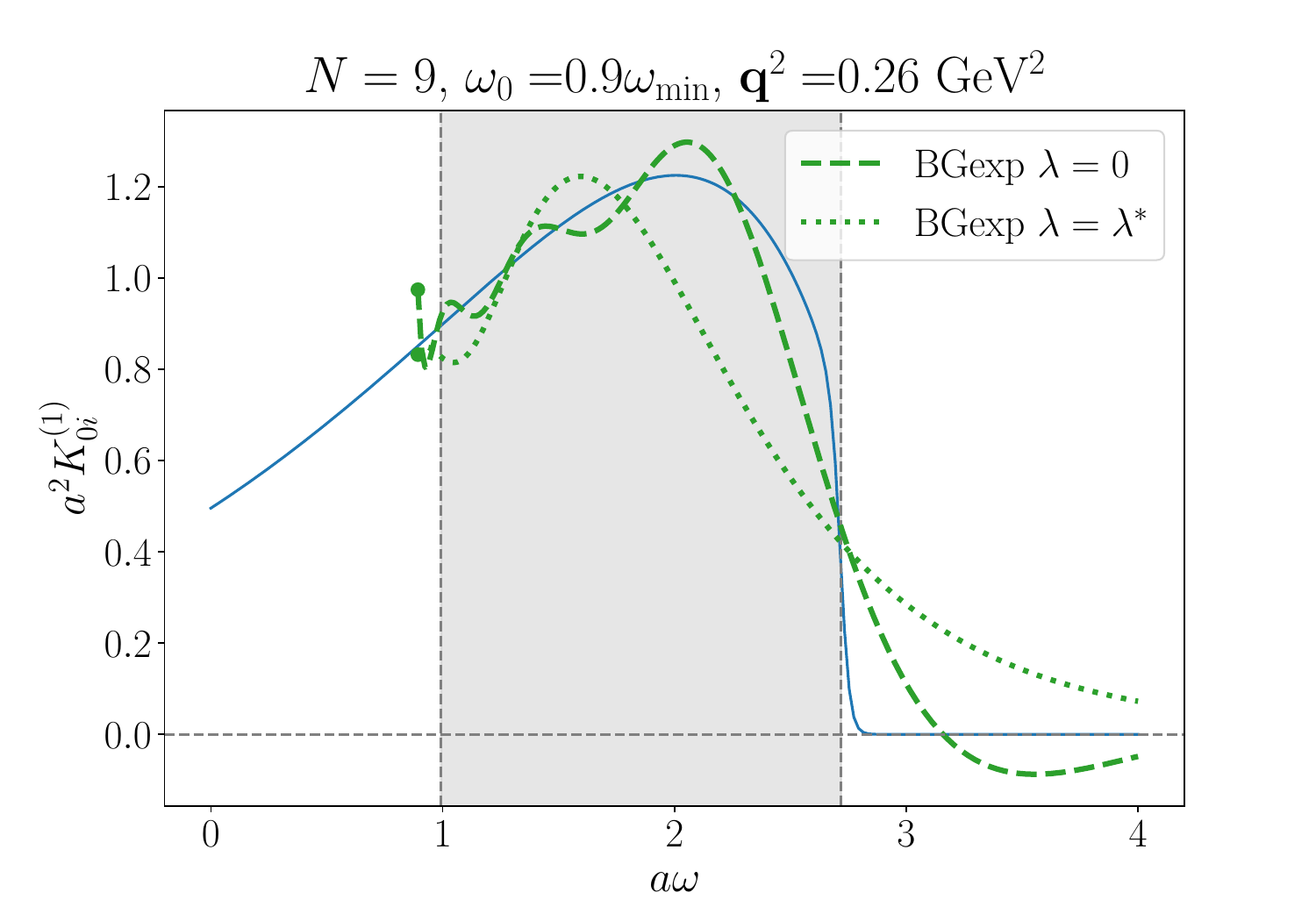}
 \hspace{-0.5 cm}
 \includegraphics[scale=0.3]{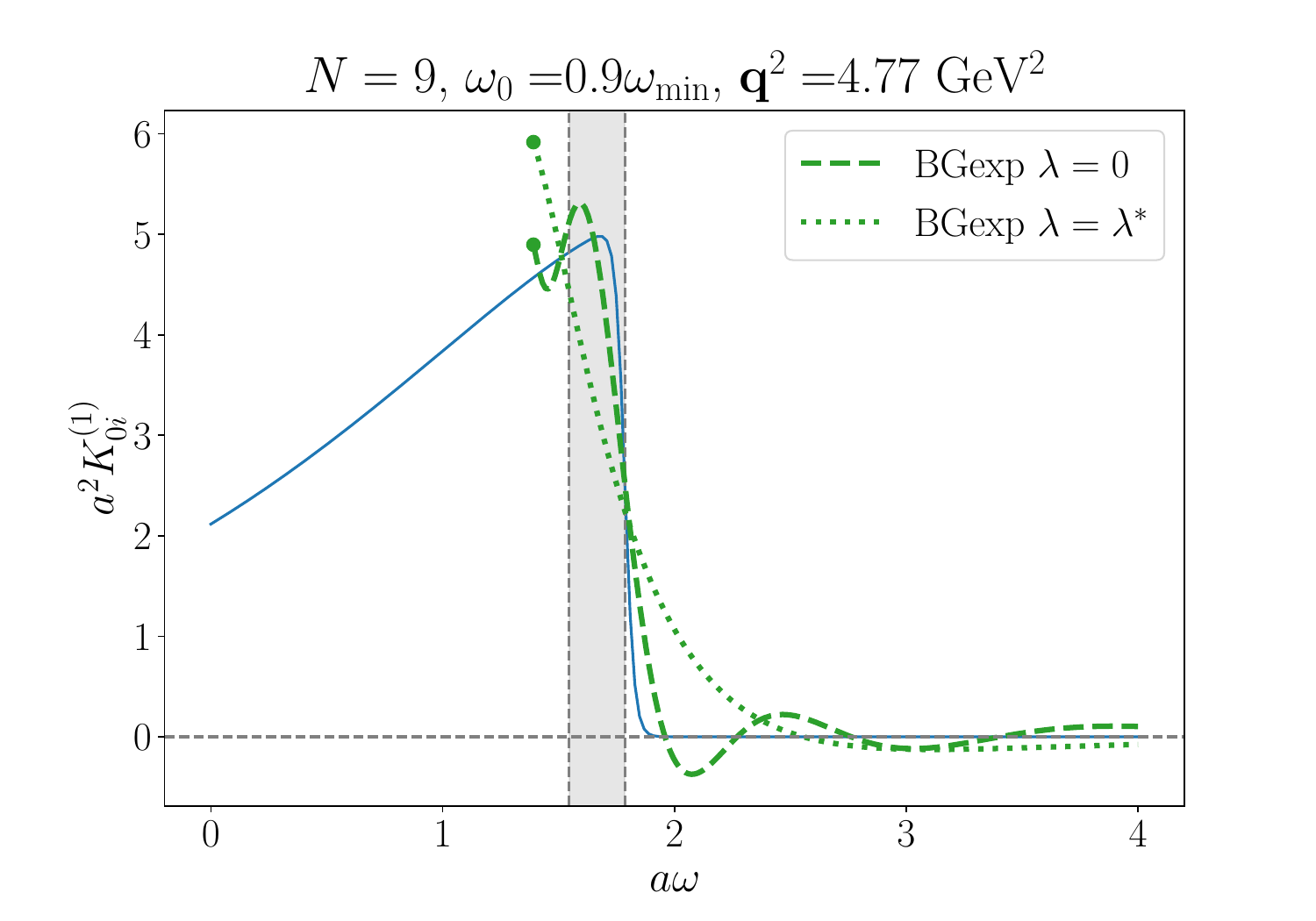}
 }
 \hbox{
 \includegraphics[scale=0.3]{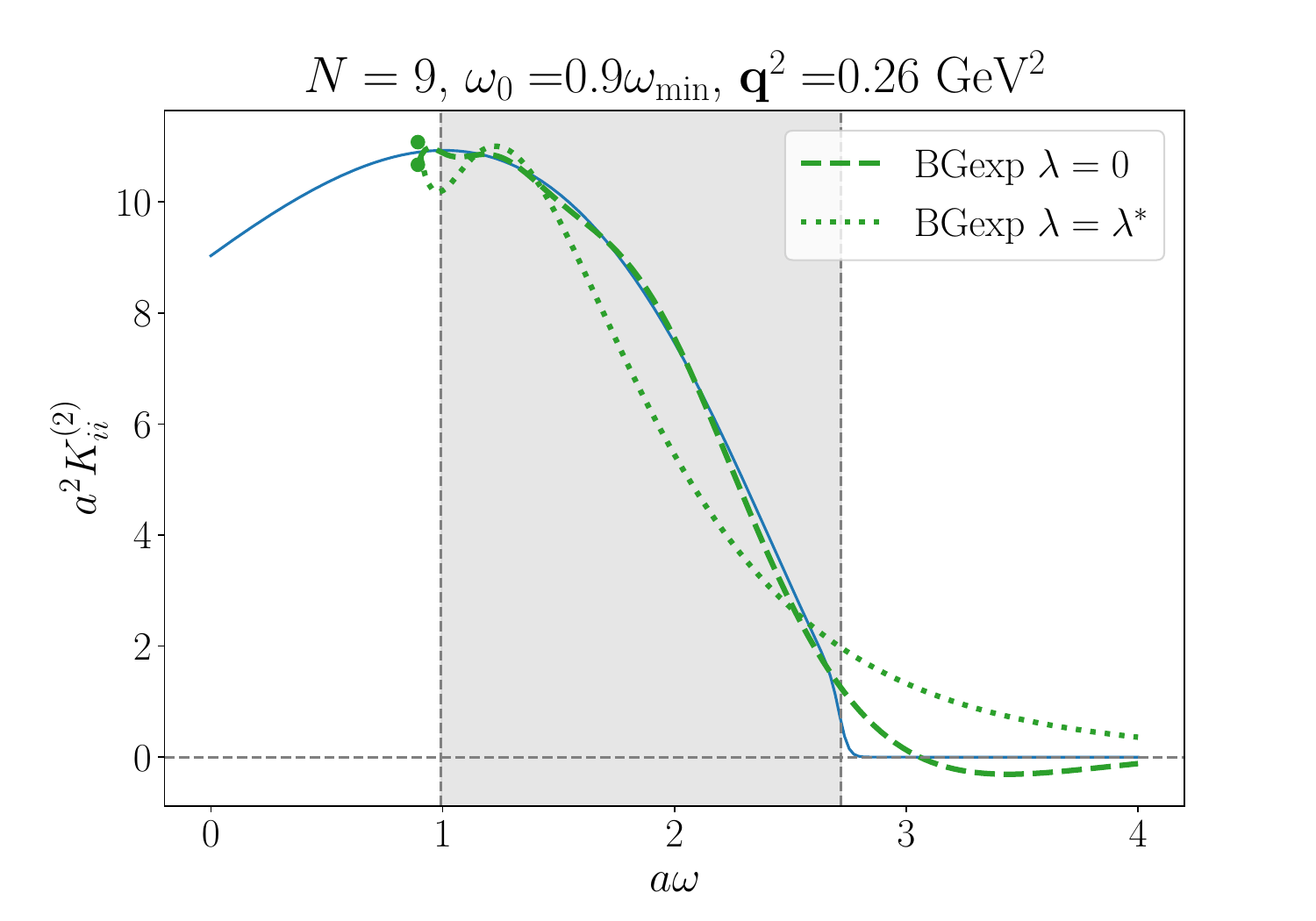}
 \hspace{-0.5 cm}
 \includegraphics[scale=0.3]{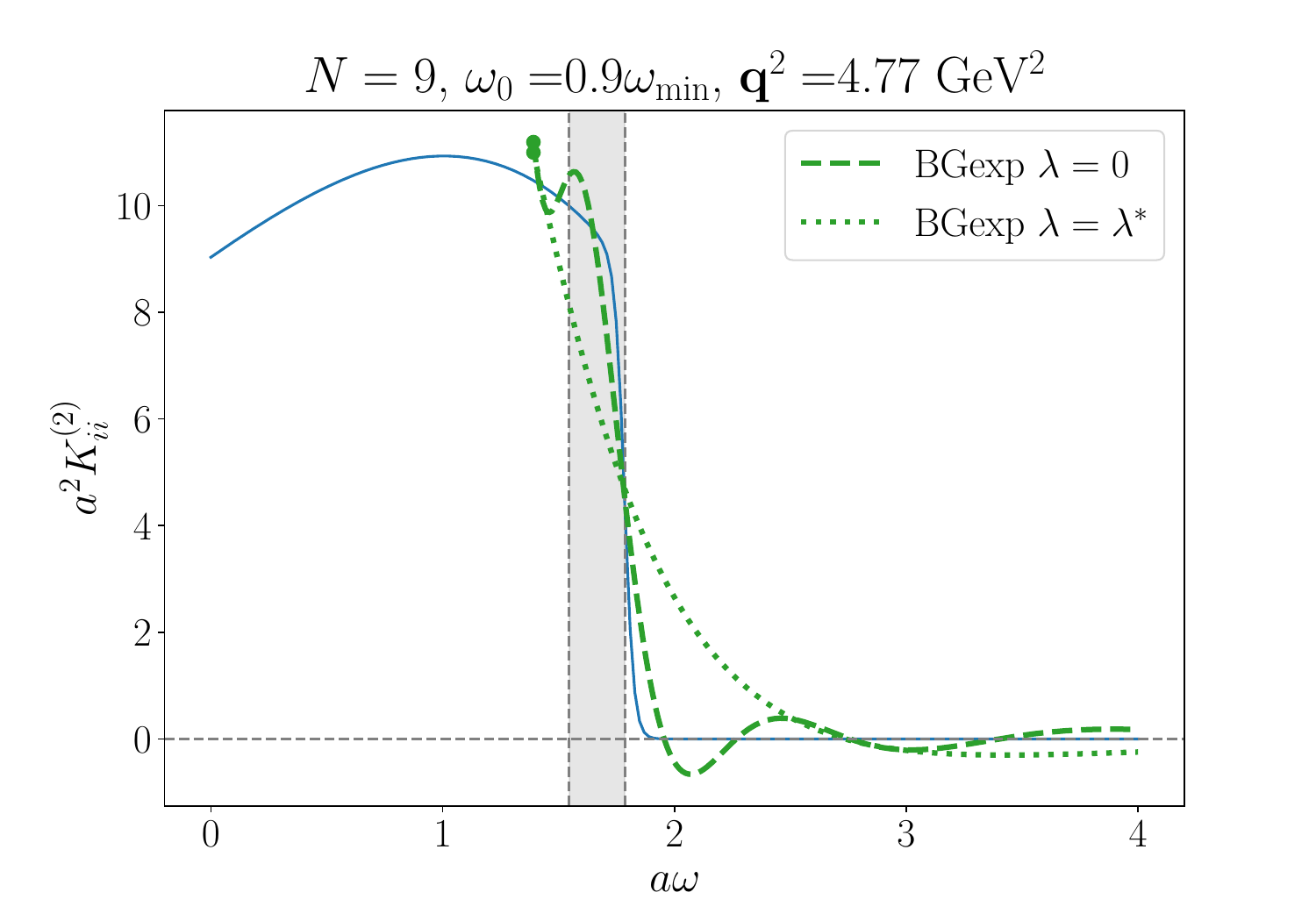}
 }
 \caption{Polynomial approximation of the kernel $K_{\mu\nu, \sigma}^{(l)}(\bm{q}, \omega; 2t_0)$, for $l=0$ (first row), $l=1$ (second row)
 and $l=2$ (third row) with $t_0=1/2$ and $\sigma=0.02$ in the case of Backus-Gilbert with exponential basis and $\lambda\neq 0$.
 The value of $\lambda$  has been chosen to be $\lambda^{*}$ for each plot, which gives
 equal weight to the statistical and systematic errors.}
 \label{fig:kernelBG}
\end{figure}

We now move to the Backus-Gilbert case, for which we have so far only considered the limit $\lambda=0$. In this limit the coefficients of the polynomial approximation are determined without reference to the data. We then consider the case $\lambda\neq 0$ and, 
by visual inspection of \reffig{fig:kernelBG},
find that the polynomial approximation of the kernel function
gets worse.
The effect of non-zero $\lambda$ can be understood as a correction to the optimal coefficients, as outlined in \refsec{sec:app_BG_diff_persp}.
In particular, if we rewrite
the coefficients as $g_{\mu\nu,k}^{*(l)} = \gamma_{\mu\nu,k}^{(l)} + \epsilon_{\mu\nu, k}^{*(l)}$ we have
\begin{align}
 \langle K^{(l)}_{\sigma} \rangle_{\mu\nu} 
 = \sum_{k=0}^{N}g^{*(l)}_{\mu\nu,k} \bar{C}_{\mu\nu}(k)
 = \sum_{k=0}^{N}\gamma^{(l)}_{\mu\nu,k} \bar{C}_{\mu\nu}(k) + \sum_{k=0}^{N}\epsilon^{*(l)}_{\mu\nu,k} \bar{C}_{\mu\nu}(k) \, , 
 \label{eq:kernelBG_epsilon}
\end{align}
where $\gamma_{\mu\nu}^{(l)}$ are the coefficients for $\lambda=0$ and $\epsilon^{*(l)}_{\mu\nu,k}$ is a correction which takes care of reducing the noise coming from the statistical error.

\subsection{The inclusive decay rate}

\begin{figure}[t!]
 \centering
 \includegraphics[scale=0.4]{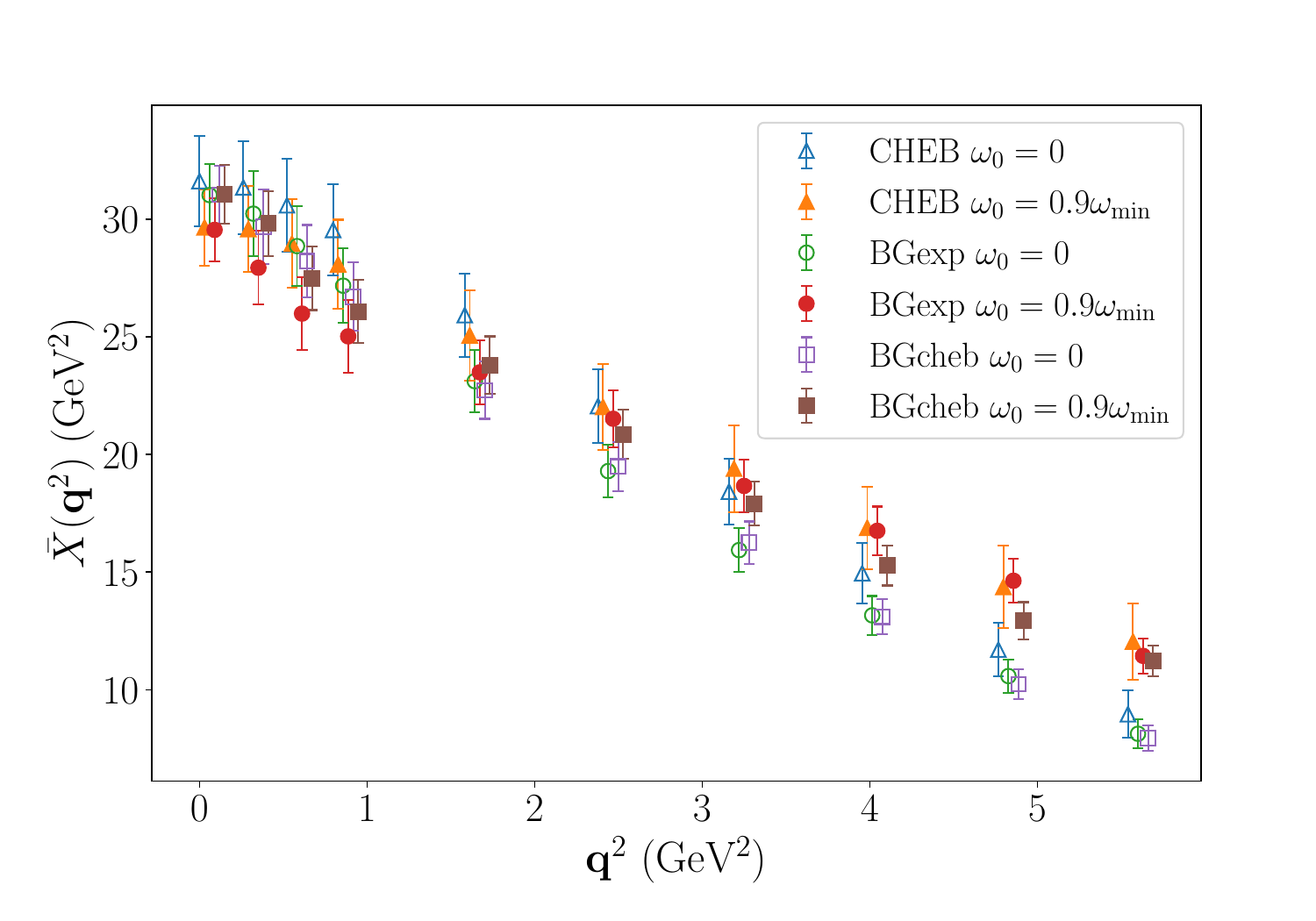}
 \caption{Estimate of $\bar{X}(\bm{q}^2)$ with the two different strategies for 10 different $\bm{q}^2$ with $N=9$ and $\bm{q}^2_{\rm max}=5.86 \, \text{GeV}^2$.
}
 \label{fig:results}
\end{figure}

In this section we present the main results of our work.
In \reffig{fig:results} we show the results of $\bar{X}(\bm{q}^2)$ for all 
the simulated values of $\bm{q}^2$. For each simulation point 
we show the results of three studied approaches, i.e., Chebyshev polynomials, exponential Backus-Gilbert and Chebyshev Backus-Gilbert, all of them for both $\omega_0=0$
and $\omega_0=0.9\, \omega_{\rm min}$. 
We  find that all sets of three points for a given value of $\omega_0$ agree very well.
However, sets with different $\omega_0$ start deviating as we increase the value of $\bm{q}^2$. As discussed in the previous section, this can be understood
in terms of the polynomial approximation of the kernel: as $\bm{q}^2$ increases, the phase space in $\omega$ shrinks, and the two approximations start differing increasingly.
Our data indicates that the approximation improves as $\omega_0 \rightarrow \omega_{\rm min}$. In order for the approximations for different $\omega_0$ to be comparable 
the order of the polynomial needs to be increased for lower $\omega_0$. It is also conceivable that other systematics like
finite-volume or cutoff effects  play a role here. These effects are beyond the scope of this work but will have to be addressed in future work.

\begin{figure}[t!]
 \centering
 \includegraphics[scale=0.4]{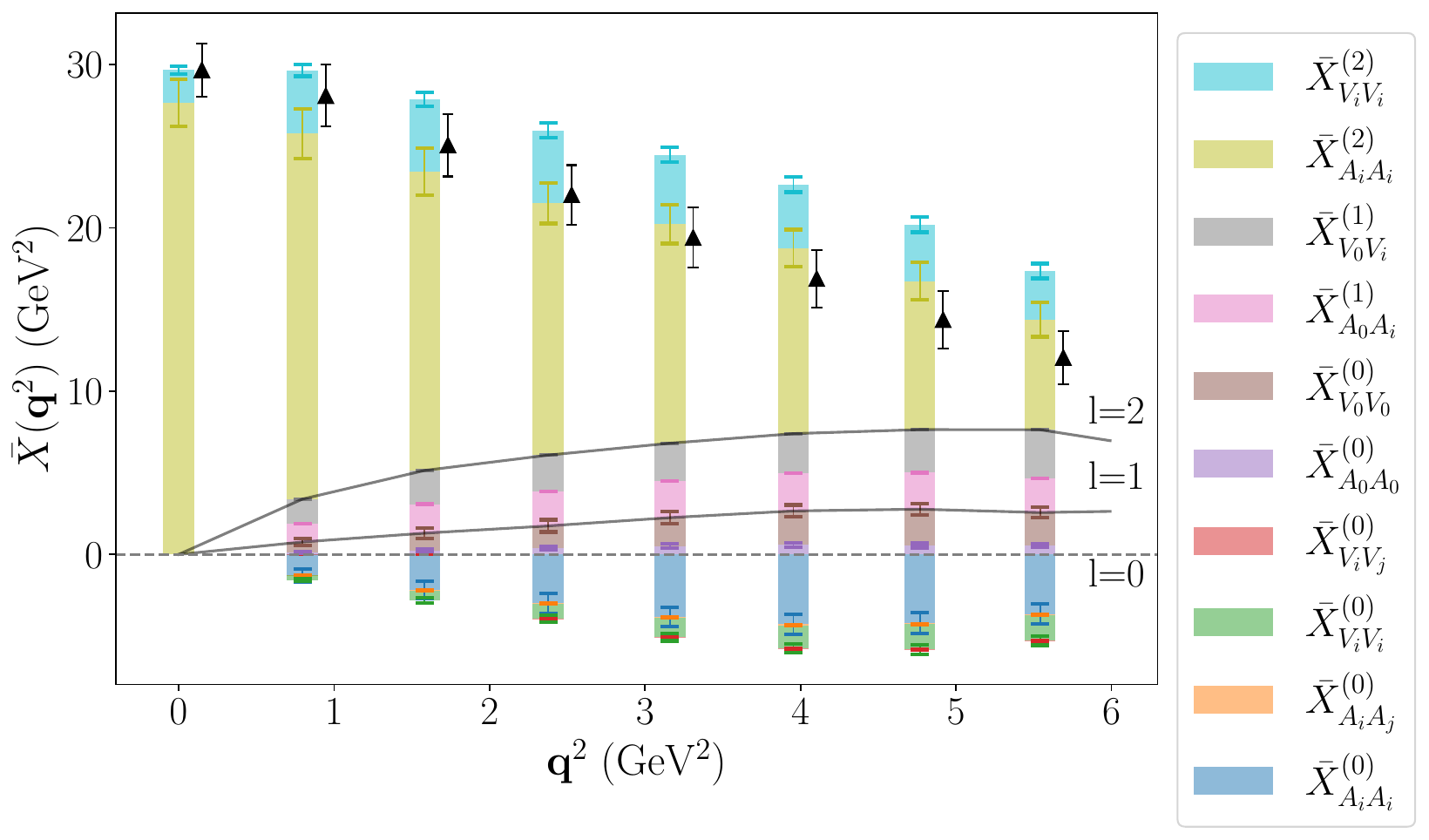}
 \caption{Contributions to $\bar{X}(\bm{q})$ from the Chebyshev-polynomial approach at $N=9$ and $\omega_0=0.9\omega_{\rm min}$ with associated error bars.
 The black triangles correspond to the final value $\bar{X}(\bm{q}^2) = \sum_{l=0}^{2}\sum_{\{\mu,\nu\}} \bar{X}^{(l)}_{\mu\nu}(\bm{q}^2)$.
 The solid black lines separate the contributions from $l=0$ (bottom), $l=1$ (middle) and $l=2$ (top).}
 \label{fig:Xbar_contribution}
\end{figure}

In the previous section we have seen that the shape of the kernel, and hence, the quality of approximation, varies substantially for different $l$ and $\bm{q}^2$. 
The degree to which this impacts the combined result $\bar{X}(\bm{q}^2)$ depends on the magnitude of each contribution, as illustrated in 
\reffig{fig:Xbar_contribution}.
The plots indicate that the largest contribution originates from the channel with $l=2$. The underlying kernel is,
at least for smaller values of $\bm{q}^2$, relatively smooth (\reffig{fig:kernel}).
We therefore expect less sensitivity to the systematics of the polynomial approximation in this 
kinematical region but more care is needed for larger $\bm{q}^2$. 

\begin{figure}[h!]
 \centering
 \hbox{
 \includegraphics[scale=0.3]{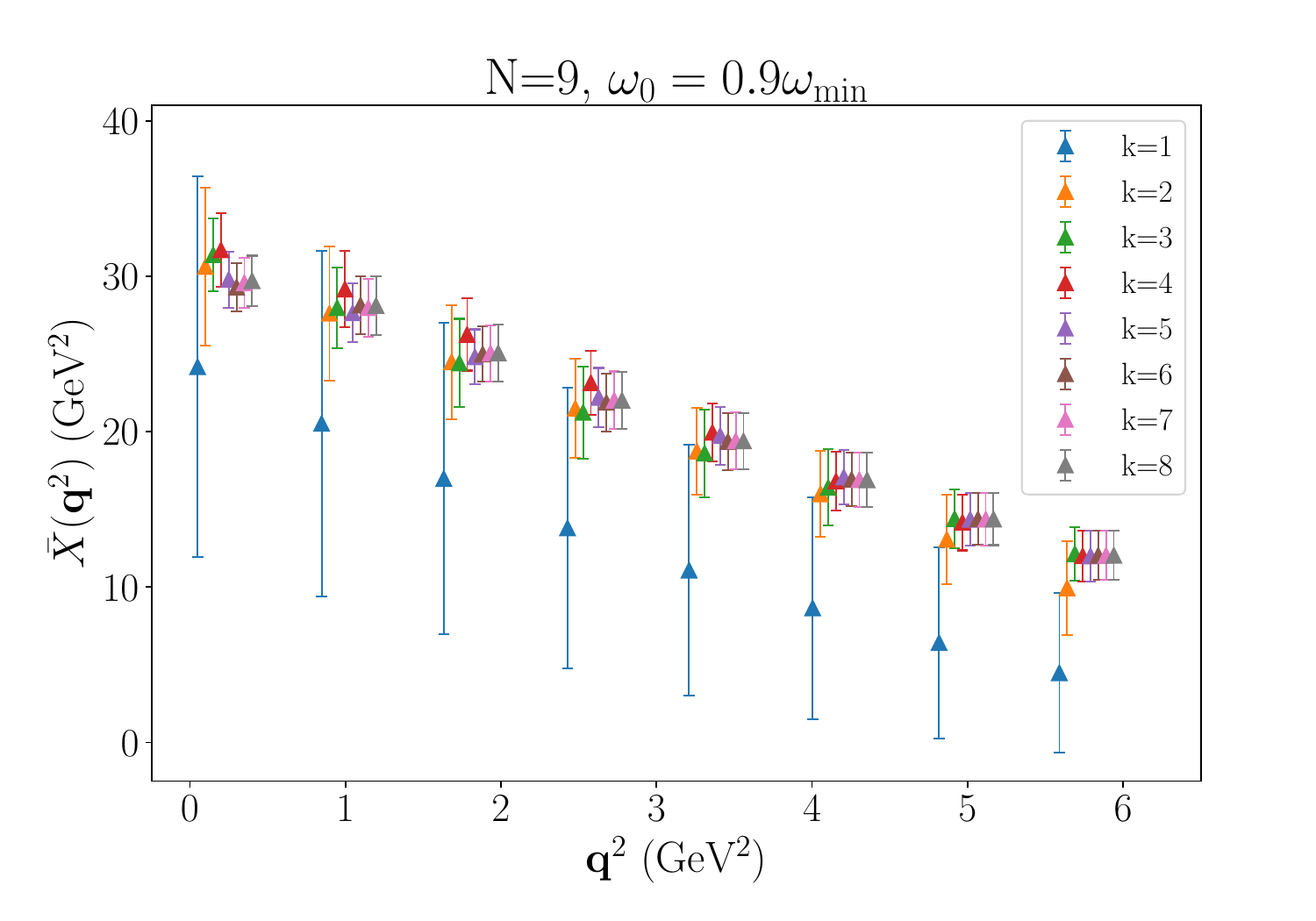}
 \hspace{-0.5cm}
 \includegraphics[scale=0.3]{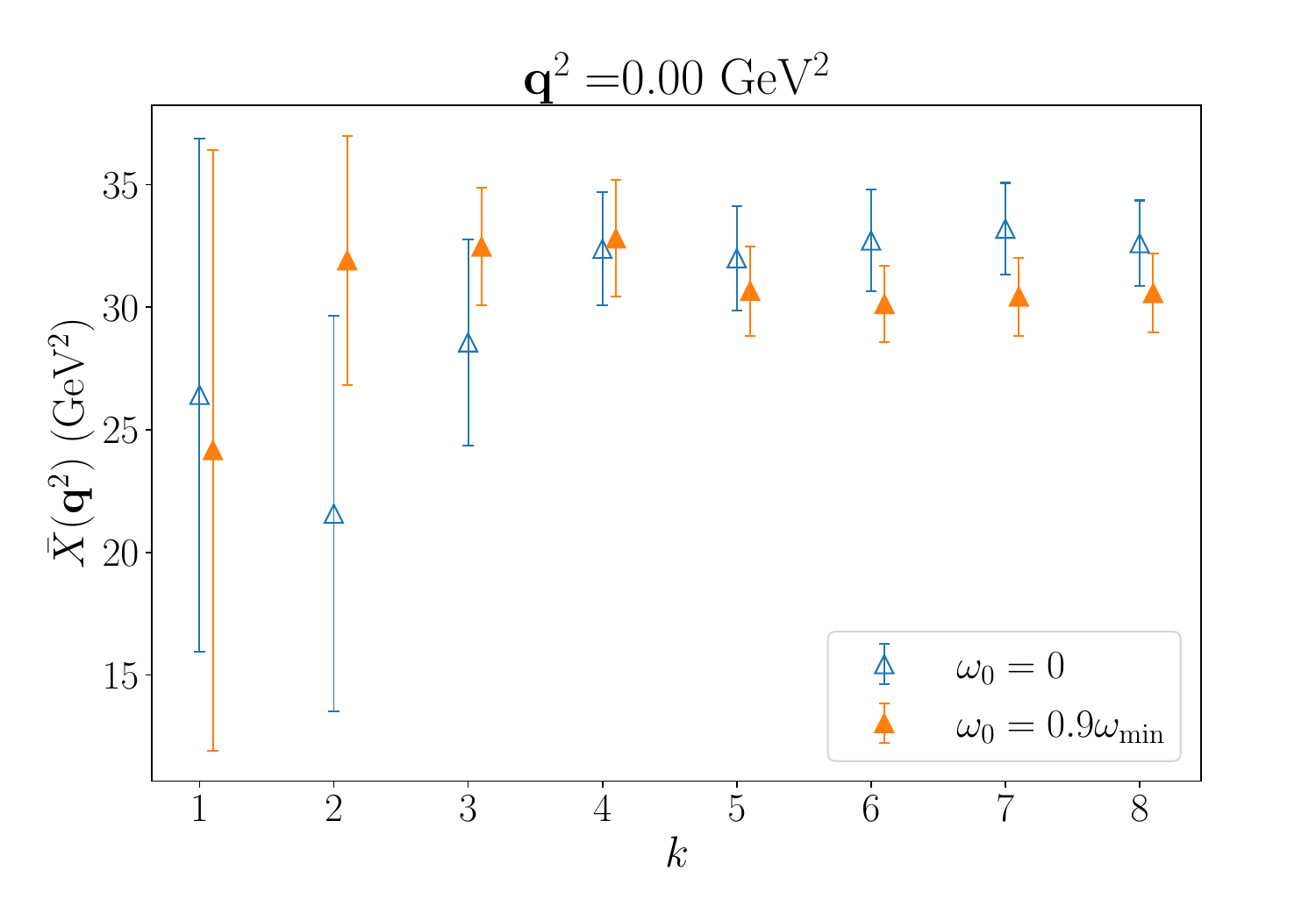}
 }
 \caption{Saturation of Chebyshev polynomial approach for different $\bm{q}^2$ and $\omega_0=0.9\omega_{\rm min}$ (left) and for case $\bm{q}^2=0$ for
 both values of $\omega_0$ as a function of $k$ (right), where
  $k$ is the number of Chebyshev matrix elements taken from the fit.}
 \label{fig:cheb_saturation}
\end{figure}

We now address the stability against the order of the polynomial $N$. Starting from the Chebyshev approach,
we study the saturation in \reffig{fig:cheb_saturation}.
We start from the fit with $N=9$. The plot  shows the result where the first $k$ Chebyshev matrix elements
(cf. legend) are taken from the fit, and the remaining $N-k$ are replaced by a flat distribution
$-1\le \langle \tilde{T}_{j}\rangle_{\mu\nu}\le 1$ with $j=k+1,\dots,N$.
We can see that the signal is dominated by small orders; for $\omega_0=0$, the signal is saturated at around $N\simeq 5$, whereas for $\omega_0=0.9\, \omega_{\rm min}$ saturation starts at  $N\simeq 3$. This is also compatible with the previous discussion on the fit of the Chebyshev matrix elements, cf. with \reffig{fig:Tn_fit_AiAi} and \reffig{fig:Tn_fit_ij}.

\begin{figure}[h!]
 \centering
 \includegraphics[scale=0.4]{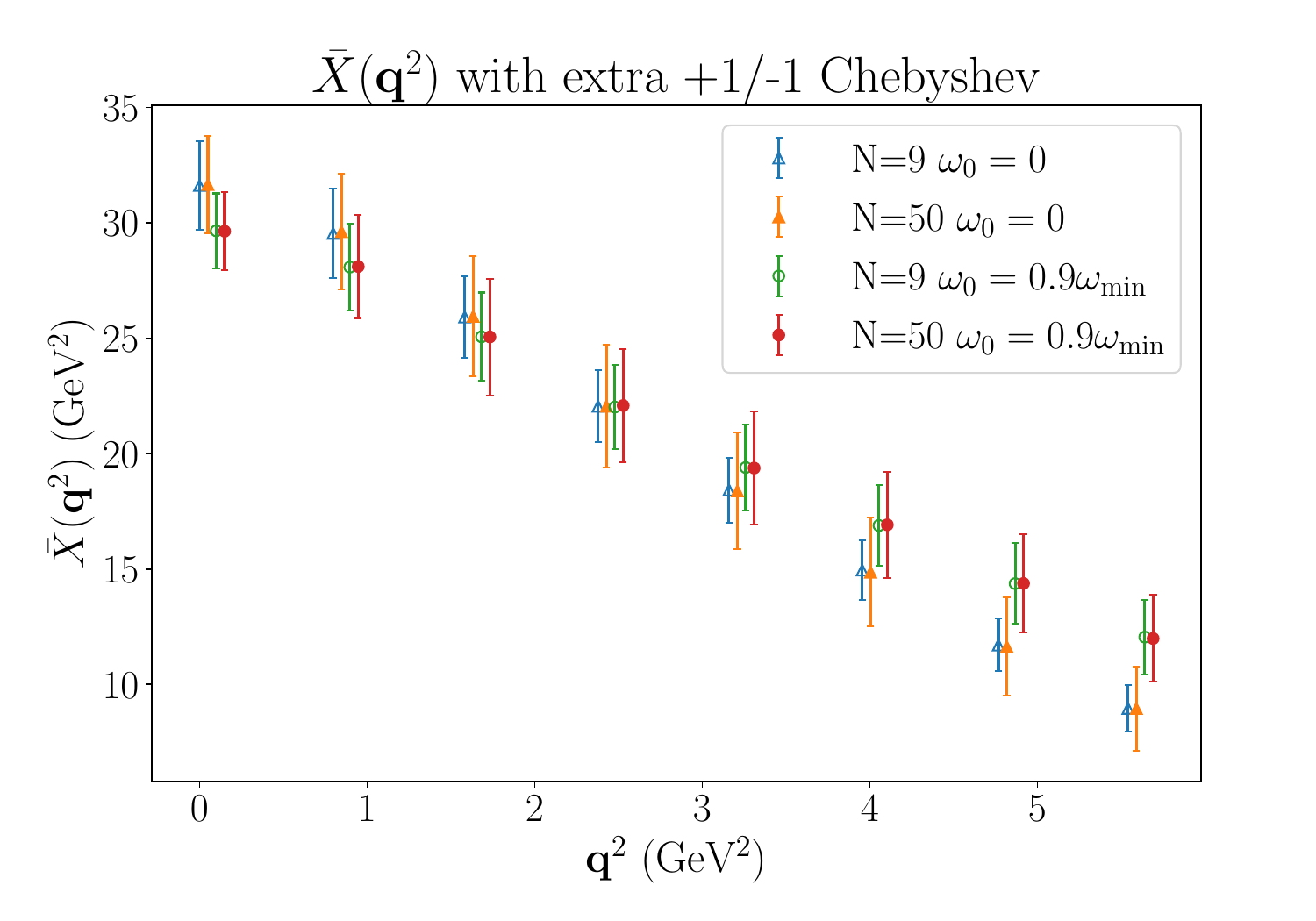}
 \caption{Saturation of Chebyshev polynomial approach, where $N=9$ is the reference case, and for $N=50$ higher-order terms are sampled from a $\mathbb{Z}_2$ distribution.}
 \label{fig:cheb_unif}
\end{figure}

In order to estimate higher-order contributions, which are not constrained by our data, we study how the results change after adding more terms in
the Chebyshev distributions on top of the $N=9$ available. In 
this way we obtain an estimate
of the approximation up to $N=50$, as in \reffig{fig:cheb_unif}.
We show in particular the case of distributions with random values in $\mathbb{Z}_2=\{-1,1\}$ for $\langle \tilde{T}_k\rangle_{\mu\nu}$ beyond $k=9$; the case
with uniform distribution with values in $[-1,+1]$ gives similar results with slightly smaller errors.
In both cases, the extra terms contribute to the final error only mildly: these observations suggest that
the results obtained do not suffer from huge systematic error from the polynomial approximation. A more complete study is however required 
for a reliable estimate of the underlying systematic effects.

\begin{figure}[h!]
 \centering
\hbox{
\includegraphics[scale=0.3]{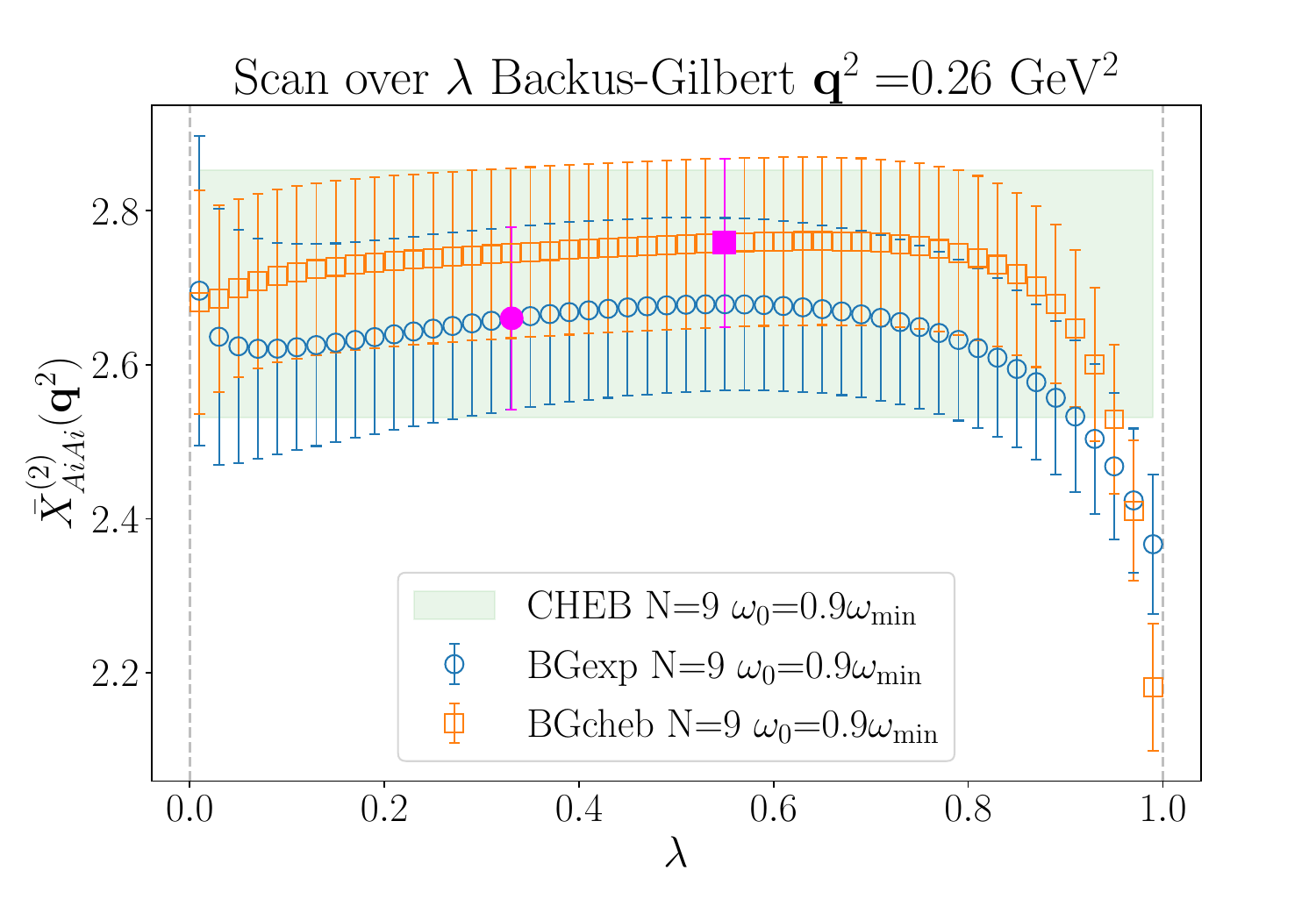}
\hspace{-0.5 cm}
\includegraphics[scale=0.3]{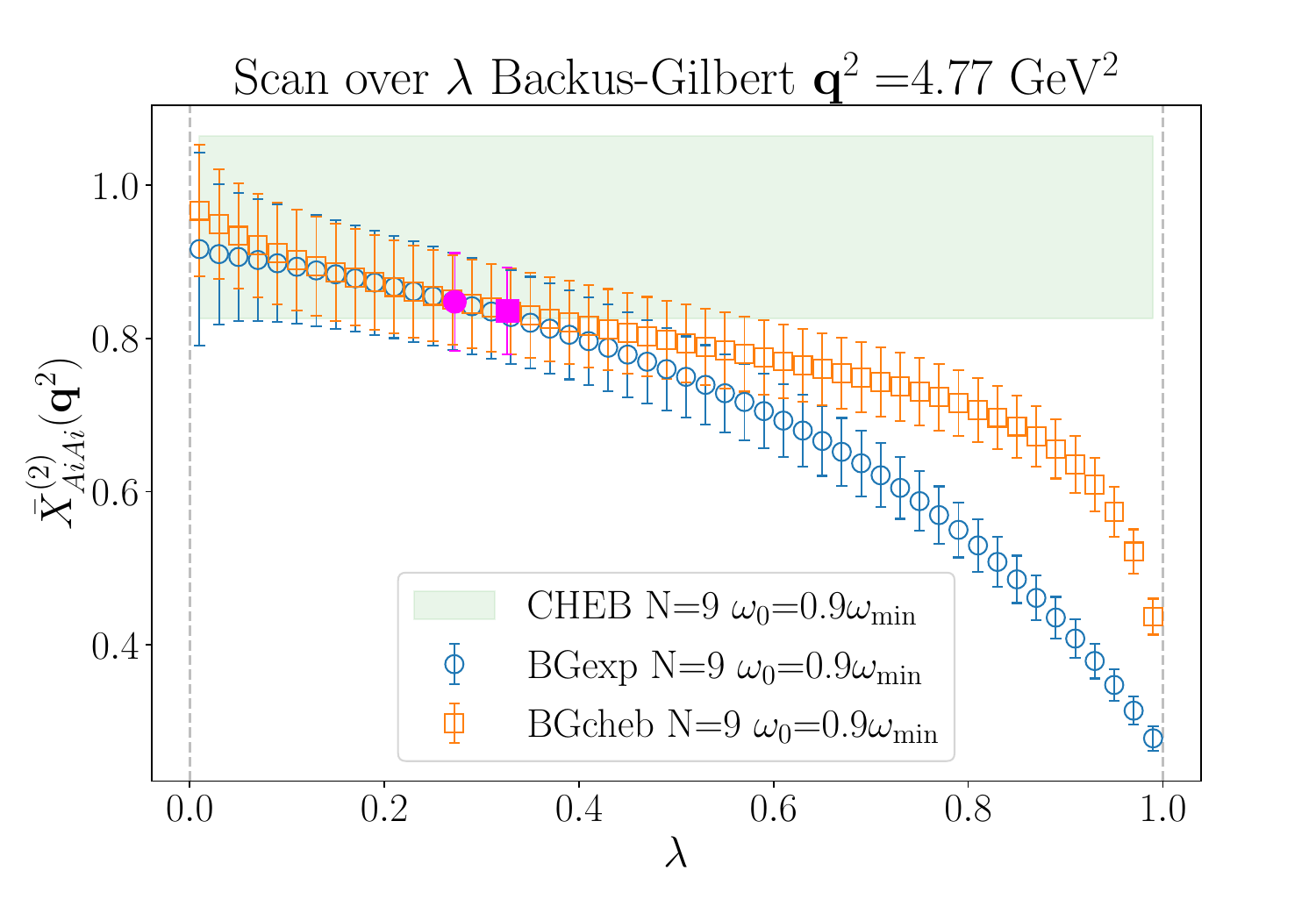}
}
\caption{Scan over $\lambda$ for $\bm{q}^2=0.26 \,\text{GeV}^2$ (left) and $\bm{q}^2=4.77 \,\text{GeV}^2$ (right) for the Backus-Gilbert method
with exponential and Chebyshev basis for $\bar{X}^{(2)}_{A_i A_i}$ with $\omega_0=0.9\,\omega_{\rm min}$.
The green shaded line is the reference from the Chebyshev; the magenta points correspond to the choice of $\lambda^*$.
Note that the points $\lambda=0$ and $\lambda=1$ (vertical grey dashed lines) are not included in this plot.}
\label{fig:BGscan}
\end{figure}

Concerning the Backus-Gilbert method, we investigate the stability around the chosen value of $\lambda^{*}$, obtained with the prescription of
\refsec{sec:analysis-BG}. We focus in particular on the channel $\bar{X}_{A_iA_i}^{(2)}$ as it is the one responsible for the largest contribution. The plot is shown in \reffig{fig:BGscan}.
We can see that for small $\bm{q}^2$ the value of $\bar{X}(\bm{q})$ is stable, which implies that statistical and systematic
errors are well balanced. For larger $\bm{q}^2$ the situation is more delicate: this can be understood in terms of the reduced phase space in $\omega$,
as shown for example in \reffig{fig:kernelBG}. 
A first attempt at mitigating the induced systematic effect could be to identify the region where the two Backus-Gilbert approaches with different bases are consistent,
to identify (where possible) a plateau, and to
estimate a value inside such region.
In the r.h.s. plot of Fig.~\ref{fig:BGscan} we see, however, that
this is not always the case: there is no clear plateau region for $\lambda$. Interestingly, the statistical error of the Chebyshev approach turns out more conservative in this case, and compatible with the result one would obtain from Backus-Gilbert.
More generally, apart from the absence of a plateau region in some cases, both choices of polynomial basis are consistent between themselves and with the Chebyshev-polynomial approach.

Coming back to the decay rate, to extract the final result we perform a polynomial fit of degree two
on $\bar{X}^{(l)}(\bm{q}^2)/ (\sqrt{\bm{q}^2})^{2-l}$. The final result is then obtained integrating these results in the physical range in $\bm{q}$.
Since this is a qualitative study, we don't report any final number; however,
the result obtained here seems to be in the right ballpark if compared with the $B_s$ meson decay rate.
Furthermore, all the approaches give compatible results, and the final statistical error is of order $5\%$.

\begin{figure}[t!]
 \centering
 \includegraphics[scale=0.4]{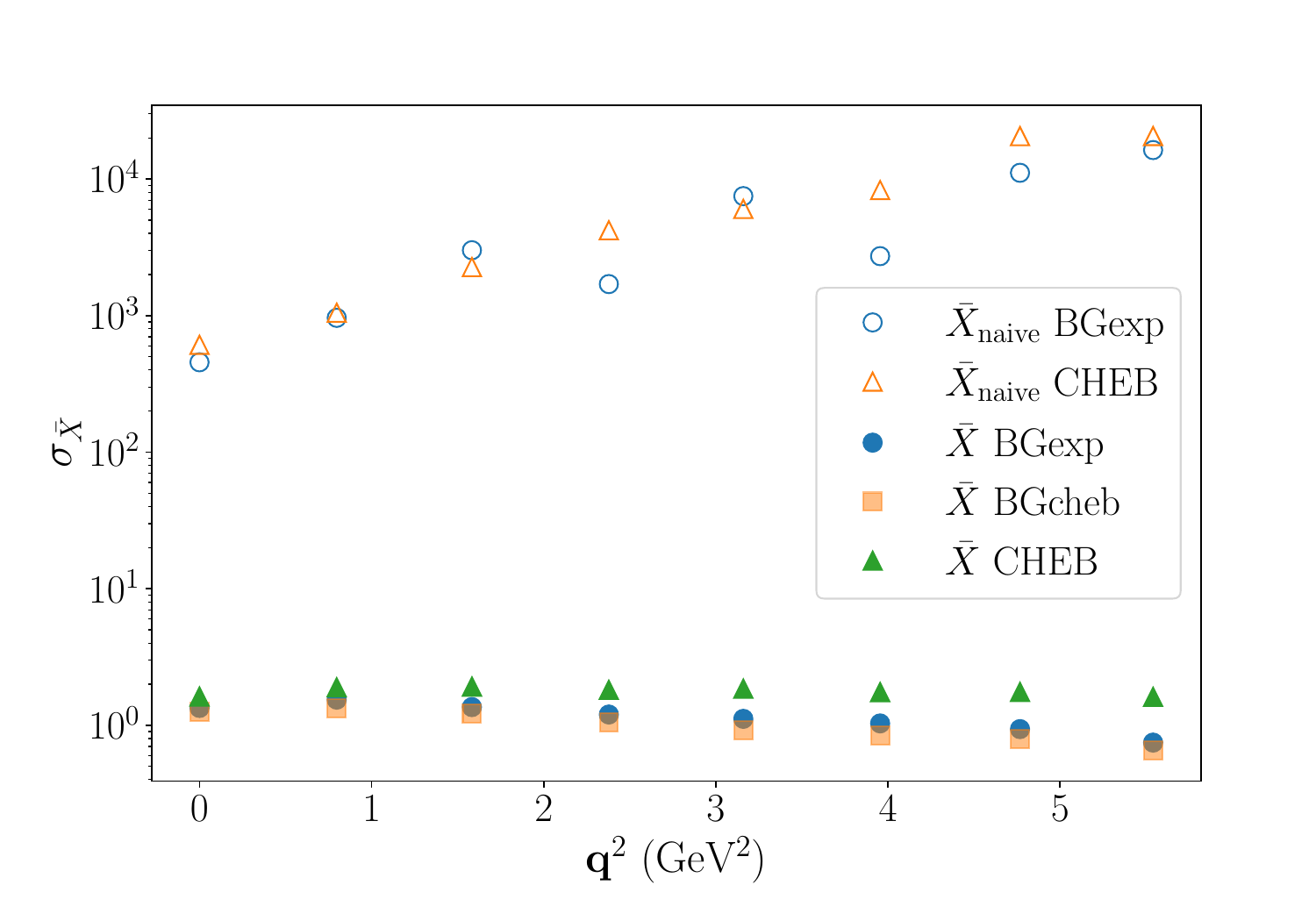}
 \caption{Effect of the variance reduction to $\bar{X}_{\rm naive}(\bm{q}^2)$ from the correction $\delta \bar{X}(\bm{q}^2)$ as in \eqrefeq{eq:XbarNaive} for
 $\omega_0=0.9 \omega_{\rm min}$ and $N=9$. The $y$ axis
 shows the standard deviation $\sigma_{\bar{X}}$ for $\bar{X}_{\rm naive}(\bm{q}^2)$ (empty symbols)
 and $\bar{X}(\bm{q}^2)=\bar{X}_{\rm naive}(\bm{q}^2) + \delta \bar{X}(\bm{q}^2)$ (filled symbols).}
 \label{fig:deltaXbar}
\end{figure}

We now address similarities and differences between the two approaches. The calculation of $\bar{X}(\bm{q}^2)$ aims to improve accuracy
by combining the naive polynomial approximation with a correction term $\delta \bar{X}(\bm{q}^2)$ that accounts for variance reduction, i.e.,
\begin{align}
 \bar{X}(\bm{q}^2) = \bar{X}^{\rm naive}(\bm{q}^2) + \delta \bar{X}(\bm{q}^2) \, ,
 \label{eq:XbarNaive}
\end{align}
where $\bar{X}^{\rm naive}(\bm{q}^2)$ would correspond to \eqref{eq:Xl_approx}.
The correction term is specific to the adopted strategy and is given by:
\begin{itemize}
 \item $\delta \bar{X}^{\rm CHEB}(\bm{q}^2) = C_{\mu\nu}(2t_0) \sum_{k=0}^{N}\bar{c}_{\mu\nu,k} \delta \bar{C}_{\mu\nu}(k)$,
 for the Chebyshev polynomials technique, where $\delta \bar{C}_{\mu\nu}(k) = \bar{C}_{\mu\nu}(k)-\bar{C}_{\mu\nu}^{\rm fit}(k)$;
 \item $\delta \bar{X}^{\rm BG}(\bm{q}^2) = C_{\mu\nu}(2t_0) \sum_{k=0}^{N} \epsilon^{*}_{\mu\nu,k} \bar{C}_{\mu\nu}(k)$,
 for the Backus-Gilbert method, which corrects the coefficients of the  polynomial approximation as in \eqref{eq:kernelBG_epsilon}.
\end{itemize}
In both cases, $\delta \bar{X}(\bm{q}^2)$ can be interpreted as a noisy zero that does not impact the naive calculation but helps with variance reduction. This is represented
in \reffig{fig:deltaXbar}, which shows  the statistical error
on $\bar X$ with and without the correction term. The reduction in statistical error is substantial. Additionally, the magnitude of the correction varies depending on $\omega_0$, where larger values result in a greater increase in $|\delta \bar{X}(\bm{q}^2)|$ as $\bm{q}^2$ increases.

\begin{figure}[t!]
 \centering
  \includegraphics[scale=0.4]{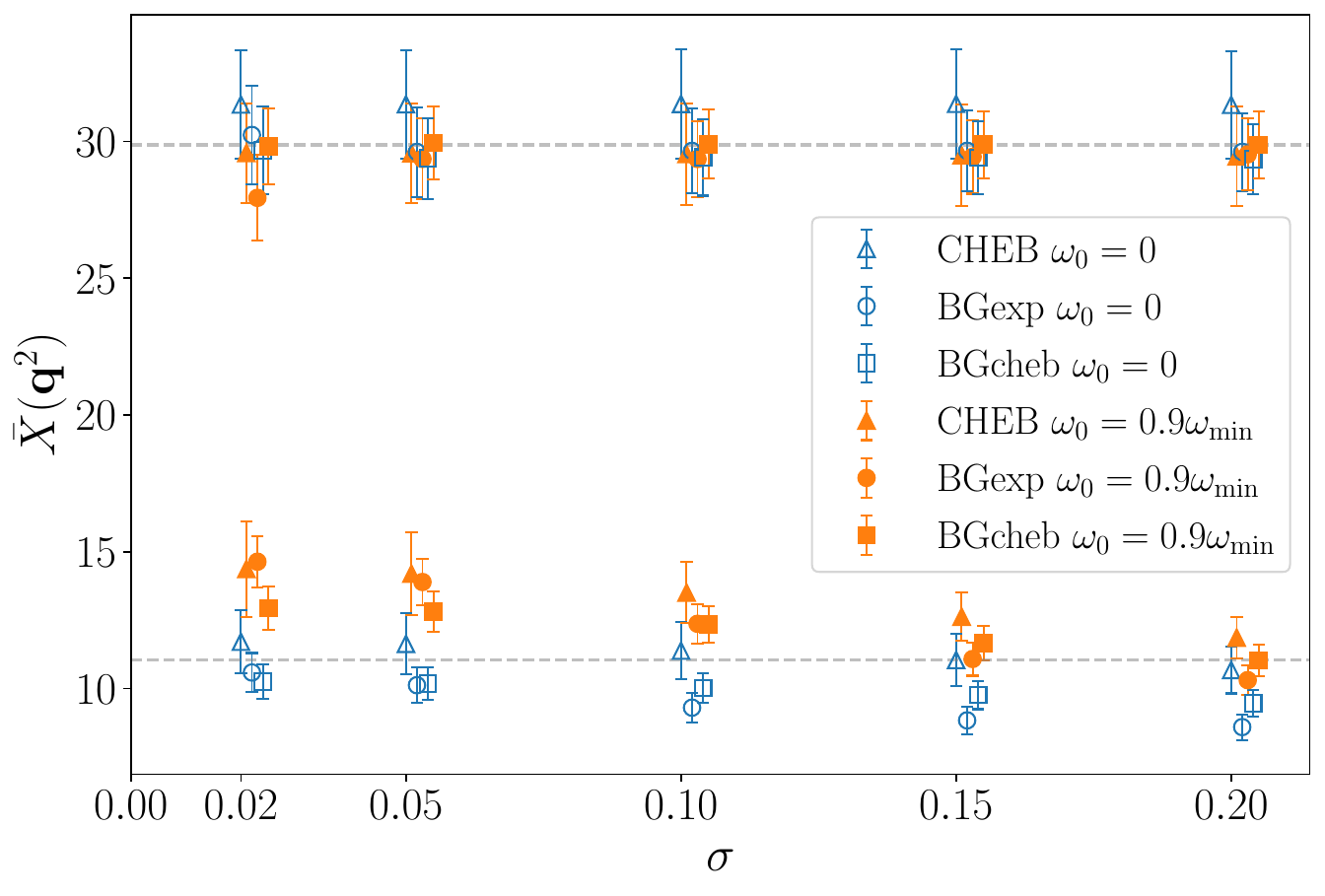}
 \caption{Dependence on the smearing parameter $\sigma$ of $\bar{X}(\bm{q}^2)$ for all the approaches at $N=9$ in the case $\bm{q}^2=0.26\, \text{GeV}^2$ (top) and $\bm{q}^2=4.77\, \text{GeV}^2$ (bottom). The horizontal dashed lines correspond to the central values for the choice $\sigma=0.02$ with $\omega_0=0.9\omega_{\rm min}$.}
 \label{fig:sigma_scan}
\end{figure}

To conclude this section we discuss some of the aspects we neglected for the purpose of this study. In particular, all the results presented here have been 
obtained with kernels smeared by a sigmoid with a fixed $\sigma=0.02$.
Eventually however, one will first have to first take the infinite-volume and continuum limits, followed by an extrapolation to $\sigma \rightarrow 0$. 
Exemplarily though, we show the $\sigma$ dependence at finite lattice spacing and volume in \reffig{fig:sigma_scan}. There, one sees that for our setup and statistical precision the dependence on $\sigma$ is mild. There is an indication that it might be more pronounced for  
larger $\bm{q}^2$.
We argue that here the extrapolation in $\sigma$ is quite delicate and could lead to misleading results. Indeed, increasing values of sigma would result in
kernel functions quite different from the target ones; on the other side, differences in small values of $\sigma$ will not be captured by a polynomial approximation with small value of $N$, as small deviations would be noticeable only for higher degrees of approximations.

\subsection{The inclusive decay rate in the ground-state limit}
\label{sec:results-ground}

\begin{figure}[t!]
 \centering
 \includegraphics[scale=0.4]{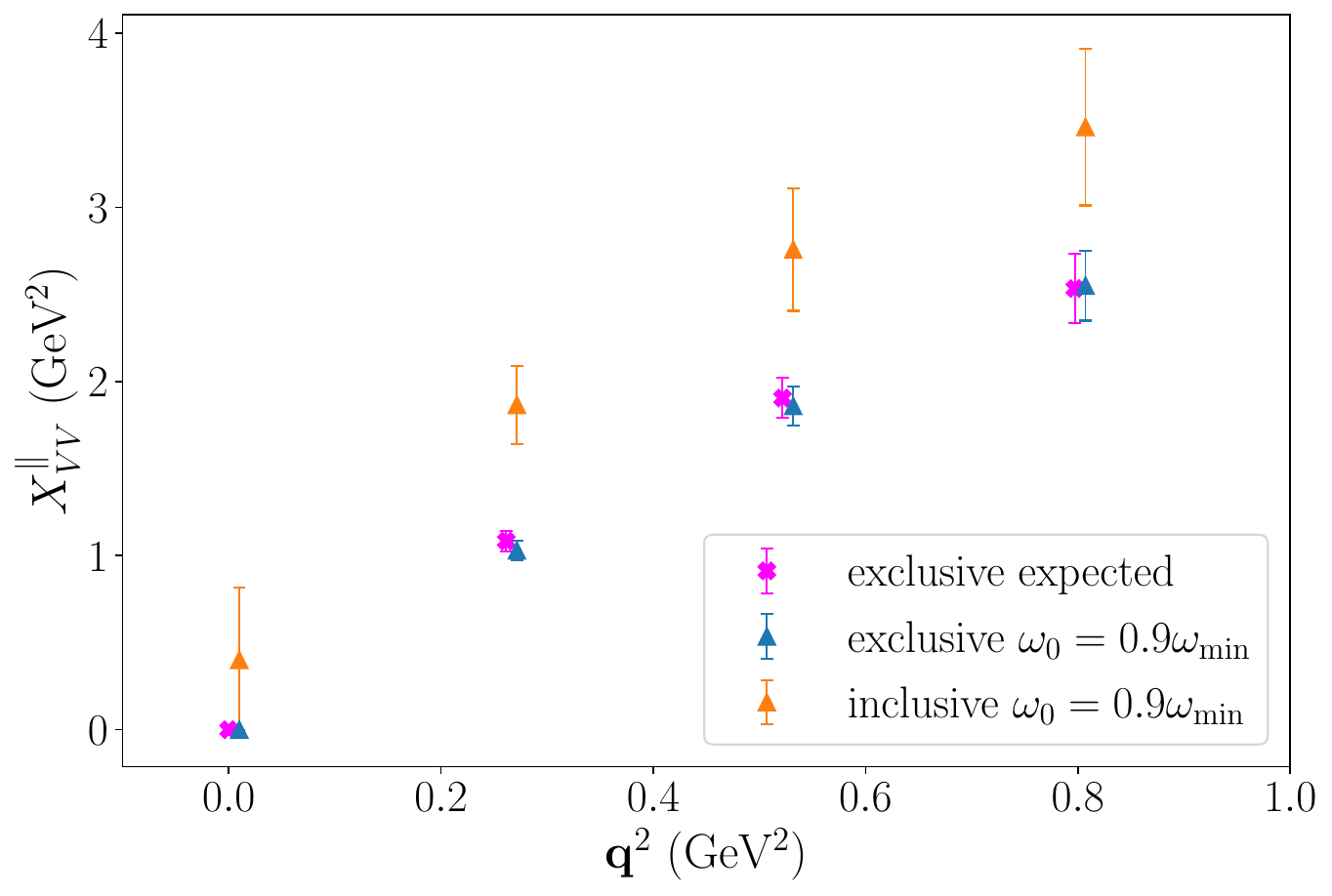}
 \caption{Ground-state limit. The ``exclusive'' labels refer to the data built from the three-point correlators as in \eqrefeq{eq:CJJ_ground},
 whereas the ``inclusive'' label refers to the full inclusive data 
 analysis starting from the four-point correlation functions.
 The analysis has been performed using the Chebyshev approach.}
 \label{fig:ground_state}
\end{figure}

We now study the ground-state limit of the inclusive approach as discussed in \refsec{sec:theory-ground},
which provides for a cross-check of the inclusive-decay analysis
strategies. The four-point function representing
the ground state can be constructed with input from lattice data for the exclusive decay $B_s \rightarrow D_s l \nu_l$.
In particular, restricting the discussion to the vector channel $VV$, the ground-state correlator
\begin{align}
 C^{G}_{\mu\nu}(t) = \frac{1}{4M_{B_s}E_{D_s}}
 \langle B_s | {{V}}_{\mu}^{\dagger} | D_s\rangle \langle D_s | {{V}}_{\nu} | B_s\rangle e^{-E_{D_s}t} \, ,
 \label{eq:CJJ_ground}
\end{align} 
can be constructed from lattice data for the ratio of three-point and two-point functions
\begin{align}
 R_{B_s D_s, \mu}(t; \bm{q}) = \sqrt{4M_{B_s} E_{D_s}} \sqrt{\frac{C^{SS}_{B_sD_s,\mu}(\bm{q}, t_{\rm snk}, t, t_{\rm src})C^{SS}_{D_sB_s,\mu}(\bm{q}, t_{\rm snk}, t, t_{\rm src})}
 {C^{SS}_{B_s}(t_{\rm snk}, t_{\rm src})C^{SS}_{D_s}(\bm{q}, t_{\rm snk}, t_{\rm src})}} \, ,\label{eq:Rfp}
\end{align}
which converges to $\mathcal{M}_{\mu} \equiv \langle D_s | V_{\mu} | B_s\rangle$ for $t \gg t_{\rm src} $ and $t<t_{\rm snk}$. 
The matrix element can be decomposed into form factors
\begin{align}
 \mathcal{M}_{\mu} = f_{+}(q^2)(p_{B_s}+p_{D_s})_{\mu} + f_{-}(q^2)(p_{B_s}-p_{D_s})_{\mu} \, .
\end{align}
Recalling that we assume $\bm{p}_{B_s}=\bm{0}$, we then extract $f_+ (q^2)$ from a constant fit to the combination
\begin{align}
  R_{f_+}(t; \bm{q}) \stackrel{\bm{q}\neq \bm{0}}{=}
   \frac{1}{2M_{B_s}} \left( R_{B_s D_s,0}(t; \bm{q}) + (M_{B_s}-E_{D_s}) \frac{\sum_{i=1}^{3} R_{B_s D_s,i}(t; \bm{q})}{\sum_{i=1}^{3} q_i}  \right) \, ,
\end{align}
which converges to $f_+(q^2)$ as $R_{B_s D_s, \mu}(t; \bm{q}) \,\rightarrow\, \mathcal{M}_{\mu}$.
We consider only the three smaller momenta to test the approach, as the signal-to-noise deteriorates rapidly with larger $\bm{q}^2$.

The result of the inclusive analysis for the channel $\bar{X}_{VV}^{\parallel}$ is reported in \reffig{fig:ground_state}. In particular, we compare the expected value
\eqref{eq:XVVpar} from the extracted values of $f_+ (q^2)$
with the inclusive analysis performed using the mock data $C_{\mu\nu}^{G}$ and the real data $C_{\mu\nu}$.
Note that for the mock data the normalised correlator corresponds simply to $\bar{C}_{\mu\nu}^{G}(t) = e^{-E_{D_s}t}$ by construction.

We find excellent agreement between the results from the conventional analysis for exclusive decay on the one side, and the one based on ground-state saturation, but using the full analysis chain adopted for the inclusive decay, on the other side.
This provides a strong test of the analysis method for
inclusive decay discussed in this paper.
The results for the full inclusive decay on the other hand
differ significantly from the exclusive case: 
while future studies will have to establish to which extend this 
could be down to systematics like finite-volume or cutoff effects, the magnitude of the effect makes appear likely to be to a large part due to contributions from the tower of finite states contributing
to the inclusive decay.
In particular, the deviation is expected to be larger for smaller $\bm{q}^2$, as the available phase space in $\omega$ is larger and may include more excited states.

\section{Conclusions and outlook}
\label{sec:concl}

In this work, we have presented a full and flexible setup for studying inclusive semileptonic decays in lattice QCD, focusing in particular on $B_{(s)}$ mesons.
We incorporate and compare Chebyshev polynomials and the Backus-Gilbert method,
both of which enable efficient and accurate calculations of the total decay rate. In particular, we improved the Chebyshev polynomial technique through the use of a
generic set of shifted polynomials in $e^{-\omega}$, and we refined the statistical analysis with a bootstrap method, fully accounting for the bounds $[-1,1]$.
We also showed how the result depends on the number of Chebyshev matrix elements and presented a possible way to take the limit $N\rightarrow\infty$ to address the systematics
associated with the polynomial approximation.
On the Backus-Gilbert side, we introduced a generalisation of the method of \cite{Hansen2019} to allow for the use of arbitrary bases of polynomials.

The two methods have been shown to be compatible, and the final results for the decay rate are in agreement.
We compared how the two techniques deal with the variance reduction of the final observable: the Chebyshev polynomials'
approach relies on trading the data with Chebyshev matrix elements that fully account for the bounds, whereas the Backus-Gilbert method
achieves the same goal by modifying the coefficients of the polynomial approximation to reduce the statistical error.
We also studied the ground-state limit, which offered a cross-check of the inclusive analysis technique and outlined the effect of excited states in the inclusive
decay with respect the corresponding exclusive process $B_s \rightarrow D_s \, l\nu_{l}$.

Overall, our work provides a solid foundation for future studies with these techniques.
However, there are still several areas that require further investigation,
including systematic errors associated with the polynomial approximation, finite-volume effects, discretisation errors, and the continuum limit.
We intend to address these issues in future works, repeating the computations on more ensembles
and also addressing similar processes involving $D_{s}$ mesons, which offer a more controlled environment.
Additionally, we plan to explore alternative observables such as hadronic and lepton moments to compare with experimental data
and to gain a deeper understanding of the ground state limit, which may provide useful insight on the physics contributing in such processes.


\acknowledgments

This work used the DiRAC Extreme Scaling service at the University of Edinburgh, operated by the Edinburgh Parallel Computing Centre on behalf of the STFC DiRAC HPC Facility (www.dirac.ac.uk). This equipment was funded by BEIS capital funding via STFC capital grant ST/R00238X/1 and STFC DiRAC Operations grant ST/R001006/1. DiRAC is part of the National e-Infrastructure.
A.B. is a JSPS International Research Fellows and received funding from the "JSPS Postdoctoral Fellowship for Research in Japan (Short-term)"
and is supported by the Mayflower scholarship in the School of Physics and Astronomy of the University of Southampton. 
The work of S.H. and T.K. is supported in part by JSPS KAKENHI Grant Number
22H00138 and 21H01085 respectively and by the Post-K and Fugaku supercomputer project through the
Joint Institute for Computational Fundamental Science (JICFuS).


\appendix
\section{Chebyshev polynomials}
\label{sec:app_cheb}

We summarise here important properties of the standard Chebyshev polynomials relevant for this work and in particular the generalisation for the shifted version
extensively used in the analysis. We refer to other sources \cite{RevModPhys.78.275} for more details.

\subsection{Standard polynomials}

The standard Chebyshev polynomials of the first kind are defined as
\begin{align}
 T_{k}: [-1,1] \ra [-1,1] \, , \quad T_{k}(x)=\cos\left(k\cos^{-1}(x)\right) \, ,  \quad k \in \mathbb{N} \, .
\end{align}
They are orthogonal with respect the scalar product
\begin{align}
 \int_{-1}^{1} T_{r}(x)T_{s}(x)\Omega(x) \, \dd x = \frac{\pi}{2}\delta_{rs} \, \left(1+ \delta_{r0}\right) \, ,
 \label{eq:chebOrthog}
\end{align}
where $\Omega(x)= 1/\sqrt{1-x^2}$ is a weight function. 
Their polynomial expansion in $x^{k}$ is given by
\begin{align}
 T_n(x) = \sum_{k=0}^{n} t^{(n)}_{k} x^{k} \, ,
\end{align}
with
\begin{align}
\begin{split}
 \begin{dcases*}
 t^{(n)}_0 =   (-1)^{n/2} & if $n$ even   \\
 t^{(n)}_k = 0  & if $n-k$  odd \\
 t^{(n)}_k = (-1)^{(n-k)/2} \, 2^{k-1} \frac{n}{\frac{n+k}{2}} \binom{\frac{n+k}{2}}{\frac{n-k}{2}}  & if  $k\neq 0$ and $n-k$  even   
 \end{dcases*} \, .
 \end{split}
 \label{eq:tn_standard}
\end{align}
A useful property involves the representation of $x^n$ in terms of the standard Chebyshev polynomial
\begin{align}
 p_n(x) \equiv x^{n} = 2^{1-n} \sideset{}{'} \sum_{\substack{k=0 \\n-k\, \text{even}}}^{n} \binom{n}{\frac{n-k}{2}} T_{k}(x) \, ,
\end{align}
where the prime indicates that the first term is halved.

\subsubsection{Expansion in Chebyshev polynomials}

Chebyshev polynomials provide the best approximation of the function  $f: [-1,1] \ra \mathbb{R}$
to any given order $N$ in terms of the L$_{\infty}$-norm. In other words, the $minmax$ error, i.e.
the maximum difference between
the target function and the reconstructed one, is minimised. In particular, for the functions considered in this work,
it is guaranteed that the Chebyshev approximation converges when $N\ra \infty$.
The polynomial approximation reads
\begin{align}
 f(x) \simeq \frac{1}{2}c_{0} T_{0}(x) + \sum_{k=1}^{N} \, c_{k} T_{k}(x) \, , \quad c_{k} = \frac{2}{\pi}\int_{-1}^{1} \dd x \, f(x) T_k(x)\Omega(x) \, ,
\end{align}
where we recall that $T_0(x)=1$ by definition. The coefficients are given by the projection of the target function $f$ on the basis
of Chebyshev polynomials.

\subsection{Shifted Chebyshev polynomials}\label{app:Shifted Chebysehv polynomials}

In general, for the purpose of this work we consider generic functions $f(x)$ defined in an interval $[a, b]$,
which we want to approximate with Chebyshev polynomials in $e^{-x}$.
To this end we can define shifted polynomials $\tilde{T}_{n}(x)$ with $x \in [a, b]$, such that their domain matches the one of the target function.
The relation to the standard polynomials is given by
\begin{align}
 \tilde{T}_{k}(x) = T_k (h(x)), \quad
\end{align}
where $h: [a, b] \rightarrow [-1,1]$ is an invertible function that maps the new domain into the domain of the standard Chebyshev polynomials,
\begin{align}
 h(x) = A e^{-x} + B \, .
\end{align}
The coefficients $A$ and $B$ can be determined by imposing $h(a)=-1$ and $h(b)=+1$, for which one obtains
\begin{align}
  A = - \frac{2}{e^{-a}-e^{-b}} \, , \qquad
  B = \frac{e^{-a}-e^{-b}}{e^{-a}+e^{-b}} \, .
  \label{eq:AB_chebCoeff_general}
\end{align}
The orthogonality relation for the shifted polynomials reads
\begin{equation}
 \int_{a}^{b} \dd x \,  \tilde{T}_{r}(x) \tilde{T}_{s}(x) \Omega_h(x) = 
 \int_{a}^{b} \dd x \, T_{r}(h(x)) T_{s}(h(x)) \Omega_h(x) 
\end{equation}
where $\Omega_h(x)$ is the new weight for the shifted $\tilde{T}_k$, which depends on the map $h$.
To show that this recover the original integral in \eqrefeq{eq:chebOrthog},
we set $x = h^{-1}(y)$ and $\dd x = \frac{1}{h'(h^{-1}(y))} \dd y$ and get
\begin{align}
 \int_{h(a)}^{h(b)} \dd y \,  T_{r}(y) T_{s}(y)   \frac{\Omega_h(h^{-1}(y))}{h'(h^{-1}(y))} \, ;
\end{align}
choosing then
\begin{align}
 \Omega_h(x) = \Omega(h(x)) |h'(x)| \, ,
\end{align}
we finally obtain
\begin{align}
 \int_{a}^{b} \dd x \,  \tilde{T}_{r}(x) \tilde{T}_{s}(x) \Omega_h(x) = \int_{-1}^{1} \dd y \,  T_{r}(y) T_{s}(y) \Omega(y) \, .
 \label{eq:cheb_scalar_product}
\end{align}
We can also generalise the polynomial expressions and their properties.
The polynomial representation reads
\begin{align}
 \tilde{T}_n(x) = \sum_{j=0}^{n} t^{(n)}_{j} h(x)^{j} = \sum_{j=0}^{n} t^{(n)}_{j} (A e^{-x}+B)^{j} = 
 \sum_{j=0}^{n} t^{(n)}_{j} \sum_{k=0}^{j} \binom{j}{k} A^{k}B^{j-k} e^{-kx}  \, .
\end{align}
We can expand this sum explicitly and re-sum it in order to isolate the coefficients of $e^{-kx}$.
We obtain
\begin{align}
 \tilde{T}_n(x) = \sum_{k=0}^{n} \tilde{t}^{(n)}_k e^{-kx} \, , \qquad
 \tilde{t}^{(n)}_k = A^{k}\sum_{j=k}^{n} \binom{j}{k} t^{(n)}_{j} B^{j-k} =
 \left( \frac{A}{B}\right)^{k}\sum_{j=k}^{n} \binom{j}{k} t^{(n)}_{j} B^{j} \, .
 \label{eq:tilde_tn}
\end{align}
In a similar way we can generalise the power representation as
\begin{align}
 \tilde{p}_n(x) \equiv h(x)^{n} = 2^{1-n} \sideset{}{'} \sum_{\substack{j=0 \\n-j\, \text{even}}}^{n} \binom{n}{\frac{n-j}{2}} \tilde{T}_{j}(x)  \,  , \quad x\in[a,b] \, .
\end{align}
Using
\begin{align}
 \tilde{p}_n(x) = (Ae^{-x}+B)^n = \sum_{k=0}^n \binom{n}{k} A^{k}B^{n-k} e^{-kx} \, ,
\end{align}
and starting from $\tilde{p}_0=1$ we can work out iteratively the general expression for $e^{-nx}$ as
\begin{align}
 e^{-nx} = \frac{1}{A^n}\left[ \tilde{p}_n(x) - \sum_{k=0}^{n-1} \binom{n}{k}A^{k}B^{n-k} e^{-kx} \right] \, .
\end{align}
We can finally collect the numerical coefficients and rewrite everything in terms of the shifted Chebyshev polynomials as
\begin{equation}
 e^{-nx} = \sum_{j=0}^{n} \tilde{a}_j^{(n)} \tilde{T}_{j}(x)  \,.
 \label{eq:e_to_n}
\end{equation}
The set of coefficients $\{ \tilde{a}_0^{(n)}, \tilde{a}_1^{(n)}, ..., \tilde{a}_n^{(n)} \}$ can be easily found numerically for each value of $n$.

\subsubsection{Expansion in Chebyshev polynomials with exponential map}

We have now all the elements necessary to proceed with the polynomial approximation of a generic function in $e^{-x}$. For the purpose of this work we will
restrict ourselves to the case $f:[x_0, \infty)\ra \mathbb{R}$. In particular, the
approximation is now
\begin{equation}
 f(x) = \frac{1}{2}\tilde{c_0} \tilde{T}_{0}(x)  + \sum_{k=1}^{N} \, \tilde{c_k} \tilde{T}_{k}(x) \, ,\quad
 \tilde{c}_k = \frac{2}{\pi}\int_{\omega_0}^{\infty} \dd x \, f(x) \tilde{T}_{k}(x) \Omega_{h}(x) \, .
\end{equation} 
The coefficients can be rewritten more explicitly as
\begin{align}
 \tilde{c}_k =  \frac{2}{\pi}\int_{0}^{\pi} \dd \theta \, f(h^{-1}(\cos\theta)) (\cos k\theta) 
 =  \frac{2}{\pi} \int_{0}^{\pi} \dd \theta\, f\left(-\ln\left( \frac{\cos\theta-B}{A} \right)\right) \cos (k\theta) \, .
 \label{eq:chebyshev_ck_projection}
\end{align}
The last equality follows from setting $y = h(x)$ and inverting
\begin{align}
  x=h^{-1}(y) = - \log \left( \frac{y-B}{A} \right) \, .
\end{align}
In this case, the coeffients $A$ and $B$ are given by 
\begin{align}
 A=-2e^{x_0} \, , \quad B=1 \, .
 \label{eq:MAPcoefficients}
\end{align}

\subsubsection{Matrix relations}

In this subsection we illustrate some useful properties that arise when setting $x_0 \neq 0$, assuming the domain of the target function
in $[x_0, \infty)$. We then explicitly consider
$B=1$ and $A=-2 e^{-x_0}$ to simplify the treatment, but the following discussion can be generalized trivially.
We start expressing \eqref{eq:tilde_tn} in matrix notation
\begin{equation}
 \begin{pmatrix}
  \tilde{T}_0(x) \\ \tilde{T}_1(x) \\ \vdots \\ \vdots \\ \tilde{T}_n(x) 
 \end{pmatrix} = 
 \begin{pmatrix}
  \tilde{t}_0^{(0)} & 0 & \cdots & \cdots & 0 \\
  \tilde{t}_0^{(1)} & \tilde{t}_1^{(1)} & 0 & \cdots &0 & \\
  \vdots & \vdots & \ddots & \ddots & \vdots  \\
  \vdots & \vdots &  & \ddots & 0 \\
  \tilde{t}_0^{(n)} & \tilde{t}_1^{(n)} & \cdots & \cdots & \tilde{t}_n^{(n)}
 \end{pmatrix}
 \begin{pmatrix}
  1 \\ e^{-x} \\ \vdots \\ \vdots \\ e^{-nx}
 \end{pmatrix} \, ,
\label{eq:matrix_eq_Tn}
\end{equation}
and \eqref{eq:e_to_n}  as
\begin{equation}
 \begin{pmatrix}
  1 \\ e^{-x} \\ \vdots \\ \vdots \\ e^{-nx}
 \end{pmatrix} =
 \begin{pmatrix}
  \tilde{a}_0^{(0)} & 0 & \cdots & \cdots & 0 \\
  \tilde{a}_0^{(1)} & \tilde{a}_1^{(1)} & 0 & \cdots &0 & \\
  \vdots & \vdots & \ddots & \ddots & \vdots  \\
  \vdots & \vdots &  & \ddots & 0 \\
  \tilde{a}_0^{(n)} & \tilde{a}_1^{(n)} & \cdots & \cdots & \tilde{a}_n^{(n)}
 \end{pmatrix} 
 \begin{pmatrix}
  \tilde{T}_0(x) \\ \tilde{T}_1(x) \\ \vdots \\ \vdots \\ \tilde{T}_n(x) 
 \end{pmatrix} \, .
\label{eq:matrix_eq_an}
\end{equation}
It is clear that these $(n+1)\times (n+1)$ matrices
$\tilde{\bm{t}}$ with $(\tilde{\bm{t}})_{ij}=\tilde{t}^{(i)}_j$ and
$\tilde{\bm{a}}$ with $(\tilde{\bm{a}})_{ij}=\tilde{a}^{(i)}_j$
are one the inverse of the other, i.e. $\tilde{\bm{a}}=(\tilde{\bm{t}})^{-1}$ and vice versa.
From \eqref{eq:tilde_tn} we can further decompose $\tilde{\bm{t}}^{(n)}$ as
\begin{align}
 \tilde{\bm{t}} = \bm{A} \bm{P} \bm{t} \, ,
\end{align}
where $\bm{A}_{kk}=A^{k}=(-2e^{x_0})^{k}$ is a diagonal matrix, $\bm{P}_{jk}= \binom{j}{k} $ is the lower triangular Pascal matrix and 
the matrix $\bm{t}$ follows from \eqref{eq:tn_standard}. This expression makes it easy to see the effect of $x_0$: considering
$\bm{A}_{kk}\bigg|_{x_0\neq 0} = e^{x_0 k} \bm{A}_{kk}\bigg|_{x_0= 0}$ it follows that
\begin{align}
 (\tilde{\bm{t}})_{nk}\bigg|_{x_0\neq 0} = \tilde{t}^{(n)}_{k} \bigg|_{x_0\neq 0} = e^{x_0 n} \, \tilde{t}^{(n)}_{k} \bigg|_{x_0=0} \, , \qquad
 (\tilde{\bm{a}})_{nk}\bigg|_{x_0\neq 0} = \tilde{a}^{(n)}_{k} \bigg|_{x_0\neq 0} = e^{-x_0 n} \, \tilde{a}^{(n)}_{k} \bigg|_{x_0= 0} \, .
 \label{eq:app_a-t_relation}
\end{align}

\section{Generalised Backus-Gilbert}
\label{sec:app_bg}

In this appendix we reformulate and generalise the modified Backus-Gilbert approach proposed in \cite{Hansen2019,Bulava2021,ExtendedTwistedMassCollaborationETMC:2022sta}. The idea is to provide a more general framework which allows for the use of an arbitrary basis and to explore the properties and numerical advantages of different
choices.

\subsection{The method}
\label{sec:app_bg_method}

The problem we want to address is the evaluation of a generic observable $O$ of the form
\begin{align}
 O = \int_{a}^{b} \dd \omega \, \rho(\omega) K(\omega) \, ,
 \label{eq:Observable_app}
\end{align}
where $K(\omega)$ is a function we will refer to as \textit{kernel} and $\rho (\omega)$ is the spectral function related to a given correlation function
\begin{align}
   C(t) = \int_{a}^{b} \dd \omega \, \rho(\omega) e^{-\omega t} \, .
\end{align}
While typically the range of integration is $a=0$ and $b=\infty$, here we chose to leave it generic
to keep the discussion general.
The idea to address the computation is to approximate the kernel in polynomial up to some degree $N$,
i.e. $K(\omega) = \sum_{j=0}^{N} g_j e^{-\omega j}$,
such that the target observable can be estimated as
\begin{align}
 O \simeq \sum_{j=0}^{N} g_j \int_{a}^{b} \dd \omega \, \rho(\omega)e^{-\omega j} = \sum_{j=0}^{N} g_j C(j).
\end{align}
For example, a typical problem consists in the extraction of the spectral density of a correlator, in which case one would consider the kernel
to be a smoothed Dirac delta $K(\omega)= \delta_{\sigma}(\omega)$ with a finite width $\sigma$, as for example a Gaussian.

The approach consists of weighting the two functionals $A[g]$ and $B[g]$ against each other, where the first one provides a measure for
the systematic effects coming from the polynomial approximation, and the second one provides a measure for the variance $\sigma_O^2$ of the observable $O$, in particular, $B[g] = \sigma^2_O = g_i \sigma_{ij} g_j$,
where we defined $\sigma_{ij} = \Cov [C(i), C(j)]$.
This is equivalent to solving a minimisation problem with constraints.
We can then define a new functional $F_{\theta}$ as
\begin{align}
 F_{\theta}[g] = A[g] + \theta^2 B[g] \, ,
\end{align}
and determine the coefficients by variational principle $\frac{\partial F_{\theta}[g]}{\partial g_j}=0$ at different values of $\theta^2$. The value $\theta^2=0$ corresponds
to addressing exclusively the polynomial approximation, as prescribed by the choice of $A[g]$, whereas the choices $\theta^2 \rightarrow \infty$ would correspond to dealing purely 
with the variance minimisation and would result in $g_j=0$.
Note that we can map $\theta^2 = \frac{\lambda}{1-\lambda}$ for simplicity, such that $\lambda \in [0,1)$ and $\theta^2\rightarrow \infty$ for $\lambda \rightarrow 1$.
Furthermore, any relative normalisation term between the two functionals can be reabsorbed into $\theta^2$.
Depending on the choice of the basis, the coefficients $g_j$ may grow over different orders of magnitude and numerical instabilities may appear. This can be addressed in practice by using arbitrary precision arithmetic.

We now discuss in detail how to generalise the modified Backus-Gilbert \cite{Hansen2019} for a generic basis of functions, starting
from the construction of $A[g]$. Following the original paper we can generalise the L$_2$-norm of the difference between the target function
and the polynomial reconstruction using an arbitrary family of basis function $P_k(x)=\sum_{j=0}^{k}p_{j}^{(k)} x^{j}$ defined in an interval $x\in [p_-, p_+]$.
As for the Chebyshev, we will deal in general with a shifted version of this family of polynomials in $e^{-x}$
defined in a generic interval $[a, b]$
\begin{align}
 \tilde{P}_k(x)=\sum_{j=0}^{k}\tilde{p}_{j}^{(k)} e^{-jx} \, , \quad x\in [a, b] \, ,
\end{align}
where $\tilde{P}_k(x)=P_k(h(x))$ and $h(x)=Ae^{-x}+B$ is an invertible map that satisfies $h(a)=p_-$ and $h(b)=p_+$.
The interval $[a, b]$ has to match the range of integration of the observable $O$ in \eqref{eq:Observable_app}.
The functional $A[g]$ now reads
\begin{align}
 A[g] = \int_{a}^{b} \dd  \omega \,  \Omega(\omega) \left[ K(\omega) - \sum_{j=0}^{N}g_j \tilde{P}_j(\omega)  \right]^2.
\end{align}
With respect to the original version we now have introduced a generic weight $\Omega(\omega)$; note that we
start the approximation at $\tilde{P}_0(\omega)$ (as long as $\Omega(\omega)$ can be integrated in $[a,b]$).

If we consider only the $A[g]$ term, the solution of the system by variational principle is given by
\begin{align}
 \bm{A} \cdot \bm{g} = \bm{K} \quad \longleftrightarrow \quad  \bm{g} = \bm{A}^{-1} \cdot \bm{K}
 \label{eq:bg_inversion}
\end{align}
where
\begin{align}
 A_{ij} &= \int_{a}^{b} \dd \omega \, \Omega(\omega)\tilde{P}_i(\omega)\tilde{P}_j(\omega) \, , \\
 K_i &= \int_{a}^{b} \dd \omega\, \Omega(\omega)\tilde{P}_i(\omega)K(\omega) \, ,
\end{align}
and $\bm{g}$ is a vector of parameters.

With this setup, the convenient choice consists in picking a set of (shifted) orthogonal polynomials
\begin{align}
 \langle \tilde{P}_i, \tilde{P}_j\rangle=\int_{a}^{b} \dd x \, \Omega(x) \tilde{P}_{i}(x)\tilde{P}_j(x)  \propto \delta_{ij} \, ,
\end{align}
with $\Omega$ being the actual weight that defines the scalar product.
The advantage is immediately clear, as the matrix $\bm{A}$ becomes
\begin{align}
 A_{ij} = \langle \tilde{P}_i, \tilde{P}_j\rangle \, ,
\end{align}
and the coefficients are given by
\begin{align}
g_i = \frac{1}{\langle \tilde{P}_i,\tilde{P}_i\rangle} \int_{a}^{b} \dd \omega \, \Omega(\omega)\tilde{P}_i(\omega)K(\omega) \, .
\end{align}
Since the matrix $\bm{A}$ is now diagonal, the inverse required to compute
\eqrefeq{eq:bg_inversion} is analytically known.
Furthermore, the solution is now equivalent to the projection on the polynomial basis.

We can now include the $B$ term, i.e. the covariance matrix of the data.
Note that in general we now need to consider a linear combination of the correlator at different time slices according to the polynomial basis, i.e.
\begin{align}
 C^{P}(k) = \int_{a}^{b} \dd \omega \, \rho(\omega) \tilde{P}_k(\omega)
   =  \int_{a}^{b} \dd \omega \, \rho(\omega) \sum_{j=0}^{k} \tilde{p}_j^{(k)} e^{-j\omega}
   = \sum_{j=0}^{k} \tilde{p}_j^{(k)} C(j) \, ,
\end{align}
such that
\begin{align}
 B[g] = \sum_{i,j}g_i\, \sigma_{ij}^{P}\, g_j \, , \qquad \sigma_{ij}^{P}= \Cov[C^{P}(i), C^{P}(j)] \, .
\end{align}
The full functional is then
\begin{align}
 F_{\theta}[g] = A[g] + \theta^2 B[g]
\end{align}
and the final solution is
\begin{align}
 \bm{g}_{\theta} =
  \bm{F}_{\theta}^{-1} \cdot \bm{K}
\end{align}
with
\begin{align}
 \bm{F}_{\theta} = \bm{A} + \theta^2 \bm{B} \, ,
\end{align}
where $B_{ij}=\sigma_{ij}^{P}$.
If $\bm{A}$ is diagonal (and possibly proportional to the identity), the inversion of the matrix $\bm{F}_{\theta}$ may be better conditioned and possible numerical instabilities
arising from an ill-conditioned matrix $\bm{A}$ may be avoided.

On top of that we could also implement some constraints that our approximation has to fulfil. In particular, following what was done for the spectral function
in \cite{Hansen2019,Bulava2021}, we can require that the polynomial approximation preserves the (weighted) area of the target function, i.e.
\begin{align}
\int_{a}^{b} \dd \omega \, \Omega(\omega)\sum_{k=0}^{N} g_{k}\, \tilde{P}_{k}(\omega) =
\int_{a}^{b} \dd \omega \, \Omega(\omega) K(\omega) \, .
\end{align}
This can be expressed as
\begin{align}
 \bm{R}^{T}\cdot \bm{g}_{\theta} = r \, ,
\end{align}
where
\begin{align}
 R_k = \int_{a}^{b} \dd \omega \, \Omega(\omega)\tilde{P}_{k}(\omega) \, , \qquad \,
 r = \int_{a}^{b} \dd\omega \, \Omega(\omega) K(\omega) \, .
\end{align}
Taking into account these constraints, the solution becomes
\begin{align}
 \bm{g}_{\theta} = \bm{F}_{\theta}^{-1} \cdot \bm{K}
  + \bm{F}_{\theta}^{-1} \cdot \bm{R} \frac{r-\bm{R}^{T}\cdot\bm{F}_{\theta}^{-1}\cdot \bm{K}}{\bm{R}^{T} \cdot \bm{F}_{\theta}^{-1}\cdot\bm{R}} \, .
\end{align}
The final observable then reads
\begin{align}
 O_{\theta} \simeq \sum_{j=0}^{N} g_{\theta,j} C^{P}(j) \, ,
\end{align}
for a given value of $\theta$. The choice of $\theta$ is in principle arbitrary.
A common choice is to take the value $\theta^{*}$ that gives equal weight to the $A$ and $B$ functional, $A[g_{\theta^{*}}]=B[g_{\theta^{*}}]$,
i.e. an equal weight to statistical and systematic error.
For a given choice of $\theta$, it is important to make sure that the value of the final observable is stable for small changes in $\theta$, in order to make sure
that the procedure did not introduce any bias.

To conclude, note that this recovers the method first proposed in \cite{Hansen2019} if we consider the following substitutions
\begin{align*}
\begin{split}
 \tilde{P}_j(\omega) \, &\rightarrow \, e^{-(j+1)\omega} \, , \\
 \Omega(\omega) \, &\rightarrow \, 1 \, ,\\
 \theta^2 \, &\rightarrow \, \lambda / (1-\lambda)  \, ,\\
 F[g] \, &\rightarrow \, (1-\lambda) F[g] \, .
 \end{split}
\end{align*}

\subsection{A different perspective}
\label{sec:app_BG_diff_persp}

The previous reformulation in \refsec{sec:app_bg_method} allows us to rely on arbitrary polynomials for the approximation. In this general picture it is useful to consider a different
perspective to the method:
we can reduce the problem to finding a suitable correction to the optimal coefficients, i.e.
\[
   g_{j}= \gamma_j + \epsilon_j \, ,
\]
where $\gamma_j$ are the coefficients of the polynomial approximation coming purely from the functional $A[\gamma]$, i.e. $\bm{\gamma} = \bm{A}^{-1}\bm{K}$ as
in \eqrefeq{eq:bg_inversion}, and $\epsilon_j$ a correction that takes into account the data.
We can then rewrite the functional as
\begin{align}
 F_{\theta}[g] = F_{\theta}[\gamma + \epsilon] = F_{\theta}[\gamma] + \delta F_{\theta}[\epsilon] \, ,
\end{align}
and explicitly
\begin{align}
\delta F_{\theta}[\epsilon] = \int_{a}^{b} \dd \omega \, \Omega(\omega) \left[ \sum_{k=0}^{N}\epsilon_k \tilde{P}_{k}(\omega) \right]^2 +
 \theta^2 \left( 2 \gamma_i \sigma^{P}_{ij} \epsilon_j + \epsilon_i \sigma^{P}_{ij} \epsilon_j \right) \, .
\end{align}
The minimisation of $\delta F_{\theta}[\epsilon]$ gives
\begin{align}
\bm{\epsilon}_{\theta} = -\theta^2 \left( \bm{A} + \theta^2 \bm{\sigma^{P}} \right)^{-1} \bm{\sigma^{P}} \bm{\gamma} \, ,
\end{align}
which is equivalent to the previous approach.
It is then clear that $\epsilon_j$ are by construction coefficients that should not modify the quality of the polynomial approximations but take care of the
reduction of the statistical noise. In practice, this will of course depend on the choice of $\theta$.

\section{Fit strategy}
\label{sec:app_fit}

We discuss the general strategy for the Bayesian fit used in the analysis. We consider only linear fits, as these are the ones directly relevant
for this work. To keep the discussion very general we consider a linear model in the form
\begin{align}
 y(\bm{p}, x) = \sum_{\alpha=1}^{M} p_{\alpha} X_{\alpha}(x) \, , \quad \bm{p} = (p_{1}, p_{2}, \cdots, p_{M}) \, ,
\end{align}
where $X_\alpha (x)$ are known coefficients (which in principle can depend on $x$) and $p_{\alpha}$ are $M$ parameters we want to determine.

\subsection{MAP with bounds}

We address the fits using Bayesian statistics, in particular using a \textit{maximum a posteriori} (MAP) probability estimate, which relies on
an \textit{augmented $\chi^2$} with Gaussian priors.
On top of that, we implement generic bounds on the parameters. The way we address this is by ``wrapping'' the parameters in a function $p_{\alpha}=f(\pi_{\alpha})$ 
which encodes the desired bounds. In this case, the fit is performed on the new parameters $\pi_{\alpha}$, and the prior is introduced accordingly. The augmented $\chi^2$ reads
\begin{align}
 \chi^2_{\rm aug} = \sum_{i,j=1}^{N} \left( y_i - \sum_{\alpha=1}^{M}f(\pi_\alpha) X_{\alpha}(x_i) \right) \Cov^{-1}_{ij} 
                                 \left( y_j - \sum_{\alpha=1}^{M}f(\pi_\alpha) X_{\alpha}(x_j) \right)
                                +\sum_{\alpha=1}^{M} \frac{(\pi_{\alpha} - \bar{\pi}_{\alpha})^2}{\bar{\sigma}^2_\alpha} \, .
\end{align}
Note that the prior distributions refer to the internal parameters $\pi_{\alpha}$ and are assumed to be Gaussians. This allows to deal with a more generic distribution for the parameters
$p_{\alpha}$, depending on the shape of the wrapping function $f$. The parameters are found as usual by 
imposing $\pdv{\chi^2_{\rm aug}}{\pi_\gamma}=0$; note that in this case the problem is no more linear due to the presence of $f$.

\subsection{MAP with bootstrap}

As outlined in the sections above, the presence of a ``wrapping'' function on the parameters $p_{\alpha}$ implies that their distribution is in general non Gaussian. This is obvious
from the fact that we assume the internal parameters $\pi_{\alpha}$ to be Gaussian and that the wrapping function implements some bounds, therefore limiting the domain of $p_{\alpha}$.
Instead of fitting the central value of the data and estimating their error from the inverse of the curvature matrix (the Hessian of the $\chi^2$ with respect to the parameters),
it is then more convenient to adopt a bootstrap approach, such that
the procedure automatically takes into account any deviation from Gaussianity. In practice, one would then fit all the bootstrap bins and reconstruct the distribution
of the parameters, treating the error accordingly.

The approach we adopt consists in assuming a normal distribution for the internal parameters $\pi \sim \mathcal{N}(\mu, \sigma)$ such that
$p_{\alpha} = f(\pi_{\alpha})$ is distributed according to our prior knowledge of the parameters.
In practice, considering a set of $N_{b}$ bootstrap bins with corresponding data $y_i^{b}$, we perform $N_b$ fits to the data where each time we use a different prior value
$\bar{\pi}^{b}_{\alpha}$ sampled from the normal distribution $\mathcal{N}(\mu, \sigma)$. This ensures that the correct prior is assumed for $p_{\alpha}$. For example, in the case
where the data contain little information and $min(\chi^2_{\rm aug})\simeq min(\chi^2_{\rm prior})$, the fit gives back the prior information we encoded by hand.

%



\providecommand{\href}[2]{#2}\begingroup\raggedright\endgroup

\end{document}